\documentclass[onecolumn,11pt]{article}
\usepackage{jheppub}
\usepackage{ifpdf}
\usepackage{graphicx}
\usepackage{amsfonts}
\usepackage{amsmath}
\usepackage{amssymb}
\usepackage{epsfig}
\usepackage{subfigure}
\usepackage{pdfpages}
\usepackage{graphicx,epstopdf}
\usepackage{caption}

\usepackage[makeroom]{cancel}

\usepackage{hyperref}
\usepackage{array}

\usepackage[utf8]{inputenc}

\allowdisplaybreaks[1]
\usepackage{amsmath,bm}
\usepackage{subfigure}
\usepackage{amssymb}
\usepackage{colortbl}
\usepackage{epsfig,bm}

\hypersetup{pdftex,colorlinks=true,allcolors=blue}
\def\pa{\partial}

\renewcommand{\(}{\left(}
\renewcommand{\)}{\right)}
\renewcommand{\[}{\left[}
\renewcommand{\]}{\right]}

\begin{document}

\title{Thermodynamically stable asymptotically flat hairy black holes with a dilaton potential: the general case}

\author[1]{Dumitru Astefanesei,}
\author[2]{Jose Luis Bl\'azquez-Salcedo,}
\author[3]{Francisco G\'omez,}
\author[1]{and Ra\'ul Rojas}

\affiliation[1]{Instituto de F\'\i sica, Pontificia Universidad Cat\'olica de Valpara\'\i so, Casilla 4059, Valpara\'{\i}so, Chile}
\affiliation[2]{Institut f\"ur  Physik, Universit\"at Oldenburg, Postfach 2503, D-26111 Oldenburg, Germany}
\affiliation[3]{Institute Lorentz of Theoretical Physics, Leiden University, 2333 CA Leiden, The Netherlands}
\emailAdd{dumitru.astefanesei@pucv.cl}
\emailAdd{jose.blazquez.salcedo@uni-oldenburg.de}
\emailAdd{f.j.gomez.serrano@umail.leidenuniv.nl}
\emailAdd{raul.rojas@pucv.cl}

\date{\today}

%\preprint{arXiv:1601.nnnnn [hep-th]}

\abstract{We extend the analysis, initiated in \cite{Astefanesei:2019mds}, of the thermodynamic stability of four-dimensional asymptotically flat hairy black holes by considering a general class of exact solutions in Einstein-Maxwell-dilaton theory with a non-trivial dilaton potential. We find that, regardless of the values of the parameters of the theory, there always exists a sub-class of hairy black holes that are thermodynamically stable and have the extremal limit well defined. This generic feature  that makes the equilibrium configurations locally stable should be related to the properties of the dilaton potential that is decaying towards the spatial infinity, but behaves as a box close to the horizon. We  prove that these thermodynamically stable solutions  are also dynamically stable under spherically symmetric perturbations.}

\maketitle

%%%%%%%%%%%%%%%%%%%%%%%%%%%%%%%%%%%%%%%%
\section{Introduction}
\label{sec:intro}
%%%%%%%%%%%%%%%%%%%%%%%%%%%%%%%%%%%%%%%%

With the discovery of black hole entropy \cite{Bekenstein:1973ur} and Hawking radiation \cite{Hawking:1974sw}, the black hole physics provides a profound connection between thermodynamics  and  gravitation \cite{Hawking:1976de}. Although black hole thermodynamics is often studied with asymptotically flat boundary conditions, this is really not suitable for discussing equilibrium configurations. It is well-known that, e.g., Schwarzschild, Reissner-Nordstr\"om, and Kerr black holes are locally thermodynamically unstable.  

An important question is, therefore, in which conditions there could exist thermodynamically stable black holes in flat spacetime? At first sight, this is not possible. To consistently put a black hole in thermal equilibrium with its Hawking radiation, we have to consider an indefinitely large reservoir of energy, which in turn implies that there is non-zero energy density out to infinity.  One expects then a behaviour rather similar to a cosmological model that contracts or expands. Indeed, for the Schwarzschild black hole the heat capacity is negative and so a thermal fluctuation can break the equilibrium and that leads to the evaporation of the black hole or its indefinite growth.  One way to circumvent this problem is to `put the black hole in a box' \cite{York:1986it}.\footnote{Another well known example is the anti-de Sitter (AdS) spacetime. In contrast with its asymptotically flat  counterpart, asymptotically AdS spacetimes are not globally hyperbolic. The conformal asymptotic boundary at infinity is timelike and, in this case, the suitable initial data must be supplemented with appropriate boundary conditions \cite{Ishibashi:2004wx}. Therefore, from a geometric standpoint, AdS spacetime behaves as a box \cite{Hawking:1982dh}.}  Due to the fact that the temperature measured locally by a static observer is blue-shifted with respect to the usual temperature that is determined at asymptotically  flat spatial infinity, the  heat  capacity  becomes  positive in a specific range of the box radius. Interestingly, there is another way to obtain thermodynamically stable asymptotically flat black holes without imposing artificial boundary conditions similar to the `box' ones proposed by York in \cite{York:1986it}. That is, one has to consider an obvious extension of the usual gravity models by considering  scalar fields with self-interaction \cite{Astefanesei:2019mds,Anabalon:2013qua,Astefanesei:2019qsg}. One of the main reasons to include this new sector in the action is coming from the fact that the scalar fields appear naturally  as moduli in string theory.  

The uniqueness theorem for the asymptotically flat, stationary black hole solutions of the Einstein-Maxwell equations is, by now, established quite rigorously. There also are several no-hair theorems in theories which couple scalar fields to gravity (see, e.g., \cite{Herdeiro:2015waa} and references therein). However, in theories with a potential for the scalar field with certain properties \cite{Sudarsky:1995zg,Bekenstein:1995un}, there exist scalar-hairy black holes. For example, in asymptotically AdS spacetimes, various exact regular scalar-hairy black holes were analyzed in \cite{Martinez:2004nb,Henneaux:2002wm,Acena:2012mr,Anabalon:2012ta,Anabalon:2013sra,Anabalon:2014fla,Anabalon:2015xvl,Feng:2013tza,Liu:2013gja,Lu:2014maa}. Interestingly, when the scalar fields are non-minimally coupled to gauge fields as in the models emerging from string theory, there exist a family of exact asymptotically flat hairy black holes \cite{Gibbons:1982ih,Gibbons:1987ps,Garfinkle:1990qj} --- in the context of no-hair theorems, the gauge field provides an effective potential for the scalar field and so the scalar field is not independent \cite{ Goldstein:2005hq} (see, also, \cite{Astefanesei:2019pfq} for a recent discussion on dilatonic versus scalarised hairy black holes), the scalar charge is fixed by the other conserved charges. 

In  this  paper  we  explore  the  thermodynamic and dynamic stability of a large class of exact asymptotically flat charged hairy black holes in a gravity model with a dilaton and its potential \cite{Anabalon:2013qua}. We find a general criterion to obtain specific regions in parameter space where these black holes are thermodynamically stable and, then, we also show explicitly that they are dynamically stable under spherically symmetric perturbations. There is some previous related work \cite{Astefanesei:2019mds, Astefanesei:2019qsg} where a particular case was studied, though this case is obtained by rescaling the metric to remove a divergent factor and so it is not a generic representative of the general class of hairy black holes presented here.

It was shown in \cite{Anabalon:2017yhv} that the dilaton potential of \cite{Anabalon:2013qua} emerges naturally in a consistent truncation of four-dimensional $N= 2$ supergravity extended with vector multiplets and deformed by a dyonic Fayet-Iliopoulos (FI) term. This potential is characterized by three independent parameters: $\nu$ is a `hairy' parameter that is related to the moduli metric of the model, $\alpha$ is related to the FI term, and the cosmological constant $\Lambda$. Its mathematical expression consists of two parts: one that is proportional with the cosmological constant and the other one is proportional to the parameter $\alpha$. In the limit $\Lambda \rightarrow 0$, the hairy solutions exist and they remain regular as in the AdS case.\footnote{The existence of asymptotically flat hairy solutions in theories with a potential for the scalar field was conjectured in \cite{Nucamendi:1995ex}, though the model is different (it can not be embedded in SUGRA) and the results are provided only numerically.} 

This model can be generalized by adding a gauge field to which the scalar field is coupled that provides an extension of the stringy model considered in \cite{Garfinkle:1990qj} by including the dilaton potential. Some important properties of hairy black hole solutions in this extended model, when $\nu \rightarrow \infty$, were presented in \cite{Astefanesei:2019qsg} and we would like to briefly describe them here. Unlike the stringy solutions in \cite{Gibbons:1982ih,Gibbons:1987ps,Garfinkle:1990qj}, these new asymptotically flat solutions have a well defined extremal limit. However, unlike the Reissner-Nordstr\"om black hole, but similar to the solutions  \cite{Gibbons:1982ih,Gibbons:1987ps,Garfinkle:1990qj}, they can be overcharged. Based on these features and a careful analysis of the dilaton potential, it was observed in \cite{Astefanesei:2019qsg} that, in fact, these solutions interpolate between Reissner-Nordstr\"om black hole and the stringy family of \cite{Gibbons:1982ih,Gibbons:1987ps,Garfinkle:1990qj}. More importantly, there is a sub-class of thermodynamically stable hairy black holes \cite{Astefanesei:2019mds} that are also dynamically stable. The dynamical robustness of these solutions was 
confirmed not just  perturbatively, but also by a fully non-linear numerical simulations with the Einstein-Maxwell-dilaton system \cite{Astefanesei:2019qsg}. Since in this particular model the potential depends only on one parameter, $\alpha$, it is interesting to check how and if some of these properties are affected when the hairy parameter $\nu$ is turned on. 

The remainder of the paper is organized as follows: in Section \ref{sec2}, we present a general class of exact hairy black hole solutions and analyze their thermodynamic properties checking that the first law of thermodynamics and quantum statistical relation are consistently satisfied. In Section \ref{sec3}, we study the thermodynamics of a particular solution when $\nu=3$ and construct a stability criterion based on the relative signs of the relevant response functions. For clarity, we compare these results  with the ones for Reissner-Nordstr\"om black holes. We continue, in Section \ref{sec4}, with a detailed analysis of the thermodynamic stability in the general case (for arbitrary values of the parameter $\nu$). In Section \ref{sec5}, we investigate the dynamical stability of hairy black holes solutions under spherically symmetric perturbations. We close in Section \ref{sec6} with a summary of our results and an extended discussion of the response functions in the particular case $\nu=3$.

%%%%%%%%%%%%%%%%%%%%%%%%%%%%%%%%%%%%%%%%
%%%%%%%%%%%%%%%%%%%%%%%%%%%%%%%%%%%%
\section{Hairy black holes and their thermodynamics}
\label{sec2}

In this section, we present a general family of exact asymptotically flat hairy electrically charged black hole solutions with a non-trivial scalar field potential \cite{Anabalon:2013qua}. We use the quasilocal formalism of Brown and York \cite{Brown:1992br} supplemented with counterterms \cite{Lau:1999dp,Mann:1999pc,Kraus:1999di,Mann:2005yr,Astefanesei:2005ad} to study their thermodynamics. We compute the quasilocal stress tensor \cite{Astefanesei:2005ad}, energy, on-shell Euclidean action (and the corresponding thermodynamic potential) and show that the first law of thermodynamics and quantum statistical relation are satisfied.

\subsection{Exact asymptotically flat solutions}
\label{sol1}

We will be studying the thermodynamic and dynamical stability in a theory where gravity couples to a dilaton, as well as the Maxwell  field (with a dilaton potential, but vanishing cosmological constant),
\begin{equation}
I=\frac{1}{2\kappa}\int_{\mathcal{M}}d^{4}x\sqrt{-g}\left[R-\frac{1}{4}e^{\gamma \phi}F^{2}-\frac{1}{2}(\partial \phi)^{2}-V(\phi)\right]
\label{action}
\end{equation}
where the fundamental constants are set to $G_N=c=1$ ($\kappa=8\pi$). As usual, the compact notation for the fields is $F^2\equiv F_{\mu\nu}F^{\mu\nu}$, $(\pa\phi)^2\equiv g^{\mu\nu}\pa_\mu\phi\pa_\nu\phi$; the parameter $\gamma$ controls the strength of the coupling of the dilaton to the Maxwell field and, as we will see, it also determines the shape of the potential.

The equations of motion are
\begin{equation}
R_{\mu\nu}-\frac{1}{2}g_{\mu\nu}R=
T_{\mu\nu}^{\phi}+T_{\mu\nu}^{EM}
\label{eins} 
\end{equation}
\begin{equation}
\pa_{\mu}\(\sqrt{-g}e^{\gamma\phi}F^{\mu\nu}\)
=0 \label{maxw} \end{equation}
\begin{equation}
\frac{1}{\sqrt{-g}}\partial_{\mu}
\(\sqrt{-g}g^{\mu\nu}\partial_{\nu}\phi\)
=\frac{dV(\phi)}{d\phi}+\frac{1}{4}\gamma{e}^{\gamma\phi}F^2
\label{klein}
\end{equation}
where the corresponding energy-momentum tensors are $T_{\mu\nu}^{\phi}\equiv \frac{1}{2}\partial_{\mu}\phi\partial_{\nu}\phi -\frac{1}{2}g_{\mu\nu} \[\tfrac{1}{2}(\partial\phi)^2+V(\phi)\]$ and $T_{\mu\nu}^{EM}\equiv\frac{1}{2}e^{\gamma\phi} \left(F_{\mu\alpha}F_{\nu}^{\,\,\alpha}
-\tfrac{1}{4}g_{\mu\nu}F^2\right)$. The general family of exact solution that we are going to consider was found in \cite{Anabalon:2013qua} for the following general self-interacting potential
\begin{equation}
V(\phi)=\frac{2\alpha}{\nu^{2}}
\left[
\frac{\nu-1}{\nu+2}
\sinh\left(\sqrt{\frac{\nu+1}{\nu-1}}\;\phi\right)
-\frac{\nu+1}{\nu+2}
\sinh\left(\sqrt{\frac{\nu-1}{\nu+1}}\;\phi\right)
+4\(\frac{\nu^2-1}{\nu^{2}-4}\)
\sinh\left(\frac{\phi}{\nu^2-1}\right)\right]
\label{dilaton}
\end{equation} 
where $\alpha$ is a real constant parametrizing the strength of the potential and $\nu$ is related to $\gamma$ by
\begin{equation}
\gamma\equiv\sqrt{\frac{\nu+1}{\nu-1}}
\label{gam}
\end{equation}
so that if $\nu\leq-1$ then $0\leq\gamma\leq 1$ and, if $\nu\geq 1$ then $\gamma\geq 1$. The limit $\nu\rightarrow\infty$ corresponds to $\gamma=1$ and was studied in great detail in \cite{Astefanesei:2019mds, Astefanesei:2019qsg}.

The equations of motion are solved by
\begin{align}
\label{metric}
ds^{2}&=\Omega(x)\left[
-f(x)dt^{2}+\frac{\eta^2dx^2}{f(x)}
+\(d\theta^2+\sin^2\theta{d}\varphi^2\)\right],
\\
& A_\mu dx^\mu=-\frac{q}{\nu x^{\nu}}dt, \qquad
\phi(x)=\sqrt{\nu^{2}-1}\ln(x)
\label{fields}
\end{align}
where the metric functions are 
\begin{equation}
\Omega(x)
=\frac{\nu^{2}x^{\nu-1}}{\eta^2(x^{\nu}-1)^2},
\label{omeg}
\end{equation}
\begin{equation}
f(x)=\frac{1}{\nu^{2}}\left[
\alpha\left(\frac{x^{\nu+2}}{\nu+2}-x^{2}+\frac{x^{2-\nu}}{2-\nu}+\frac{\nu^{2}}{\nu^{2}-4}\right)
+\eta^{2}\(
1-\frac{q^2}{2\nu x^{\nu}}\frac{x^\nu-1}{\nu-1}\)(x^{\nu}-1)^{2}x^{2-\nu}\right]
\end{equation}
With this special choice of the conformal factor $\Omega(x)$, the equation of motion for the dilaton can be easily integrated leading to a simple result. The constants $q$ and $\eta$ are the integral constants that define the conserved charges of the solutions. We notice that there is no integration constant related to the scalar field.

The exact solution presented so far is characterized by having two different branches of solutions, corresponding to the domains $x\in[0,1)$ and $x\in(1,\infty]$. The former is usually called the negative branch since the scalar field takes negative values, and the latter is called the positive branch. They actually correspond to two  distinct families of solutions because the boundary condition at $x=1$ for the scalar field are different (for more details see \cite{Anabalon:2013qua}). In the remaining of the paper, we are going to present a detailed thermodynamic analysis only for the positive branch for which the thermodynamically stable black holes exist.\footnote{The negative branch contains only thermodynamically unstable black holes for any value of the parameter $\nu$ and that is why the thermodynamics of negative branch is not presented in what follows. This observation is also consistent with our previous work \cite{Astefanesei:2019mds}, where it was found that the black holes of negative branch, for the particular case of $\gamma=1$, are thermodynamically unstable.}
%%%%%%%%%%%%%%%%%%%%%%%%%%%%%%%%%%%%%%%%%%%%%%%%%%%%%%%%%%%%%%%%%%%%%%%%%

\subsection{Quasilocal formalism and conserved energy}

According to the formalism of Brown and York \cite{Brown:1992br}, the conserved quantities are obtained provided a hypersurface with an isometry generated by a Killing vector $\xi^\mu$ exists. All the observers living on this hypersurface measure the same conserved quantities. Using this specific foliation for the spacetime, the quasilocal stress tensor $\tau_{ab}$ can be defined as
\begin{equation}
\label{stress}
\tau_{ab}=\frac{2}{\sqrt{-h}}\frac{\delta I}{\delta h^{ab}}
\end{equation}
where $I=I_{bulk}+I_{GH}+I_{ct}$ is the total action consisting of the bulk part of the action, given by (\ref{action}), supplemented with the Gibbons-Hawking boundary term and the gravitational counterterm that cancels the infrared divergences of the theory. For asymptotically flat spacetimes in four dimensions, the gravitational counterterm is \cite{Lau:1999dp,Mann:1999pc,Kraus:1999di}
\begin{equation}
I_{ct}=-\frac{1}{\kappa}\int_{\partial \mathcal{M}}d^{3}x\sqrt{-h}\sqrt{2\mathcal{R}^{(3)}}
\end{equation}
where  $\mathcal{R}^{(3)}=h^{ab}\mathcal{R}^{(3)}_{ab}$ is the Ricci scalar on the boundary. We choose the foliation $x=const$ with the induced metric on each surface $h_{ab}$, whose trace is $h$. For the Killing vector $\xi=\pa/\pa{t}$, the conserved quantity is the total energy of the black hole (including the hair) \cite{Brown:1992br}:
\begin{equation}
E=\oint_{s^2_\infty}{d^2\sigma\sqrt{\sigma}n^a\tau_{at}\xi^t}
\end{equation}
where $s^2_\infty$ is the spherical surface at infinity with $t=const$, given by the metric $ds^2=\sigma_{ab}dx^adx^b$, with $n^a$ the time unit normal vector.
The concrete expression for the regularized quasilocal stress tensor in this case was obtained in \cite{Astefanesei:2005ad}:
\begin{equation}
\tau_{ab}=\frac{1}{\kappa}\left[K_{a b}-h_{a b}K+\(\frac{1}{2}{\mathcal{R}^{(3)}}\)^{-1/2}\left(\mathcal{R}^{(3)}_{a b}-h_{a b}\mathcal{R}^{(3)}\right)+h_{a b}\Box \Psi-\Psi_{;{a b}}\right]
\end{equation}
where $\Psi=\sqrt{2/\mathcal{R}^{(3)}}$.\footnote{This method was extensively used for various black hole/ring solutions \cite{Mann:2005yr,Astefanesei:2005ad,Astefanesei:2006zd,Astefanesei:2009wi,Compere:2011db,Compere:2011ve, Astefanesei:2010bm}.} 
By using the exact solution presented in the previous section, the non-zero components of the quasilocal stress tensor are
\begin{equation}
\tau_{tt}=-{\frac{(\Omega f)^{1/2}}{8\pi\eta}}
\left(2\eta f^{1/2}+\frac{\Omega'}{\Omega}f\right),
\end{equation}
\begin{equation}
\tau_{\theta \theta}=\frac{\tau_{\phi \phi}}{\sin^2\theta}
=\frac{\Omega^{1/2}}{8\pi\eta{f}^{1/2}}
\left(\frac{1}{2}f'+\eta{f}^{1/2}+\frac{\Omega'}{\Omega}f\right)
\end{equation}
Since the normal unit to $t=const$ can be written as $n_a=(f\Omega)^{1/2}\delta_a^t$, the total (conserved) energy is computed at the boundary $x=1$
\begin{equation}
\label{M}
E=\frac{1}{2\eta}\lim_{x\rightarrow 1}{\left(2\eta\Omega{f^{1/2}}+f\Omega'\right)}
=\frac{q^2}{4\eta(\nu-1)}
-\frac{1}{6\eta^3}
\left(\alpha+3\eta^2\right)
\end{equation}

It is straightforward to verify that, in this case, the conserved energy equals the Arnowitt-Deser-Misner (ADM) mass obtained by expanding the $g_{tt}$ component in the canonical coordinates \cite{Arnowitt:1960es,Arnowitt:1960zzc,Arnowitt:1961zz,Arnowitt:1962hi}.\footnote{Since the asymptotics should be preserved, when the dilaton potential is non-trivial the asymptotic value of the scalar is fixed. However, when the dilaton can vary at the boundary, the total energy receives a new contribution \cite{Astefanesei:2018vga,Mejias:2019aio} and it does not match the ADM mass.}

\subsection{First law of thermodynamics}
\label{quantities}

Before computing the action on-shell, let us obtain the thermodynamic quantites for the solution presented in Section \ref{sol1}. As explained before, we shall focus only on the positive branch, $x>1$. 

The Hawking temperature is
\begin{equation}
T
=\frac{\eta(x_{+}^\nu-1)}{2\pi\nu}
\[
\frac{\(x_{+}^{\nu}-1\)}{4\nu x_{+}^{\nu-1}}\(
\frac{{3q}^{2}}{\nu-1}
-\frac{2\alpha}{\eta^2}+2\nu-4\)
-\frac{q^2\(x_{+}^\nu-1\)^2}
{2\nu^{2}x_{+}^{2\nu-1}}
-x_{+}\]
\label{temp}
\end{equation} 
where the horizon location is obtained from the horizon equation $f(x_+)=0$.
%and\begin{equation}\omega(x_{+})=\frac{1}{4\nu x_{+}^{\nu-1}}\(\frac{{3q}^{2}}{\nu-1}-\frac{2\alpha}{\eta^2}+2\nu-4\)-\frac{q^2\(x_{+}^\nu-1\)}{2\nu^{2}x_{+}^{2\nu-1}}\end{equation}
The entropy is, as usual in gravity theories without higher derivative terms in the action, $S=A/4=\pi\Omega(x_+)$, where $A$ is the area of the event horizon and $\Omega(x)$ is given by (\ref{omeg}).
The electric charge and its conjugate potential are
\begin{equation}
\label{chargpot}
Q\equiv\frac{}{}\oint_{s^2_\infty}{\star F}=-\frac{q}{4\eta},\qquad
\Phi\equiv A_t(x=1)-A_t(x=x_+)
=\frac{q}{\nu}\(x_{+}^{-\nu}-1\)
\end{equation}
where $F=\frac{1}{2}F_{\mu\nu}dx^\mu\wedge dx^\nu$ and $\star$ is the hodge dual.
Finally, the mass equals the conserved energy of the system, computed by the quasilocal formalism, $M=E$, where $E$ is given by the equation (\ref{M}).
It follows straightforwardly that they satisfy the first law of black hole thermodynamics
\begin{equation}
dM=TdS+\Phi dQ
\end{equation}
with no independent contribution from the scalar field, which is secondary hair.

\subsection{Quantum statistical relation}

By taking the trace of the Einstein's equation (\ref{eins}) and replacing the Ricci scalar into the bulk part of the action, and by adding it to the Gibbons-Hawking boundary term, one gets that, on the Euclidean section, they add up to
\begin{equation}
I^E_{bulk}+I^E_{GH}=\beta\left(-ST-\Phi Q\right)-\frac{\beta}{\eta(x-1)}-
\beta\left[\frac {2(\nu-1)(3\eta^2+\alpha)-3\eta^2q^2}
{6\eta^3\left(\nu-1\right)}\right]
\end{equation}
The gravitational counterterm contributes to the total action by
\begin{equation}
I^E_{ct}=\beta M+\frac{\beta}{\eta(x-1)}+\beta
\left[\frac {2(\nu-1)(3\eta^2+\alpha)-3\eta^2q^2}
{6\eta^3\left(\nu-1\right)}\right]
\end{equation}
and so the divergent term (and, also, the finite contribution) are canceled out. The total on-shell action satisfies the quantum statistical relation
\begin{equation}
I^E=\beta(M-TS-\Phi Q)\equiv\beta\mathcal{G}
\label{gibbs}
\end{equation}
where $\mathcal{G}=\mathcal{G}(T,\Phi)$ is the thermodynamic potential associated to the grand canonical ensemble, where $\Phi$ is fixed as the consequence of the boundary condition $\left.\delta A_\mu \right|_{\partial \mathcal{M}}=0$.

In order to obtain the thermodynamic potential associated to the canonical ensemble, for which this time $Q$ is fixed as a consequence of the boundary condition $\left.\delta(e^{\gamma\phi}\star F)\right|_{\pa\mathcal{M}}=0$, one would have to add a new boundary term to the action, $I\rightarrow I+I_A$, so that the action principle is well defined:
\begin{equation}
I_{A}=\frac{1}{2\kappa}\int_{\partial \mathcal{M}}d^{3}x\sqrt{-h}e^{\gamma \phi}n_{\mu}F^{\mu \nu}A_{\nu}
\label{Scan}
\end{equation}
The (geometrical) boundary term (\ref{Scan}) in the action corresponds, from a thermodynamic point of view, to the Legendre transform from the grand canonical ensemble to the canonical ensemble. The new contribution is $I_{A}^E=\beta Q\Phi$ and, therefore, the on-shell action for the canonical ensemble is
\begin{equation}
I=\beta(M-TS)\equiv\beta\mathcal{F}
\end{equation}
where $\mathcal{F}=\mathcal{F}(T,Q)$ is the thermodynamic potential associated to the canonical ensemble.

%%%%%%%%%%%%%%%%%%%%%%%%%%%%%%%%%%%%%%%%%%%%%%%%%%%%%%%%%%%%%%%%%%%%%%%%

\section{A general criterion for the local thermodynamic stability}
\label{sec3}
In this section, we first review the conditions under which a black hole equilibrium configuration is thermodynamically stable against small fluctuations in the temperature and either the electric charge or the conjugate potential. We perform first the analysis  for a particular case ($\nu=3$) and, then, we develop a consistent criterion to seek for stable configurations that is based on a study of relative signs of the response functions in the parameter space of the solutions. Since it is technically easier to work with, this general criterion is going to be used in the next section when we study the general case. For simplicity, in this section we show how our criterion works in the particular case $\nu=3$. We shall also compare the local stability of this hairy black hole with its non-hairy counterpart, the Reissner-Nordstr\"om black hole.

We end this part with Section \ref{criterion} where we collect all the relevant definitions and make a summary with the main steps we are going to follow for the general analysis in the next section.

\subsection{Local thermodynamic stability conditions}

Local thermodynamic stability follows from studying the heat capacity and electric permittivity,
\begin{equation}
C_Q\equiv T\(\frac{\pa S}{\pa T}\)_Q, \qquad
\epsilon_T\equiv\(\frac{\pa Q}{\pa\Phi}\)_T
\end{equation}
Concretely, these response functions should be positively defined \cite{callen1998thermodynamics}. By imposing that the energy is a minimum at the thermodynamic equilibrium (or, equivalently, that the entropy is a maximum), and by performing small fluctuations in $T$ and $Q$ around that configuration, it follows that the local stability conditions are:
\begin{equation}
\(\frac{\pa^2 M}{\pa S^2}\)_{Q}
=TC_Q^{-1}\geq 0 \quad\rightarrow \quad 
C_Q\geq 0\,, \label{CQ1}
\end{equation}
\begin{equation}
\(\frac{\pa^2 M}{\pa Q^2}\)_{S}
=\epsilon_{S}^{-1}\geq 0 \quad\rightarrow \quad 
\epsilon_S\geq 0\,, \label{CQ2}
\end{equation}
\begin{equation}
\left(\frac{\partial^2 M}{\partial Q^2}\right)_{S}\left(\frac{\partial^2 M}{\partial S^2}\right)_{Q}-\left[\left(\frac{\partial}{\partial S}\right)_{Q}\left(\frac{\partial M}{\partial Q}\right)_{S}\right]^{2}={T}C_{Q}^{-1}
\(\epsilon_{S}^{-1}-{TC_Q^{-1}\alpha_{Q}^{2}}\)
\geq 0 \label{CQ3}
\end{equation}
The last condition is associated to the physical situation when both fluctuations are turned on simultaneously. We can use the well known thermodynamic relations
\begin{equation}
\label{relations}
C_{\Phi}=C_{Q}+T\epsilon_T\alpha_Q^2
,\qquad 
\epsilon_S=\epsilon_T-\frac{T\alpha_\Phi^2}{C_{\Phi}}
,\qquad 
\alpha_{\Phi}=-\epsilon_{T}\alpha_{Q}
\end{equation}
to show that the three conditions (\ref{CQ1}), (\ref{CQ2}), and (\ref{CQ3}) are equivalent to 
\begin{equation}
\label{conditions}
C_Q\geq 0,\qquad
\epsilon_T\geq 0, \qquad
{C_Q\epsilon_T}
{\(C_Q+T\epsilon_T\alpha_Q^2\)^{-1}}\geq 0
\end{equation}
Notice that $C_Q\geq 0$ and $\epsilon_T\geq 0$ implies $C_\Phi\geq 0$ and $\epsilon_S\geq 0$. 

In order to understand the local thermodynamic stability conditions in a given ensemble from a physical point of view, consider first an equilibrium configuration in the canonical ensemble, where both the electric charge $Q$ and Hawking temperature $T$ are kept fixed. The only quantities we can freely vary are the mass of the black hole $M$ and the conjugate potential $\Phi$. Since the charge is fixed, the conjugate potential is a function of mass, $\Phi=\Phi(M)$, and so a variation of the conjugate potential comes as a consequence of variations of the mass, which, in turn, defines the sign of $C_Q$. Thus, in the canonical ensemble, the only requirement for local thermodynamic stability is $C_Q\geq 0$.\footnote{This condition is widely used in the literature. However, we would like to point out a subtlety that is not considered when studying black holes in a box. In this specific case, in principle, one can vary independently the chemical potential by moving the walls of the box and so one could also consider the permittivity as a response function.} In the grand canonical ensemble, there is more freedom since a variation of the entropy could come from variations of $M$ and $Q$.

\subsection{The particular solution: $\nu=3$}

Let us consider now a particular case, namely the exact solution with $\nu=3$ of the family presented in Section \ref{sol1}. In this theory, the scalar field is governed by the following potential:
\begin{equation}
\label{potsqrt3}
V(\phi)=\frac{4\alpha}{45}
\left[
\sinh\left(\sqrt{2}\phi\right)
-2
\sinh\left(\frac{\phi}{\sqrt{2}}\right)
+16
\sinh\left(\frac{\phi}{8}\right)\right]
\end{equation}
The metric functions are
\begin{equation}
\Omega(x)=\frac{9x^2}{\eta^2(x^3-1)^2},
\end{equation}
\begin{equation}
f(x)=\frac{\alpha\(x^6-5x^3+9x-5\)}{45x}
+\frac{\eta^2(x^3-1)^2}{9x}
\[1-\frac{q^2(x^3-1)}{12x^3}\]
\end{equation}

To understand the meaning of the coordinate $x$, let us consider the relation with the canonical coordinate $r$ near the boundary, given by $\Omega(x)\approx r^2$. That is
\begin{equation}
x=\frac{2^{1/3}\[
	(\eta r)^{3/2}+\sqrt{(\eta r)^3- 4}
	\]^{2/3}+ 2}
{2^{2/3}\sqrt{\eta r}\[(\eta{r})^{3/2}+	\sqrt{(\eta r)^3- 4}\]^{1/3}}
=1+\frac{1}{\eta{r}}-\frac{1}{3(\eta r)^3}
+\frac{1}{3(\eta r)^4}+\mathcal{O}(r^{-6})
\end{equation}
that is consistent with the domain  $x>1$ for $r>0$\footnote{We emphasize that the solution comes up with a `negative branch' given by the domain $0\leq x<1$, which is not considered in this work.}. Without loss of generality, we can assume $\eta>0$. The boundary defined in the limit $x=1$ corresponds to the limit $r\rightarrow\infty$ in canonical coordinates. Note also that the scalar field, $\phi(x)=2\sqrt{2}\ln(x)$, has the following asymptotic form
\begin{equation}
\phi(r)=\frac{\Sigma}{r}+\frac{\sigma}{r^2}
+\mathcal{O}(r^{-3})
=\frac{2\sqrt{2}}{\eta r}-
\frac{\sqrt{2}}{\eta^2r^2}+\mathcal{O}(r^{-3})
\label{falloff1}
\end{equation}

The first observation is that in flat spacetime, in general, the scalar field has the fall-off
\begin{equation}
\phi=\phi_\infty+\frac{\Sigma}{r}+\mathcal{O}(r^{-2})
\end{equation}
However, due to the presence of the potential, the asymptotic value of the scalar field is fixed to $\phi_\infty=0$. The next term in the expansion provides the scalar charge \cite{Gibbons:1996af} $\Sigma$ that, in fact, is not an independent parameter of the solution,\footnote{For recent discussions on the scalar charges and their role on the first law of thermodynamics, see \cite{Astefanesei:2018vga,Hajian:2016iyp,Naderi:2019jhn}.} but is determined by the constants of the solution, namely, $M$ and $Q$,
\begin{equation}
M=\frac{4\sqrt{2}Q^2}{\Sigma}
-\frac{\sqrt{2}\Sigma(\alpha\Sigma^2+24)}{192}
\label{mqsigma}
\end{equation}
This is consistent with what we have claimed before, namely that the scalar field is `secondary hair' and its degrees of freedom live outside the horizon . 

The study of thermodynamic stability for a particular value of $\nu$ can be properly done by a graphical analysis. As a concrete example, in what follows we prove that there exist configurations which are thermodynamically stable for $\nu=3$. In order to make the analysis reliable, we are going to use the parameter of the theory $\alpha$ (assumed to be positive), which has dimension of length$^{-2}$, to re-scale the thermodynamic quantities as
\begin{equation}
\label{dimensionless}
\eta\rightarrow\sqrt{\alpha}\eta,\ \ \ M\rightarrow \sqrt{\alpha}M,\ \ \ T\rightarrow \frac{T}{\sqrt{\alpha}},\ \ \ S\rightarrow \alpha S,\ \ \ Q\rightarrow \sqrt{\alpha}Q
\end{equation}
so that all of them become dimensionless.

\subsubsection{Thermodynamic stability at $\Phi$ fixed}

\begin{figure}[t!]
	\centering
	\subfigure[$Q$ vs $\Phi$ -- $\nu = 3$]
	{\includegraphics[width=5.8cm]{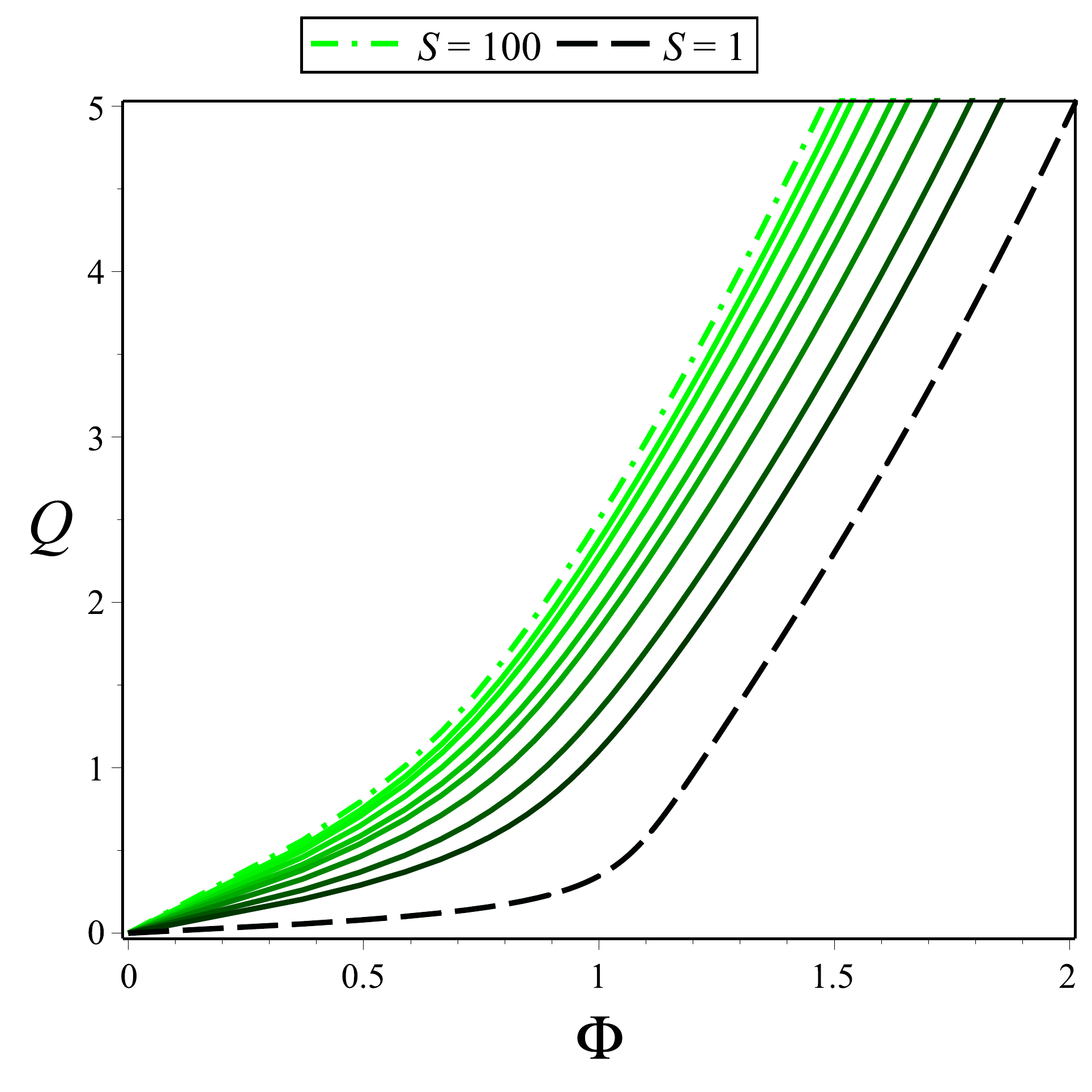}}\quad\quad\quad\quad
	\subfigure[$\sqrt{S}$ vs $T$ -- $\nu = 3$]
	{\includegraphics[width=5.8cm]{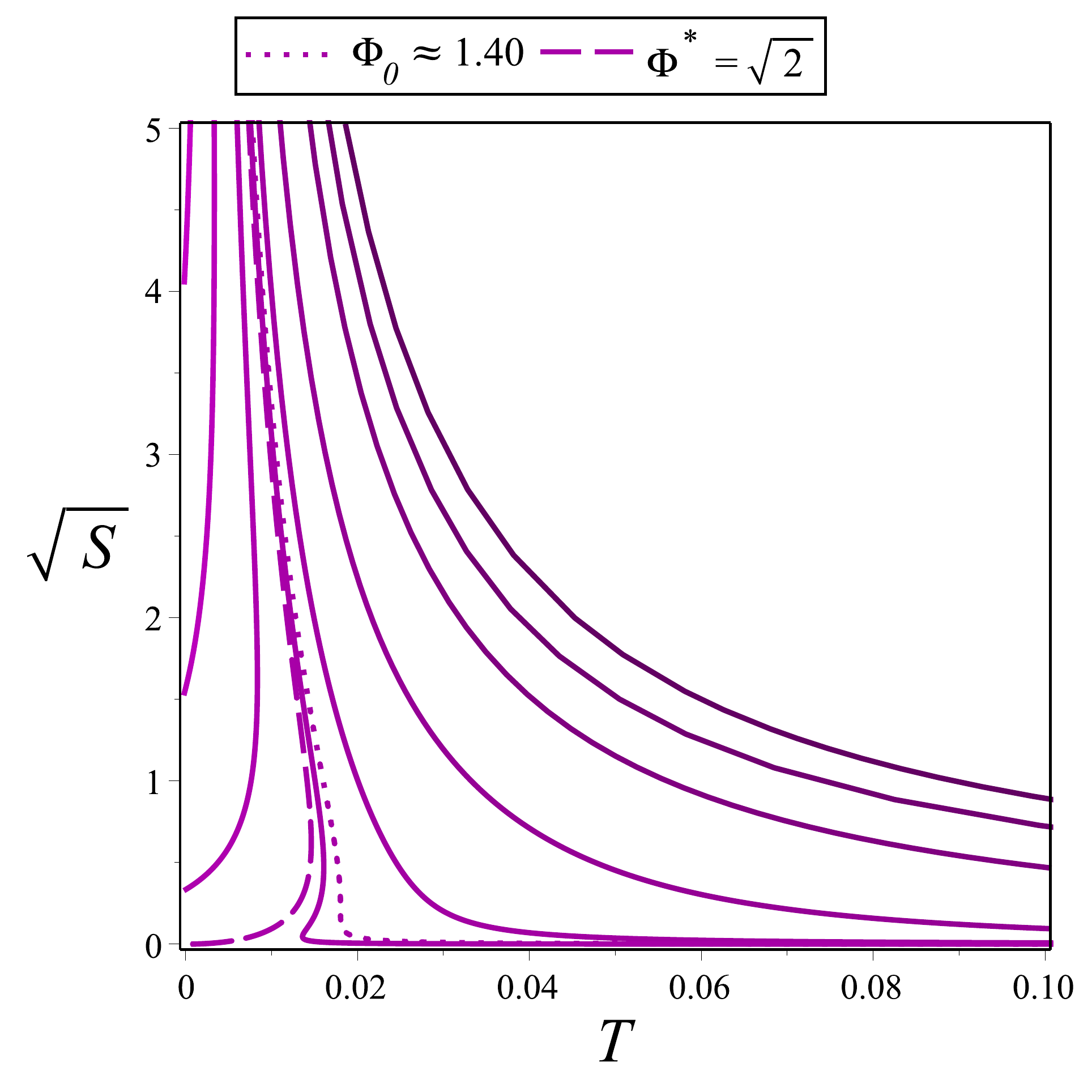}}
	\subfigure[$Q$ vs $\Phi$ -- Reissner-Nordstr\"om]
	{\includegraphics[width=5.8cm]{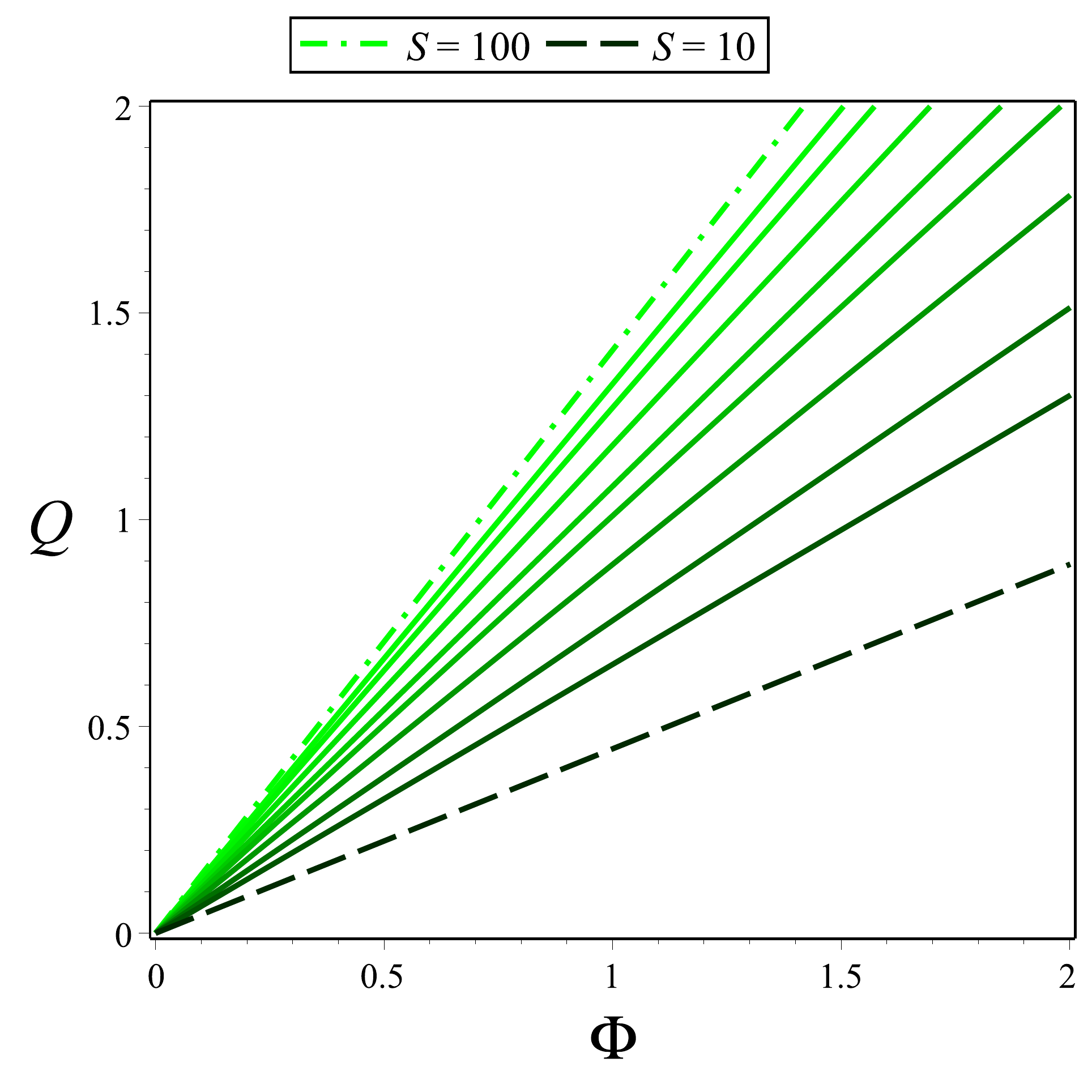}}\quad\quad\quad\quad
	\subfigure[$S$ vs $T$ -- Reissner-Nordstr\"om]
	{\includegraphics[width=5.8cm]{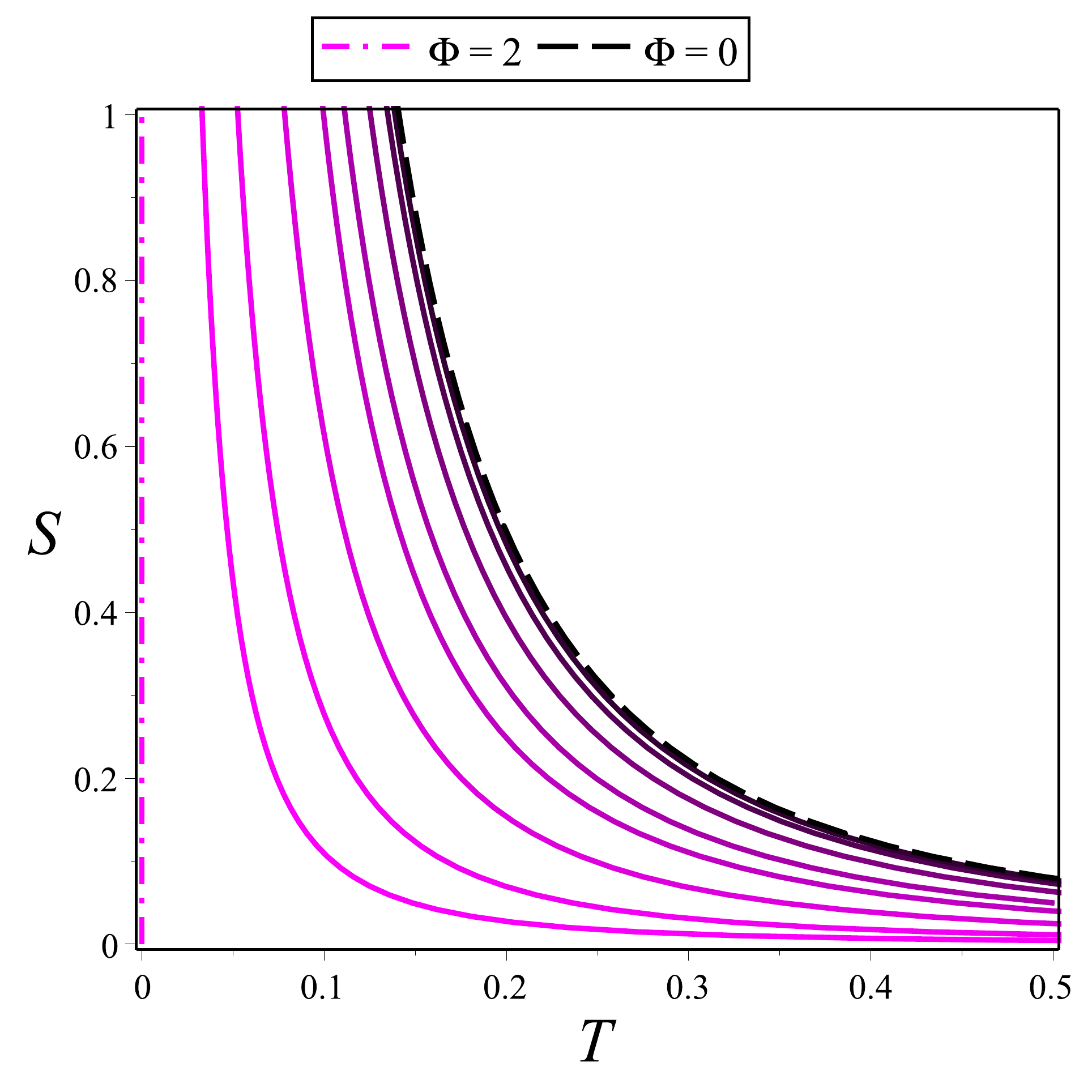}}
	\caption{\small \emph{Top:} $\nu = 3$. {\bf (a)} Electric charge as a function of its conjugate potential for constant values of entropy. {\bf(b)} Entropy as a function of temperature for constant values of the chemical potential. It has been plotted the square root of the entropy in order to make the curves behaviour clearer. \emph{Bottom:} Reissner-Nordstr\"om. {\bf (c)} Electric charge as a function of the chemical potential for constant entropy. {\bf (d)} Entropy as a function of temperature for constant values of the chemical potential.}
	\label{fig:nu3_gc}
\end{figure}

From equations (\ref{relations}), it follows that the criterion for the local thermodynamic stability in this ensemble is that heat capacity $C_{\Phi}$ and electric permittivity $\epsilon_{S}$ are simultaneously positive. In Fig. \ref{fig:nu3_gc}, it is depicted $Q$ vs $\Phi$ and $S$ vs $T$, where the corresponding slopes indicate the signs of the relevant response functions. Since the slope of Fig. \ref{fig:nu3_gc}{\bf a} is positive for all positive values for $S$, we only have to check the positivity of $C_\Phi$ from Fig. \ref{fig:nu3_gc}{\bf b}.

The comparison with the Reissner-Nordstr\"om black hole allows us to spot some important differences between the two systems. The most significant one is precisely the existence of positive slopes in the plot of the entropy as a function of temperature for the hairy black hole inside $\Phi_{0}\approx 1.40<\Phi<2$. Moreover, inside the sub-interval  $\Phi^{*}\equiv\sqrt{2}\leq\Phi\leq 2$, these thermally stable black holes have a well defined extremal limit. It is interesting to notice that $\Phi= \Phi_{0} \approx 1.40$ and $T = T_{\infty} \approx 0.018$ characterize a critical point which satisfies $(\pa T/\pa S)_{\Phi_0}=(\pa^2 T/\pa S^2)_{\Phi_0}=0$, depicted in Fig. \ref{fig:nu3_gc}{\bf b}.\footnote{The nomenclature $T_\infty$ comes from the fact that the heat capacity $C_\Phi$ diverges at that temperature. Its importance resides in the existence of stable configuration for $T<T_{\infty}$.}. A more detailed thermodynamic description of this feature is going to be presented in \cite{Astefanesei:2020toappear}

\subsubsection{Thermodynamic stability at $Q$ fixed}
%%%%%%%%%%%%%%%%%%%%%%%%

\begin{figure}[t!]
	\centering
	\subfigure[$\Phi$ vs $Q$ -- $\nu = 3$]
	{\includegraphics[width=5.8cm]{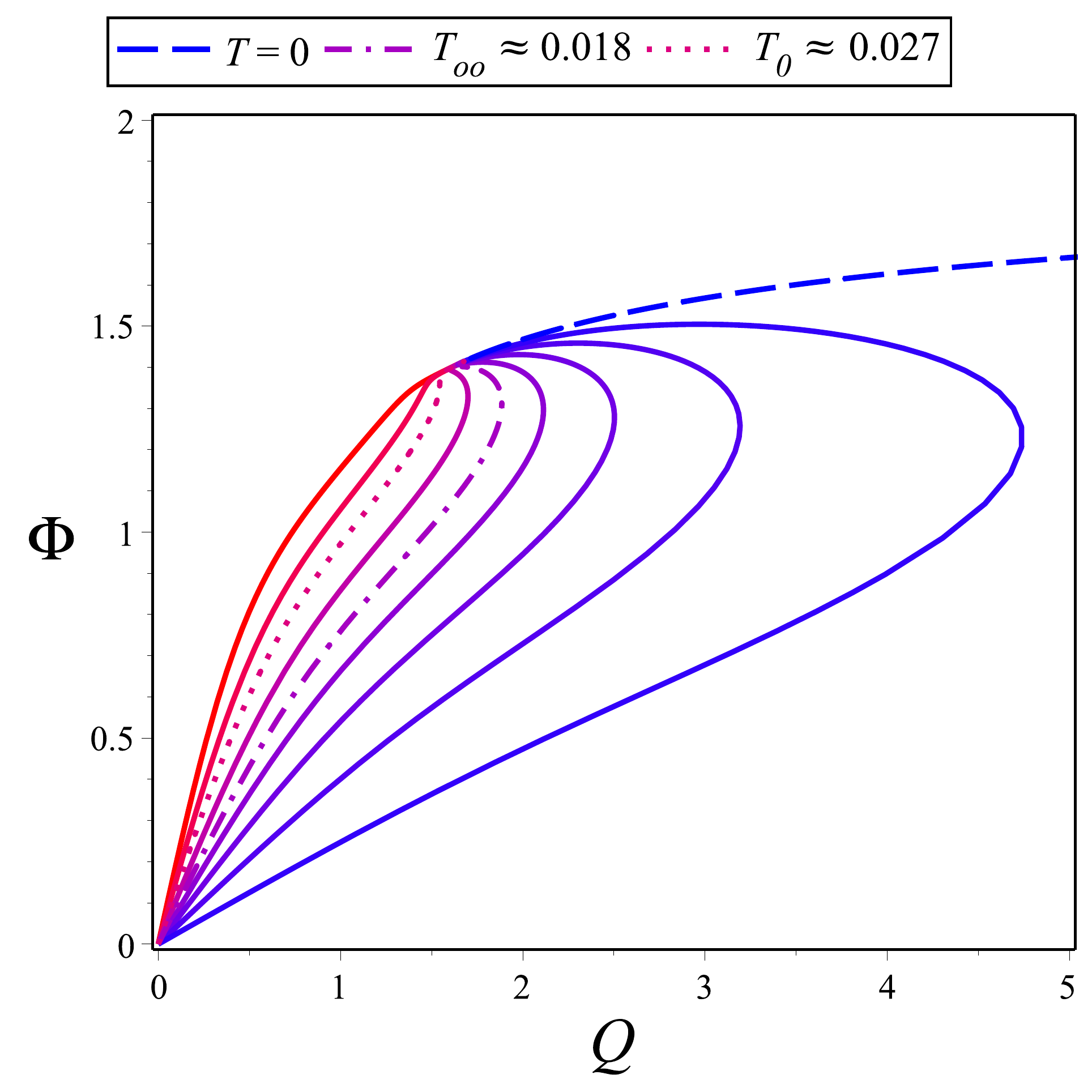}}\quad\quad\quad\quad
	\subfigure[$\sqrt{S}$ vs $T$ -- $\nu = 3$]
	{\includegraphics[width=5.8cm]{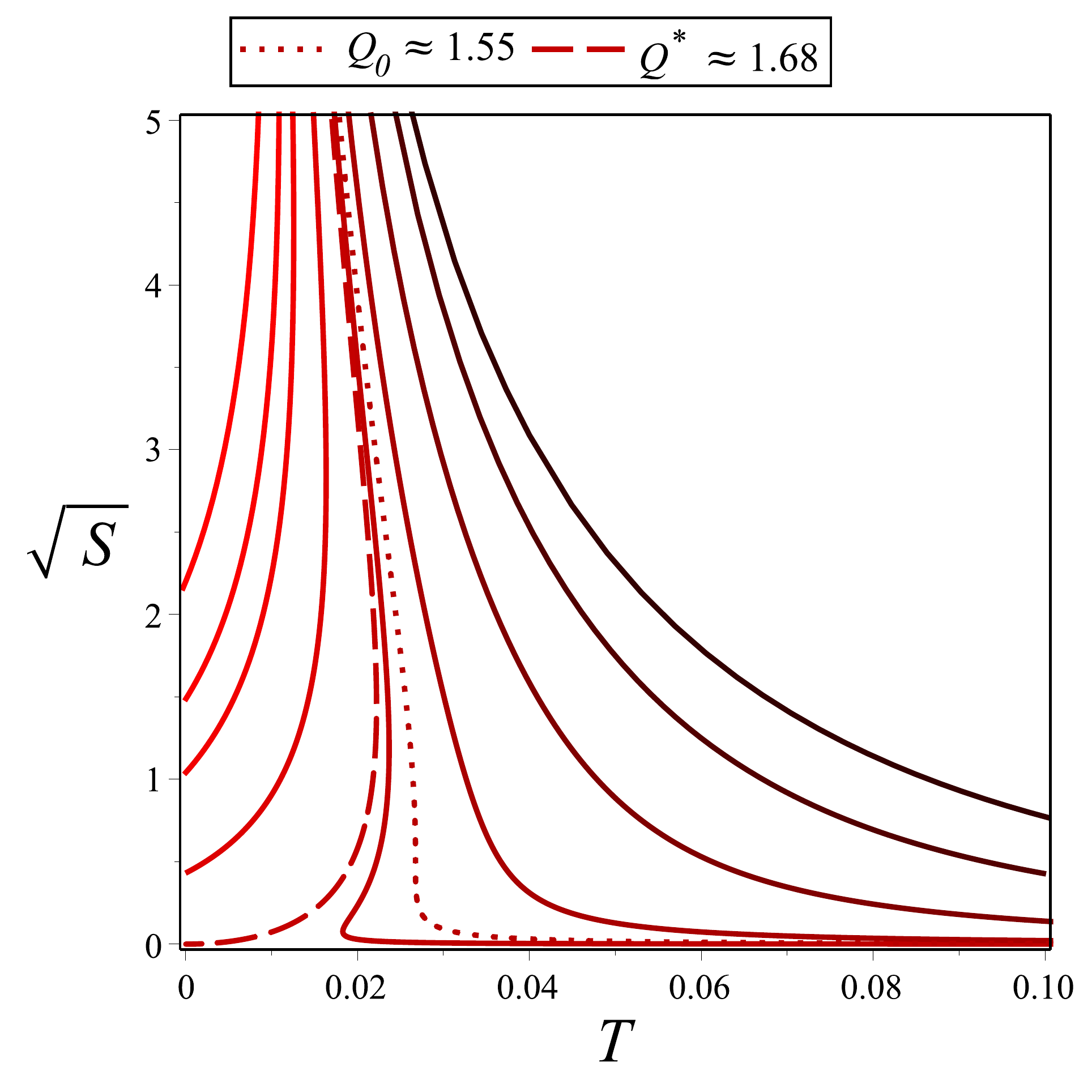}}
	\subfigure[$\Phi$ vs $Q$ -- Reissner-Nordstr\"om]
	{\includegraphics[width=5.8cm]{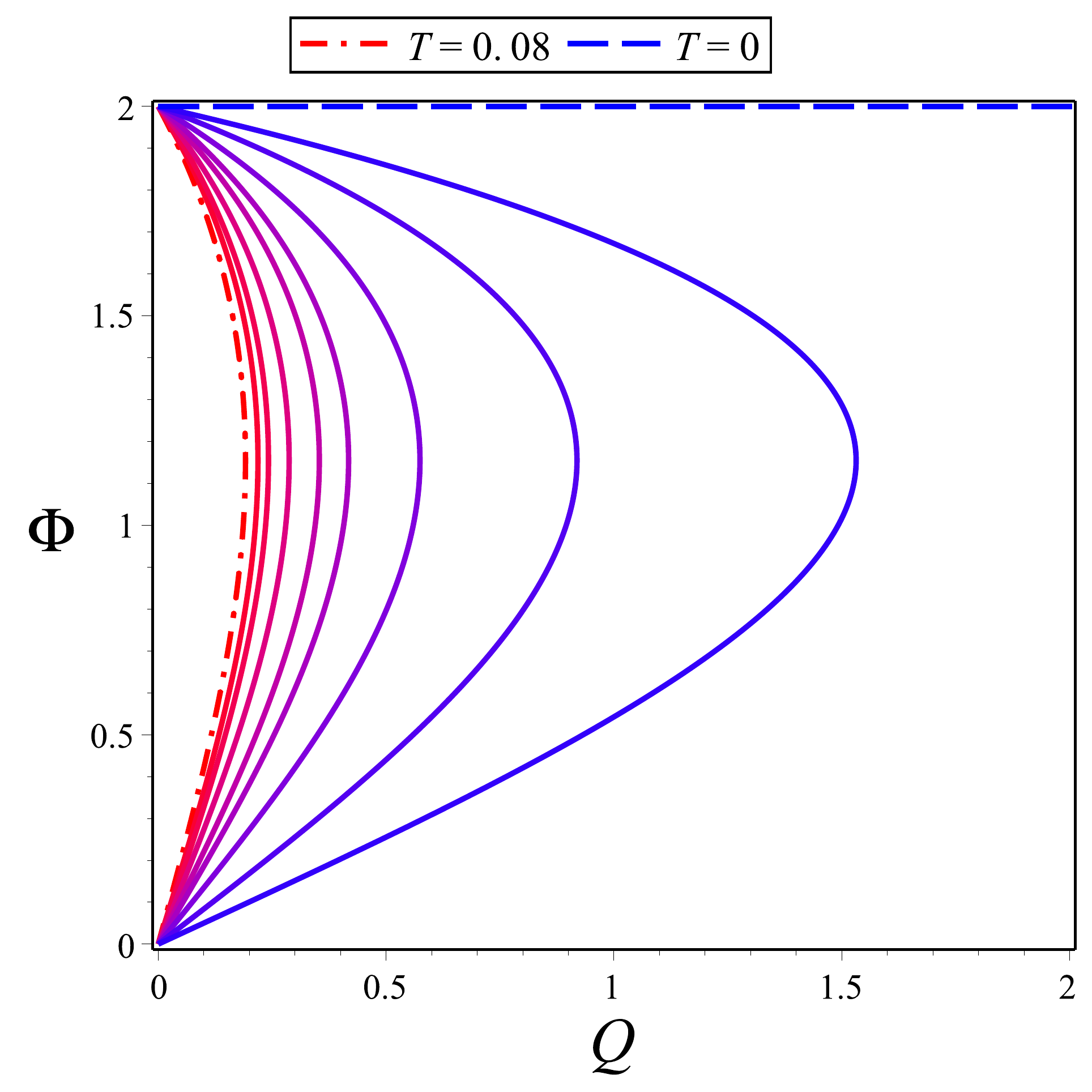}}\quad\quad\quad\quad
	\subfigure[$S$ vs $T$ -- Reissner-Nordstr\"om]
	{\includegraphics[width=5.8cm]{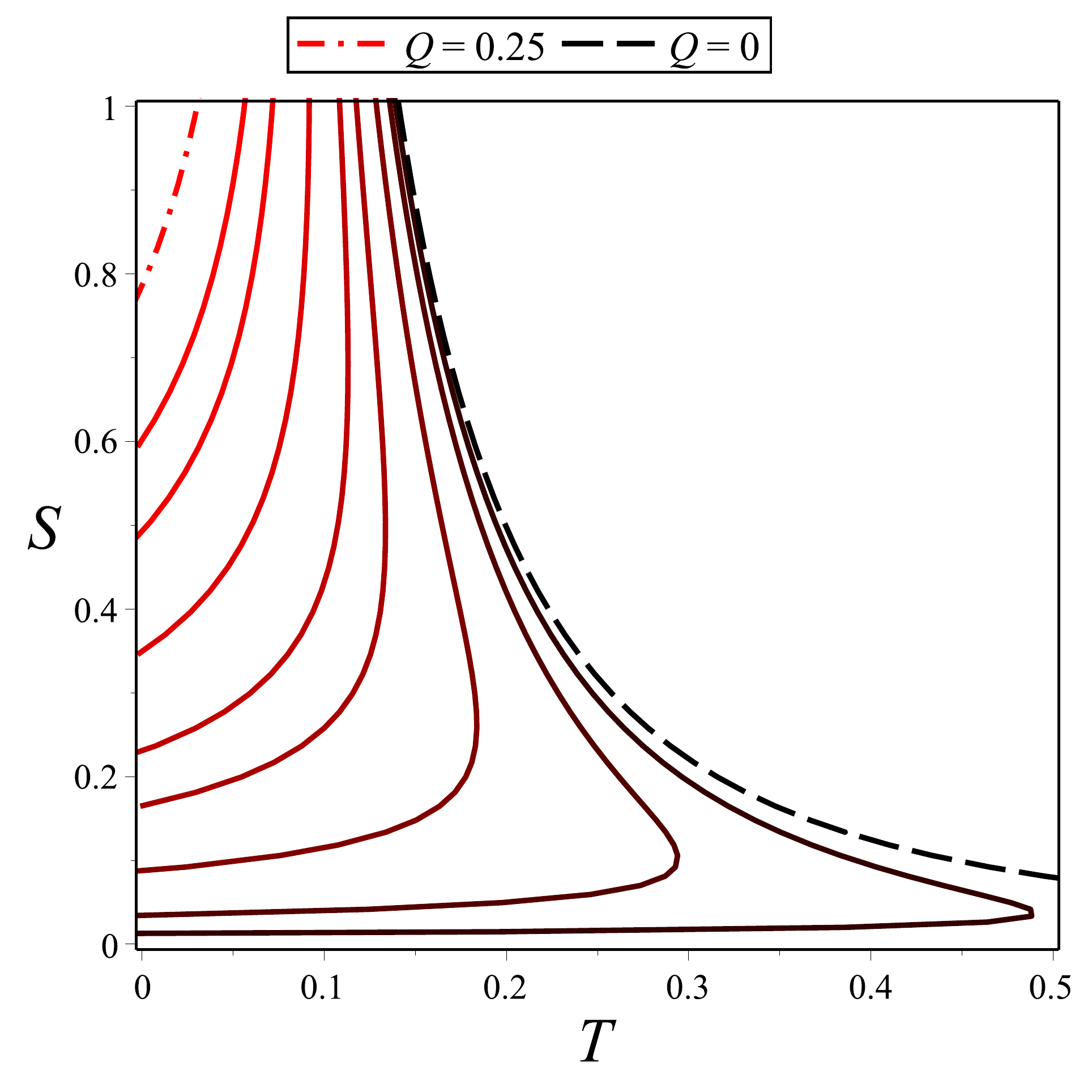}}
	\caption{\small Equations of state for the canonical ensemble. \emph{Top:} $\nu =3$. \emph{Bottom:} Reissner-Nordstr\"om. The relevant response functions, $\epsilon_T$ and $C_Q$, can be read off from the corresponding slopes. A critical behaviour is again found in $S$ vs $T$ inside the interval $Q_0<Q<Q^{*}$, similar to the critical behaviour in the grand canonical ensemble.}
	\label{fig:nu3_c}
\end{figure}

Let us now proceed with the canonical ensemble. The relevant plots, from where we can read off the sign of both $\epsilon_T$ and $C_Q$, are depicted in Fig. \ref{fig:nu3_c}. In Fig. \ref{fig:nu3_c}{\bf a}, it was plotted the equation of state, where two relevant isotherms have been highlighted, $T_\infty\approx 0.018$ and $T_0\approx 0.027$.\footnote{They actually correspond to the finite temperatures where two different critical points appear, given by
	\begin{equation}
	\(\frac{\pa Q}{\pa \Phi}\)_{T_\infty}=\(\frac{\pa^2Q}{\pa \Phi^2}\)_{T_\infty}=0, \qquad
	\(\frac{\pa\Phi}{\pa Q}\)_{T_0}=\(\frac{\pa^2\Phi}{\pa Q^2}\)_{T_0}=0
	\end{equation}
	respectively. An appropriate zoom in the plot $Q$ vs $\Phi$ for the isotherms $T_\infty$ and $T_0$ is going to be shown later, in Fig. \ref{fig:bosquejo1}.} The critical point at $T=T_0$ is located on the vertical line $Q=Q_0\approx 1.55$, which can also be observed from Fig. \ref{fig:nu3_c}{\bf b} (the dotted curve at $Q_0$).

The main observation here is that, while for the Reissner-Nordstr\"om black hole there is no configuration with both $\epsilon_T>0$ and $C_Q>0$, the hairy black hole develops a region where $\epsilon_T>0$ at the top of the plot $\Phi$ vs $Q$, that is for $\Phi>\Phi_0$, within the interval $0<T<T_\infty$, as can be seen from Fig. \ref{fig:nu3_c}{\bf a}. This new branch in the equation of state contains black holes with both response functions positive definite.

%%%%%%%%%%%%%%%%%%%%%%%%%%%%%%%%%%%%%%%%%%%%%%%%%%%%%%%%%%%%%%%%%%%%%%%%%%%%%%%%%%%%%%
\subsection{The general criterion}
\label{criterion}
Inspired by the previous results, we are ready to introduce a criterion that can be used to analyze the general case for arbitrary $\nu$. This consists in a careful graphical analysis of the equation of state corresponding to each ensemble. In order to facilitate the analysis, it is convenient to introduce a set of variables defined in Table \ref{table:def_var}, which were of great help to study the case $\nu=3$. We should also use the dimensionless physical quantities defined by the equations (\ref{dimensionless}).

\begin{table}[t!]
	\centering
	\begin{tabular}{|c|c|c|}
		\hline
		Name & Definition & Motivation \\
		\hline
		$\Phi_{0}$ & $\left(\frac{\partial T}{\partial S}\right)_{\Phi = \Phi_{0}} = \left(\frac{\partial^{2} T}{\partial S^{2}}\right)_{\Phi = \Phi_{0}} = 0$ & {\small Critical point - grand canonical ensemble}\\ 
		\hline
		$\Phi^{*}$ & $\lim_{x_{+}\rightarrow +\infty} \Phi|_{T = 0}$ & {\small An end point of the extremal isoterm}\\
		\hline
		$Q_{0}$ & $\left(\frac{\partial T}{\partial S}\right)_{Q = Q_{0}} = \left(\frac{\partial^{2} T}{\partial S^{2}}\right)_{Q = Q_{0}} = 0$ & {\small Critical point - canonical ensemble}\\
		\hline
		$Q^{*}$ & $\lim_{x_{+}\rightarrow +\infty} Q|_{T = 0}$ & {\small An end point of the extremal isotherm}\\
		\hline
		$T_{0}$ & $\left(\frac{\partial Q}{\partial \Phi}\right)_{T = T_{0}} = \left(\frac{\partial^{2} Q}{\partial \Phi^{2}}\right)_{T = T_{0}} = 0$ & {\small Critical point - canonical ensemble}\\
		\hline
		$T_{\infty}$ & $\left(\frac{\partial \Phi}{\partial Q}\right)_{T = T_{\infty}} = \left(\frac{\partial^{2} \Phi}{\partial Q^{2}}\right)_{T = T_{\infty}} = 0$ & {\small Critical point - grand canonical ensemble}\\
		\hline
	\end{tabular}
	\caption{Definition of the set of variables used to characterize distinct features in the plots of equation of state for each ensemble.}
	\label{table:def_var}
\end{table}

The general analysis, that is going to be performed for arbitrary values of $\nu$ in the potential, consists basically of the following two steps:

{\bf {1)}} Analize the diagrams $Q$ vs $\Phi$ (by keeping $T$ fixed) and $S$ vs $T$ (by keeping either $Q$ or $\Phi$ fixed, depending on the ensemble) and identify critical curves where the behaviour changes.

{\bf {2)}} Translate the relevant points and curves from one diagram to another. That will allow us to recognize those points and regions in both diagrams at the same time and to read the corresponding slopes that represent the response functions.

\section{Thermodynamic stability analysis for arbitrary $\nu$}
\label{sec4}
In this section, we present a detailed analysis of local thermodynamic stability for an arbitrary value of $\nu$ by following the general criterion proposed in  Section \ref{criterion}. The main result is that there exists a sub-class of asymptotically flat hairy black holes that are thermodynamically stable in both ensembles and for every finite value of the hairy parameter $\nu$.

\subsection{Grand canonical ensemble, $\Phi$ fixed}
\label{grand}

The relevant quantities required to study the thermodynamic stability are the electric permittivity at constant entropy, $\epsilon_{S}=(\partial\Phi/\partial Q)_{S}$, and the heat capacity at constant conjugate potential, $C_{\Phi}=T(\partial S/\pa T)_{\Phi}$. We are going to read off the signs of these response functions by studying general properties of the corresponding phase diagrams.  

\subsubsection{$Q$ vs $\Phi$, $S$ fixed}

\begin{figure}[t!]
	\centering
	\subfigure[]{\includegraphics[width=6cm,height=6cm]{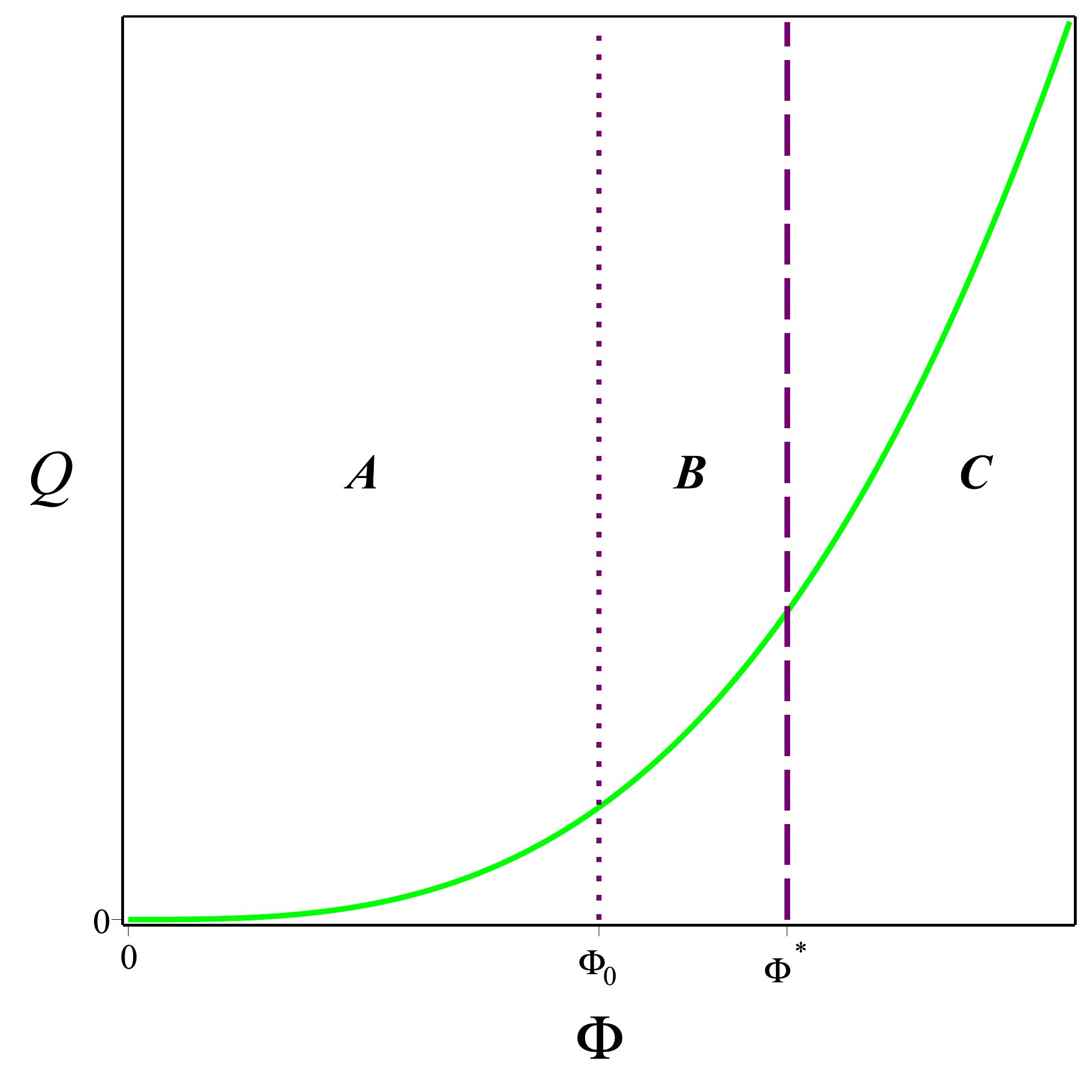}}\qquad\qquad\qquad
	\subfigure[]{\includegraphics[width=6cm,height=6.5cm]{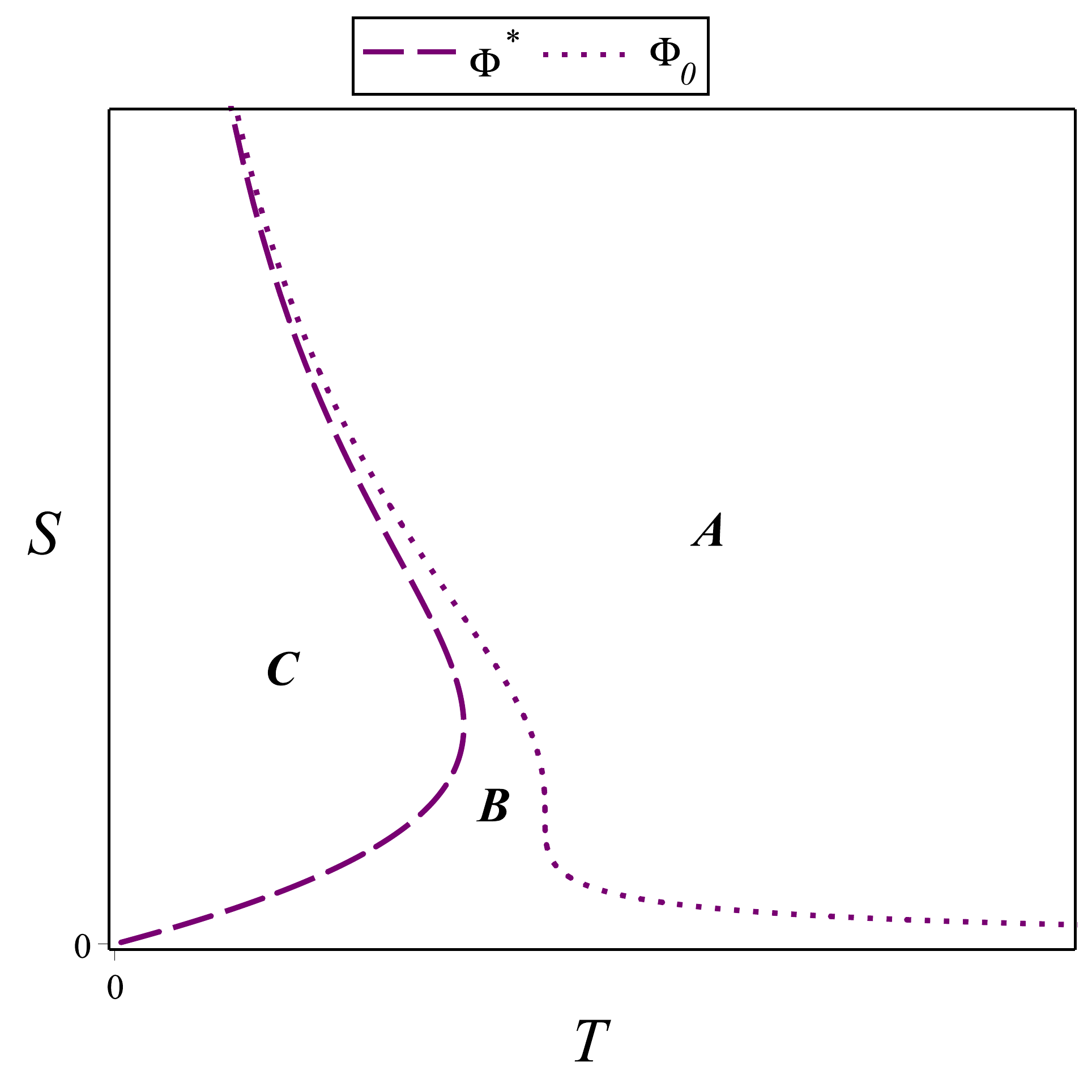}}
	\caption{\small Sketch of $Q$ vs $\Phi$ at constant entropy for any $\nu>1$ and $S$ vs $T$ at constant conjugate potential.}
	\label{fig:RegionesGC}
\end{figure}

To initiate, let us use the horizon equation and the expressions for the electric charge, conjugate potential, and entropy from Section \ref{quantities} to obtain the following parametric equations:
\begin{equation}
Q(S,x_+)=
\frac{\sqrt{2S}}{4\pi}
\[\frac {\left( x_{+}^{\nu}-1 \right)  \left( \nu-1 \right)A_{S}}{ \left( {
		\nu}^{2}-4 \right) {\nu}^{3}}\]^{1/2}
,\qquad
\Phi=\[\frac{2\(\nu-1\)\(x_{+}^{\nu}-1\)A_{S}}{\pi\,{\nu}^{3}\({\nu}^{2}-4\) x_{+}^{1+\nu}}\]^{1/2}
\label{eq:QP_Sx}
\end{equation}
where
\begin{equation}
A_{S}=A_{S}(S,x_{+})=
\[\(\nu-2\)x_{+}^{\nu+2}
-\({\nu}^{2}-4\)x_{+}^{2}-\(\nu+2\)x_{+}^{2-\nu}
+{\nu}^{2}\,\] S+\pi{\nu}^{2}\({\nu}^{2}-4\)x_{+}
\end{equation}
In order to study the isentropic behaviour, that is, the relation $Q=Q(\Phi)$ with fixed $S$, first note that, in the limit $x_{+} \rightarrow +\infty$, both the electric charge and conjugate potential diverge.\footnote{To be more exact, the limit $x_+\rightarrow\infty$ exist for $\Phi<\sqrt{2}$. For $\sqrt{2}<\Phi<2$, the entropy reaches a minimum value, which tends to infinity when $\Phi=2$. Therefore, the domain for the conjugate potential is $0\leq \Phi<2$.} In the limit $x_{+}=1$, on the contrary, both vanish. Furthermore, there are no points for which one or both of the following equations are satisfied
\begin{equation}
\(\frac{\partial \Phi}{\partial Q}\)_{S} = 0, 
\qquad\qquad
\(\frac{\partial^{2} \Phi}{\partial Q^{2}}\)_{S} = 0
\label{Scrit}
\end{equation}
which allows us to conclude that $\epsilon_S$ is a positive definite quantity for all equilibrium configurations. A sketch for $Q$ vs $\Phi$ at constant $S$, which holds for any value of $\nu>1$, is presented in Fig. \ref{fig:RegionesGC}{\bf a}.

\subsubsection{$S$ vs $T$, $\Phi$ fixed}

\begin{figure}[t!]
	\centering
	\subfigure[Notice $\Phi_{0}<\Phi^{*}$]{\includegraphics[width=6.2cm,height=6.2cm]{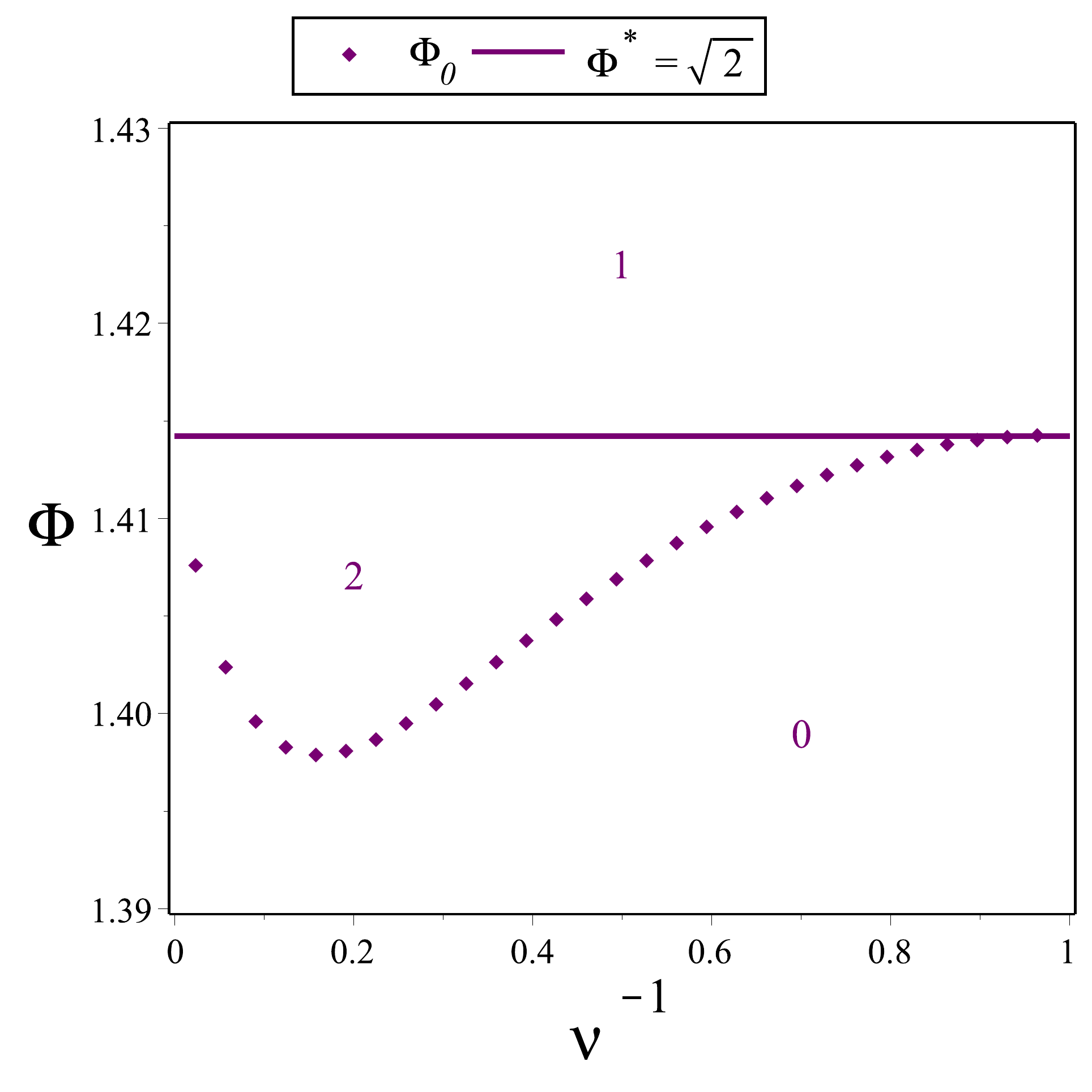}}\qquad\qquad\qquad
	\subfigure[$\Phi_{A}<\Phi_{0}<\Phi_{B}<\Phi^{*}<\Phi_{C}$]{\includegraphics[width=6cm]{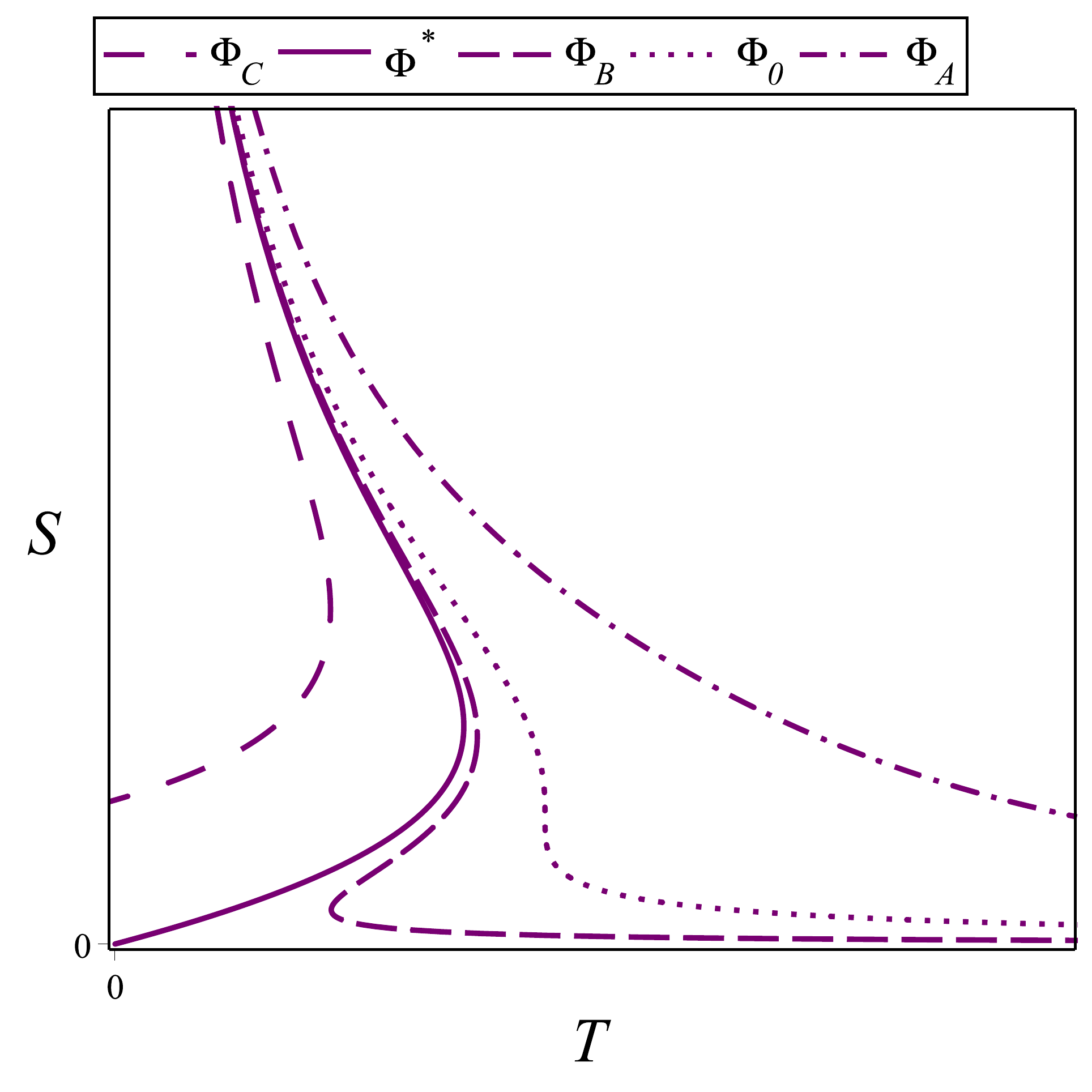}}
	\caption{\small \textbf{(a)} $\Phi_{0}$ y $\Phi^{*}$ as functions of $\nu^{-1}$. The numbers inside the plot label the number of times that $C_{\Phi}$ changes its sign inside the corresponding region. \textbf{(b)} Sketch of $S$ vs $T$ for different values of $\Phi$.}
	\label{fig:Phicrit_ST}
\end{figure}

Now, let us study the heat capacity, $C_\Phi$.  First, let us write the entropy and temperature in the following parametric form 
\begin{equation}
S={\frac { \pi{\nu}^{2}x_{+}^{\nu-1}
		\left( {\nu}^{2}-4 \right) A_{\Phi} }{2   \left( 
		\nu-1 \right)  \left( x_{+}^{\nu}-1 \right)E_{\Phi} }}
,\quad
T=\pm{\frac {\sqrt {2} \left[ 2\left( 
		\nu-1 \right)  \left( x_{+}^{\nu}-1 \right)B_{\Phi} -{\Phi}^{2} C \right] x_{+}^{1-\nu}}{8\pi\sqrt {{ {A_\Phi \left( \nu-1 \right)  \left( x_{+}^{\nu}-1 \right)  \left( {\nu}^{2}-4 \right)E_{\Phi}  } }}} }
\label{T_Phi}
\end{equation}
where we have defined
\begin{eqnarray}
\nonumber A_{\Phi} &=&2\, \left( x_{+}^{\nu}-1 \right)  \left( \nu-1 \right) x_{+}^{2-\nu}-{
	\Phi}^{2}\nu\,x_{+}^{2}
,\\
\nonumber B_{\Phi}&=&\left( \nu\,x_{+}^{2}-2\,x_{+}^{2}-\nu-2 \right) x_{+}^{\nu}+\nu\,x_{+}^{2}+2
\,x_{+}^{2}-\nu+2
,\\
\nonumber C_{\Phi}&=&
\({\nu}^{2}x_{+}^{2}-3\nu{x}_{+}^{2}-{\nu}^{2}+2x_{+}^{2}-2\nu\) x_{+}^{2\nu}+2 \left[(x_{+}^{2}-1) \nu+2x_{+}^{2} \right]  \left( \nu-1 \right) x_{+}^{\nu}+\(\nu+2 \)x_{+}^{2}
,\\
\nonumber E_{\Phi}&=&\left( \nu+2 \right) x_{+}^{2-\nu}+ \left( 2-\nu \right) x_{+}^{\nu+2}+
\left( x_{+}^{2}-1 \right) {\nu}^{2}-4\,x_{+}^{2}
\end{eqnarray}
The $\pm$ sign in the expression for the temperature (\ref{T_Phi}) must be consistently assigned according to whether $1<\nu<2$ or $2<\nu$, so that $T\geq 0$. A study of the limits $x_+=1$ and $x_+\rightarrow\infty$ on the expressions of entropy and temperature above yields the following results:
\begin{eqnarray}
\nonumber \lim_{x_{+}\rightarrow ^+\infty} S&=&0,\qquad \lim_{x_{+}\rightarrow ^+\infty} T \rightarrow \lbrace-\infty,0,+\infty\rbrace,\\
\nonumber \lim_{x_{+}\rightarrow 1^{+}} S &\rightarrow & +\infty,\ \ \lim_{x_{+}\rightarrow 1^{+}} T= 0
\end{eqnarray}
We emphasize that the physical analysis of these results is made only for positive values of the temperature. However, the existence of a divergent negative temperature indicates that there also exists a configuration with zero temperature, which corresponds to an extremal black hole. In particular, the $\lim_{x_{+}\rightarrow +\infty} \,T=-\infty$ is consistent only when $\Phi>\Phi^{*}$. Therefore, it means that there must exist extremal black holes with non-zero entropy in this interval, as can be explicitly checked in Fig. \ref{fig:Phicrit_ST}{\bf b}. Indeed, the curve characterized by $\Phi_C>\Phi^{*}$ reaches the extremality at finite entropy.

Having the asymptotic value of the curves, the next step is to check whether there exist a critical point $\Phi_0$ for some value of $\nu>1$. This will allow us to determine possible changes of the concavity and/or the slope. As shown in Fig. \ref{fig:Phicrit_ST}{\bf a}, there actually exists a finite $\Phi_0$ for any $\nu$ in the interval of interest. Now, from Fig. \ref{fig:Phicrit_ST}{\bf b}, changes of sign in $C_\Phi=T(\pa S/\pa T)_\Phi$ can be identified. 

\subsubsection{Thermodynamic stability}

Armed with the information obtained in the previous analysis, it is useful to consider three regions (noted A, B, and C, respectively) where the heat capacity $C_\Phi$ changes its sign in a particular way. Those regions are identified by some specific values of $\Phi$, as shown in Fig. \ref{fig:RegionesGC}{\bf a}. Since $\epsilon_S>0$, the thermodynamic stability for black holes in grand canonical ensemble is  basically guaranteed provided $C_\Phi>0$.

\subparagraph*{Region A with $\Phi<\Phi_0$:} in this region, $C_{\Phi}<0$, as can be seen from Fig. \ref{fig:RegionesGC}{\bf b}, implying that there are no thermodynamically stable equilibrium configurations.
\subparagraph*{Regions B with $\Phi_0<\Phi<\Phi^*$ and region C with $\Phi^*<\Phi$:} since $C_{\Phi}$ changes its sign at least one time, there must exist at least one region where $C_\Phi>0$. Accordingly, we conclude that thermodynamically stable hairy black holes can be found in the range $2>\Phi\geq \Phi_0$.

\subsection{Canonical ensemble, $Q$ fixed}

The relevant quantities are the electric permittivity at constant temperature, $\epsilon_T\equiv(\pa\Phi/\pa{Q})_T$, and the heat capacity, $C_Q\equiv T(\pa S/\pa T)_Q$. As before, we are going to use the corresponding diagrams to read off the sign of the slopes and, accordingly, split the parameter space into different regions.

\subsubsection{$\Phi$ vs $Q$, $T$ fixed}

\begin{figure}[t!]
	\centering
	\subfigure[$0=T_{extremal}<T_{0}<T_{\infty}$]
	{\includegraphics[width=6.75cm]{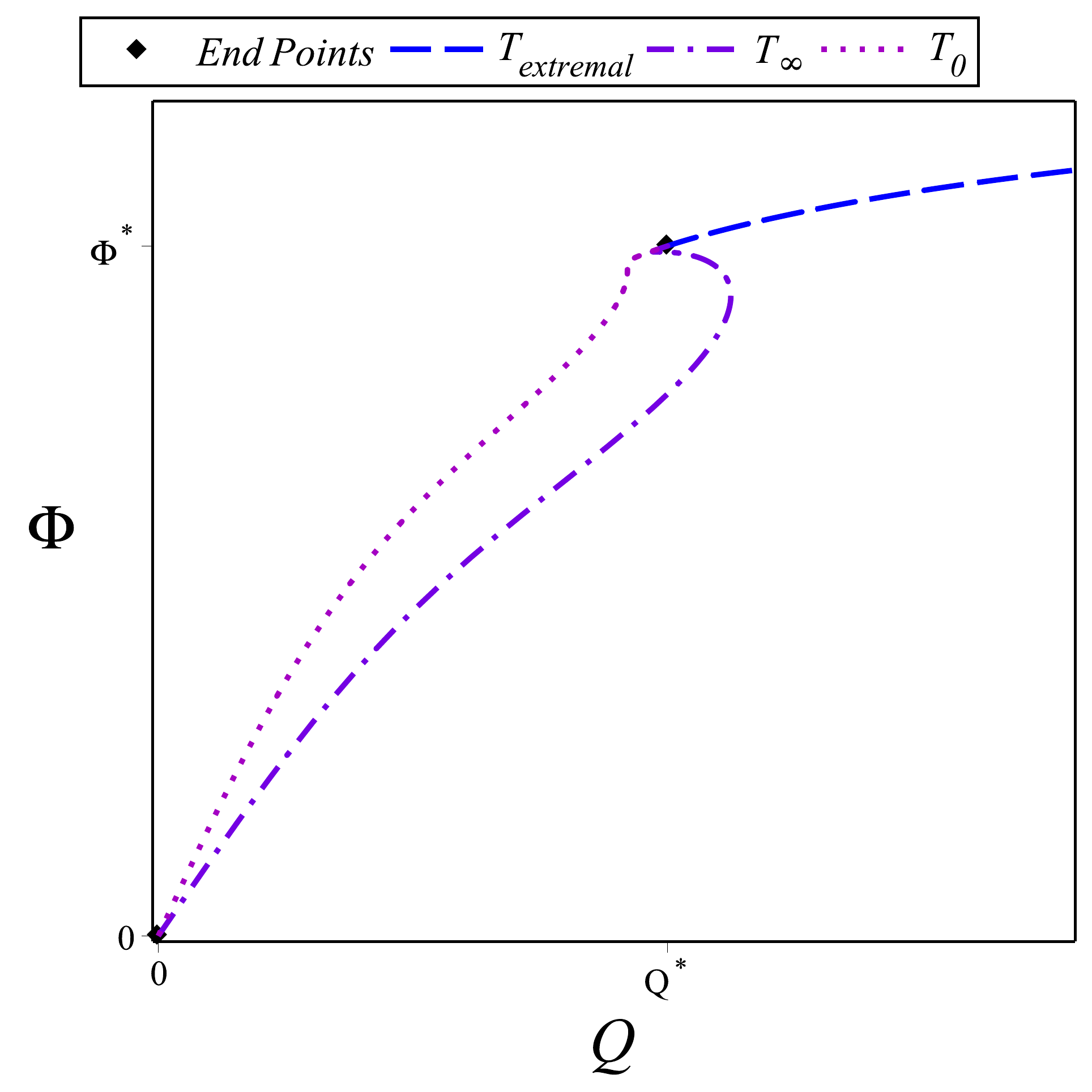}}
	\qquad
	\subfigure[Zoom]{\includegraphics[width=6.75cm]{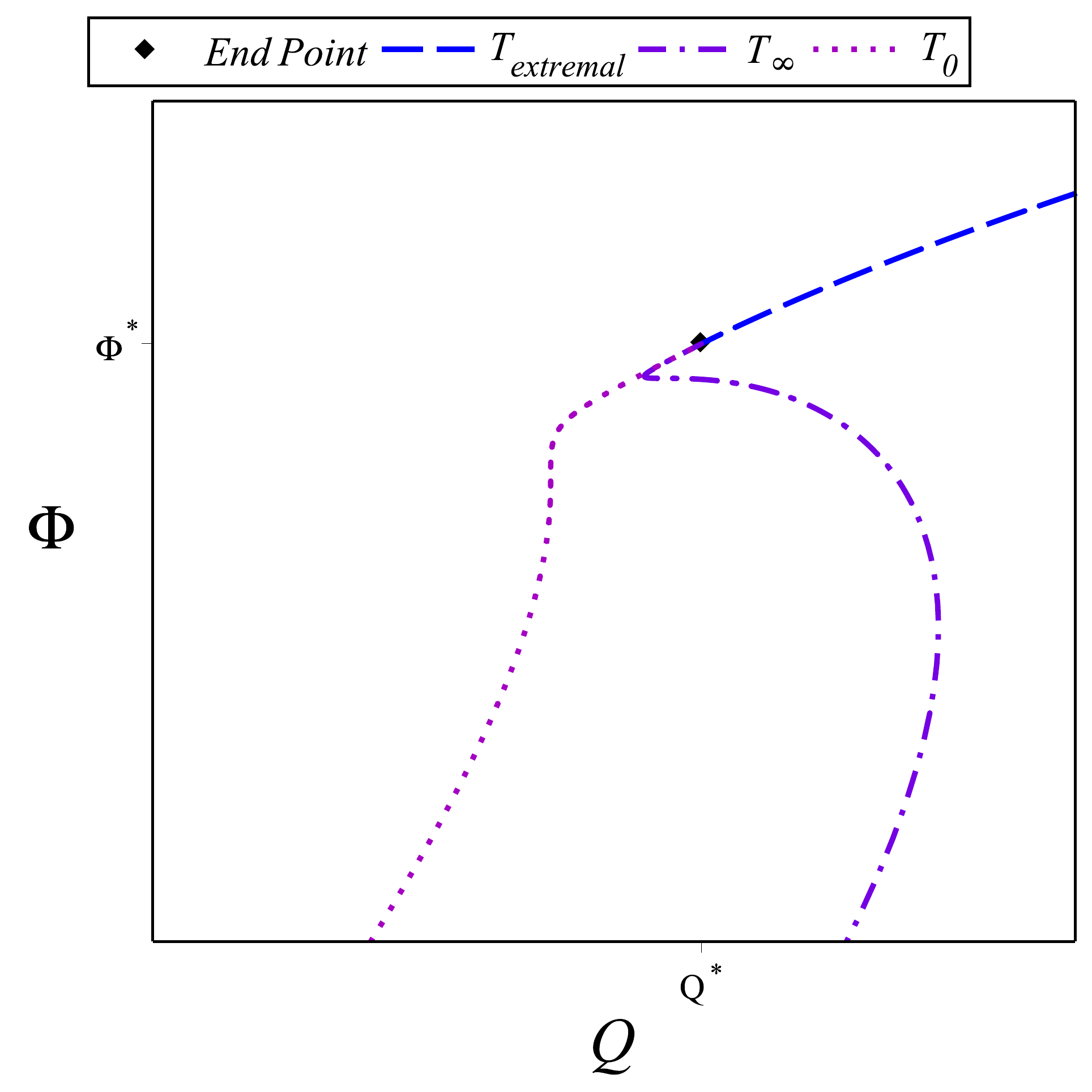}}
	\caption{\textbf{(a)} Sketch of the relevant isotherms $\Phi$ vs $Q$. \textbf{(b)} Zoom near one end point for the isotherms.}
	\label{fig:bosquejo1}
\end{figure}

By using the equations (\ref{chargpot}), for the charge and conjugate potential, equation (\ref{temp}) for the temperature, and the horizon equation, we obtain the following parametric compact expressions,
\begin{eqnarray}
{\Phi}^{2}&=&{\frac { 8\pi T\left[ 2\pi \nu x_{+}^{\nu} \left( {\nu}^{2}-4
		\right){T}\mp\sqrt { \left( {\nu}^{
				2}-4 \right) A_{T}(x_{+})} + B_{T}(x_{+})\right]
		\left( {\nu}^{2}-4 \right) }{ \left(\nu-1\right)^{-1}\left( x_{+}^{\nu}-1 \right)^{-1}\left[ 2 \pi \nu x_{+}^{\nu} \left( {\nu}^{2
		}-4 \right) T\mp\sqrt { \left( {\nu}^{2}-4 \right) A_{T}(x_{+})} \right]^{2}}},
\label{Phi_T}\\
Q^2&=&{\frac {\Phi ^{2} \nu^{2} \left( {\nu}^{2}-4 \right)^{2} {x}^{2\nu+2}}{16
		\left(2 \pi \nu x_{+}^{\nu}\left( {\nu}^{2}-4 \right) T\mp
		\sqrt { \left( \nu^{2}-4 \right) A_{T}(x_{+},T)}\right)^{2}}}
\label{Q_T}
\end{eqnarray}
where
\begin{equation}
B_{T}(x_{+}) = (x_{+}^{2}-x_{+}^{\nu})(\nu+2)+(x_{+}^{\nu+2}-1)(\nu-2)
\end{equation}
\begin{equation}
\begin{split}
A_{T}(x_{+})=& 4 \pi^{2} \nu^{2} \left( \nu^{2}-4 \right) T^{2} x_{+}^{2\nu
}+ \left( x_{+}^2(\nu-1)(\nu-2)-\nu(\nu+1)
\right) x_{+}^{2\nu}+ \\
&+2\, \left(  \left( x_{+}^{2}-1
\right) \nu+2 x_{+}^{2} \right)  \left( \nu-1 \right) x_{+}^{\nu}+x_{+}^{2
} \left( \nu+2 \right)
\end{split}
\end{equation}
The minus/plus signs in the equations (\ref{Phi_T}) and (\ref{Q_T}) correspond to the interval $1<\nu<2$ (the upper sign) and $\nu>2$ (the lower sign), respectively.
To analyze the equation of state, $Q$ vs $\Phi$ at constant temperature, let us first study some relevant limits regarding the extremal black holes,
\begin{eqnarray*}
	\Phi^{*}:= \lim_{x_{+}\rightarrow +\infty} \Phi|_{T=0}&=&\sqrt{2},\qquad \ \ \ Q^{*}:=\lim_{x_{+}\rightarrow +\infty} Q|_{T=0} =\frac{\nu\sqrt{2}\sqrt{\nu+2}}{4\sqrt{\nu-1}},\label{xoo}\\
	\lim_{x_{+}\rightarrow 1^{+}} \Phi|_{T=0}&=&2,\qquad \ \ \ \ \ \ \ \ \ \ \ \ \ \ \ \lim_{x_{+}\rightarrow 1^{+}} Q|_{T=0}\rightarrow +\infty \label{x_1}.
\end{eqnarray*}
Therefore, the isotherm $T=0$ has one ending point at $(Q^{*},\Phi^{*})$. This behaviour is shown in Fig. \ref{fig:bosquejo1}{\bf a}.
For non-extremal black holes $T\neq 0$, the limit $x_{+}\rightarrow +\infty$ gives the same result, which means that the point $(Q^{*}, \Phi^{*})$ is an end point for all the isotherms, while the limit $x_{+} \rightarrow 1$ yields $(Q,\Phi) \rightarrow (0,0)$, as shown also in Fig. \ref{fig:bosquejo1}{\bf a}. The critical temperatures $T_0$ and $T_\infty$, defined in Table \ref{table:def_var}, allow us to split the phase space into relevant regions which will be used to compare the relative signs of the electric permittivity $\epsilon_T$ and heat capacity $C_Q$. It is also worth emphasizing  that $T_\infty$ and $T_0$ depend on $\nu$, as can be seen in Fig. \ref{fig:CritF}{\bf a}.

\subsubsection{$S$ vs $T$, $Q$ fixed}
\begin{figure}[t!]
	\centering
	\subfigure[Critical temperatures]{\includegraphics[width=6cm]{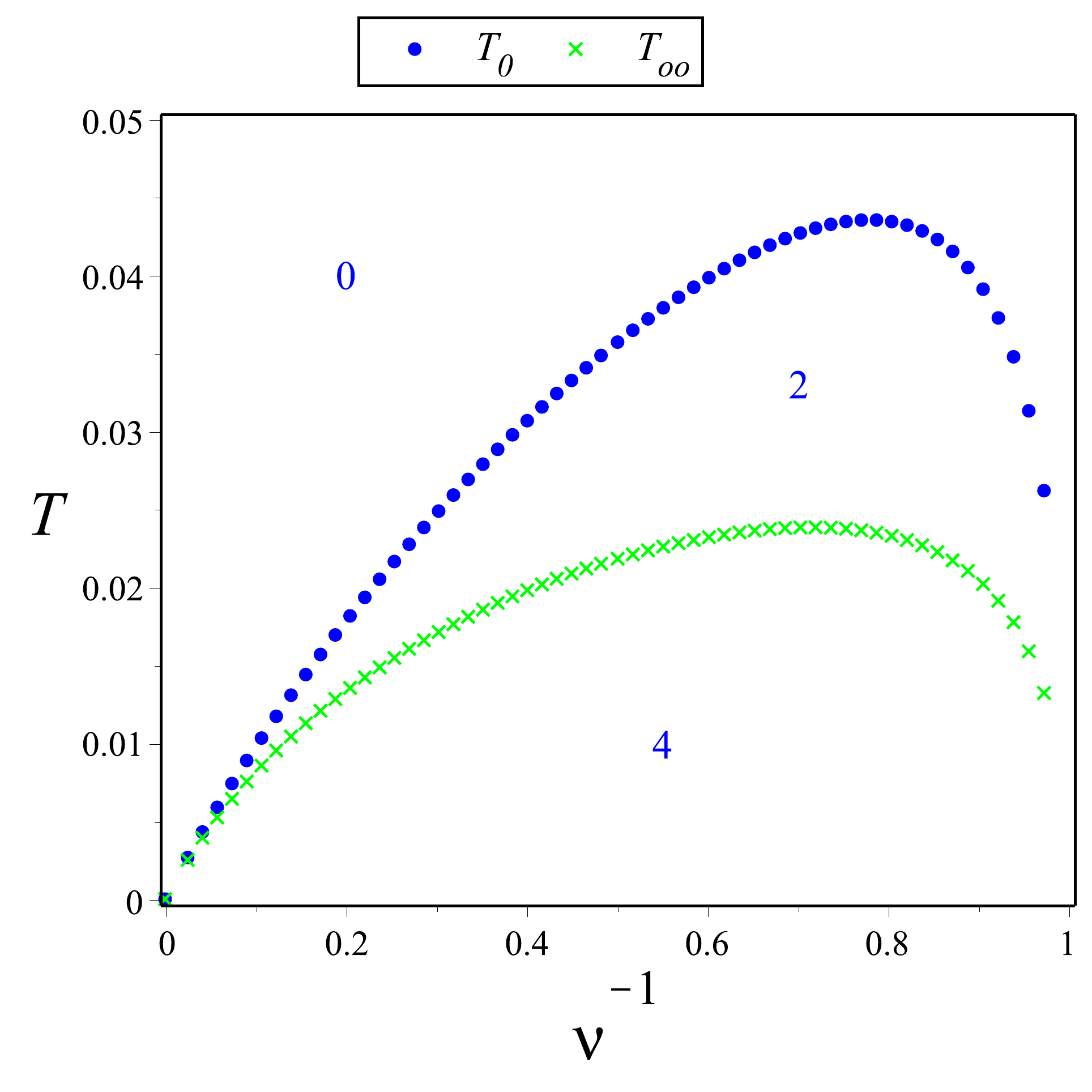}}
	\qquad\qquad
	\subfigure[Notice $Q_{0}<Q^{*}$]{\includegraphics[width=6.05cm]{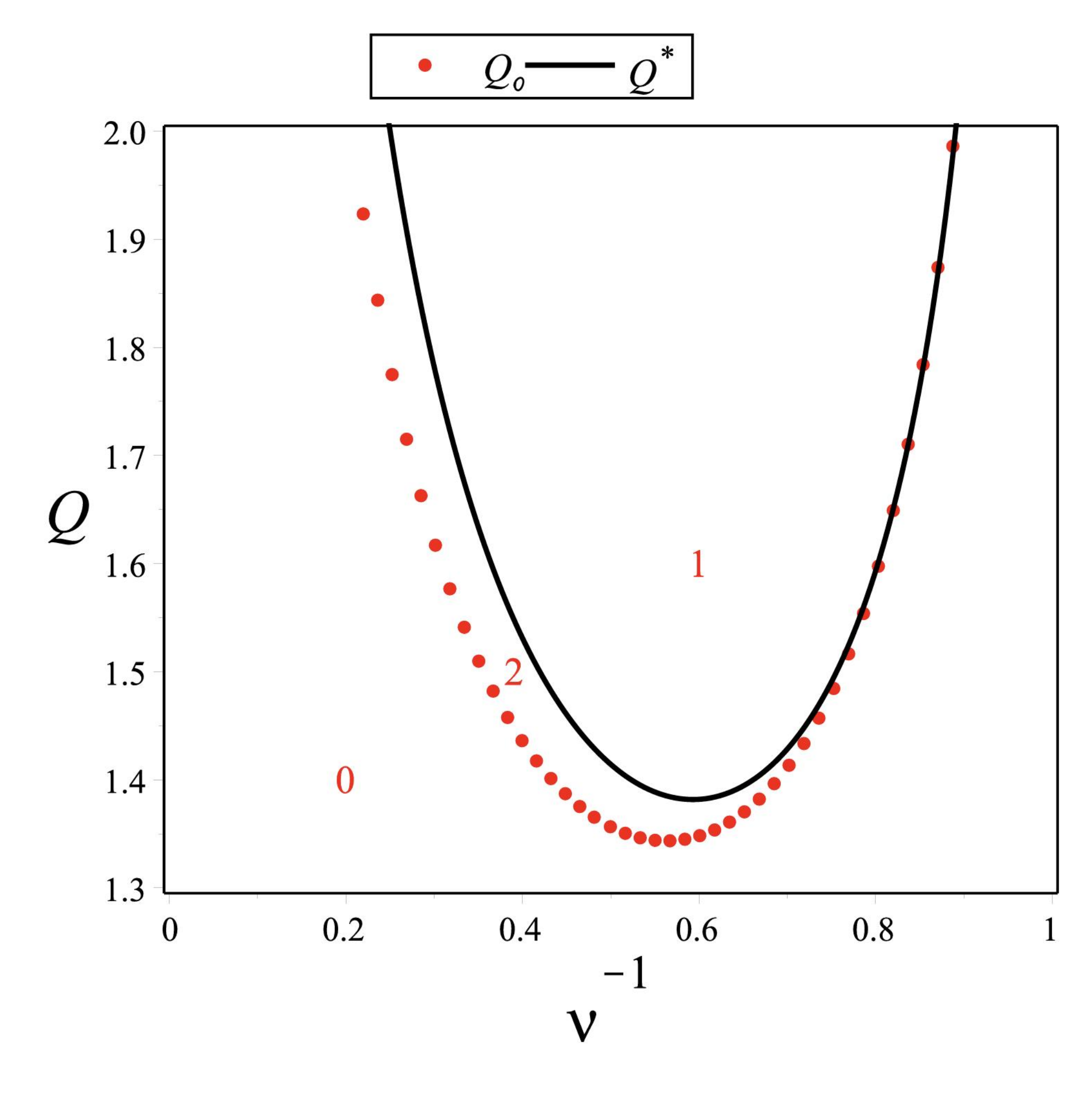}}
	\caption{\small \textbf{(a)} Critical temperatures. \textbf{(b)} Critical electric charge $Q_0$ and $Q^{*}$. The numbers indicate the maximum number of times that $\epsilon_{T}$ and $C_{Q}$ change their signs respectively.}
	\label{fig:CritF}
\end{figure}

We rewrite the relevant physical quantities in parametric form
\begin{equation}
S={\frac {16\pi \,{\nu}^{2}{Q}^{2} \left( {\nu}^{2}-4 \right)\left( x_{+}^{\nu}-1 \right) }{B_{Q} \left( x_{+},Q \right) }},
\end{equation}
\begin{equation}
T=\pm{\frac {16\left(x_{+}^{\nu}-1 \right) {Q}^{2}A_{Q}(x_{+})-x_{+}^{1+\nu}B_{Q} \left( x_{+},Q \right) }{16\nu\,\pi \,Q\sqrt {x_{+}^{3\,\nu+1} \left( {\nu}^{2}-4 \right)  \left( x_{+}^{\nu}-1 \right) B_{Q} \left( x_{+},Q \right) }}}
\label{T_Q}
\end{equation}
where
\begin{eqnarray*}
	A_{Q}(x_{+})&=&
	\[\left(\nu-2\right)x_{+}^{2\nu+2}
	-2\nu x_{+}^{\nu}\]\left(\nu-1\right)
	+\[2\left(\nu-1\right)x_{+}^{\nu+2}+x_{+}^{2}-\nu x_{+}^{2\nu}
	\]\left(\nu+2\right), \\
	B_{Q}(x_{+},Q)&=&\nu\,x_{+} \left( x_{+}^{\nu}-1 \right) ^{2} \left( \nu-1 \right)  \left( {\nu}^{2}-4 \right) \mp \sqrt {x_{+}^{-\nu}\nu\, \left( \nu-1 \right) \left( {\nu}^{2}-4 \right) \left( x_{+}^{\nu}-1 \right)^{3}E_{Q}(x_{+},Q)},\\
	E_{Q}(x_{+},Q)&=&32\left[  \left( \nu-2 \right) x_{+}^{2 \nu + 2}- \left( {\nu}^{2}-4 \right)x_{+}^{\nu+2}+{\nu}^{2}x_{+}^{\nu}-\left(\nu+2 \right) x_{+}^{2} \right]Q^{2}\\
	&&+x_{+}^{\nu+2}\nu \left( \nu-1 \right)  \left( {\nu}^{
		2}-4 \right)  \left( x_{+}^{\nu}-1 \right)
\end{eqnarray*}
Again, the minus sign corresponds to the interval $1<\nu<2$ and the plus sign is used in the range $\nu>2$. The limits of interest are now
\begin{eqnarray*}
	\lim_{x_{+}\rightarrow +\infty} S&=&0,\qquad\ \ \ \lim_{x_{+}\rightarrow +\infty} T \rightarrow \lbrace -\infty,0,+\infty\rbrace,\label{xxooST}\\
	\lim_{x_{+}\rightarrow 1^{+}} S &\rightarrow & \infty,\qquad\ \ \lim_{x_{+}\rightarrow 1^{+}} T= 0.\label{xx_1ST}
\end{eqnarray*}
The $\lim_{x_+\rightarrow\infty}{T}=-\infty$ occurs when $Q>Q^{*}$, and $\lim_{x_+\rightarrow\infty}{T}=+\infty$ when $Q<Q^{*}$. This suggests the existence of a critical line at $Q_{0}<Q^{*}$, see Fig. \ref{fig:CritF}{\bf b}.

\begin{figure}[t!]
	\centering
	\subfigure{\includegraphics[width=7cm,height=7cm]{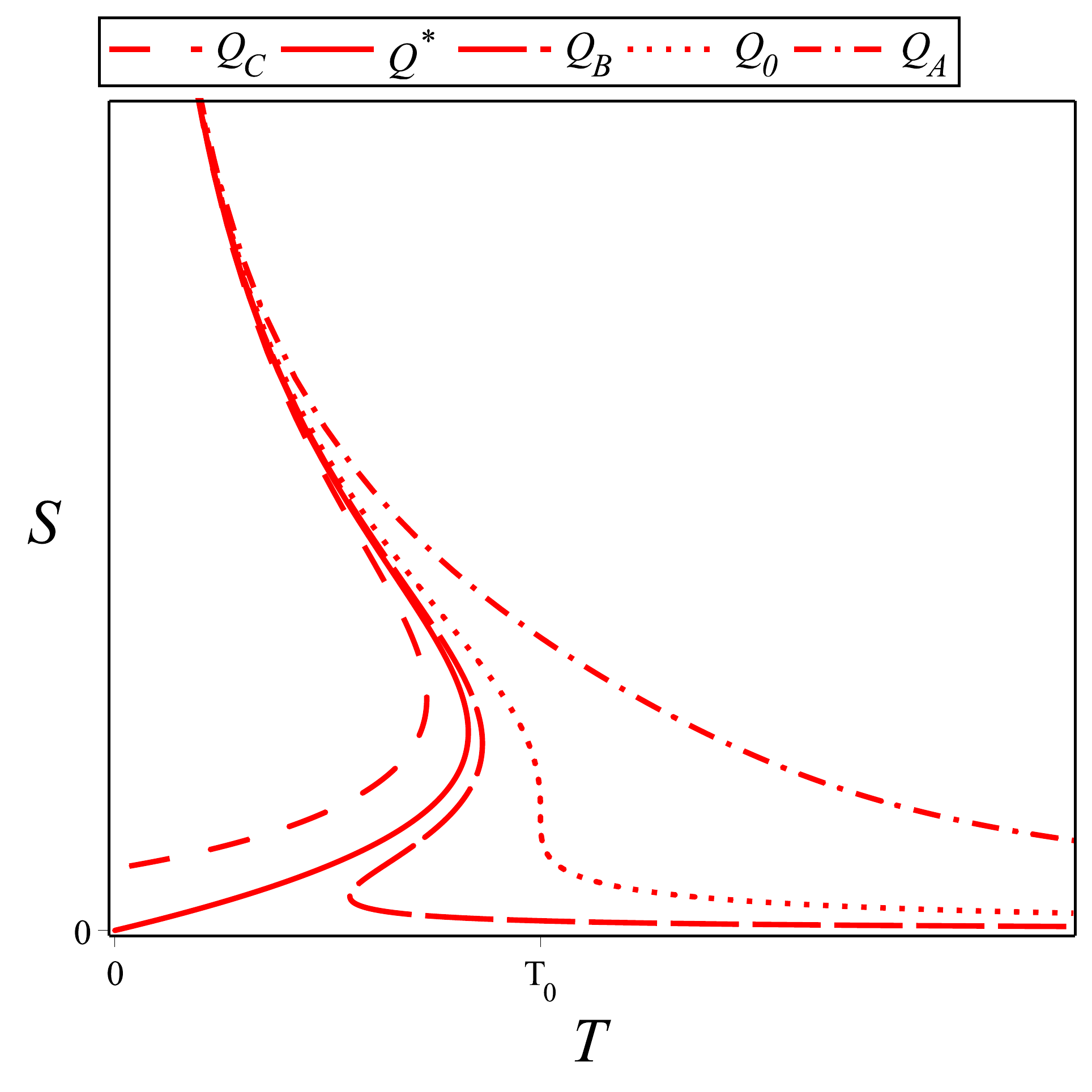}}
	\caption{\small Sketch of $S$ vs $T$ for some relevant values of the charge: $Q_{A}<Q_{0}<Q_{B}<Q^{*}<Q_{C}$.}
	\label{fig:Sketch_ST_Q}
\end{figure}

The information obtained so far allows us to plot $S$ vs $T$ for a fixed $Q$, see Fig. \ref{fig:Sketch_ST_Q}. With respect to the stability, while it is clear that $C_{Q}<0$ for all $Q<Q_{0}$, the case $Q>Q_0$ must be carefully analyzed. 

\subsubsection{Thermodynamic stability}

\begin{figure}[t!]
	\centering
	\subfigure[The phase space $S$ vs $T$ separated into the relevant regions]{\includegraphics[width=13cm]{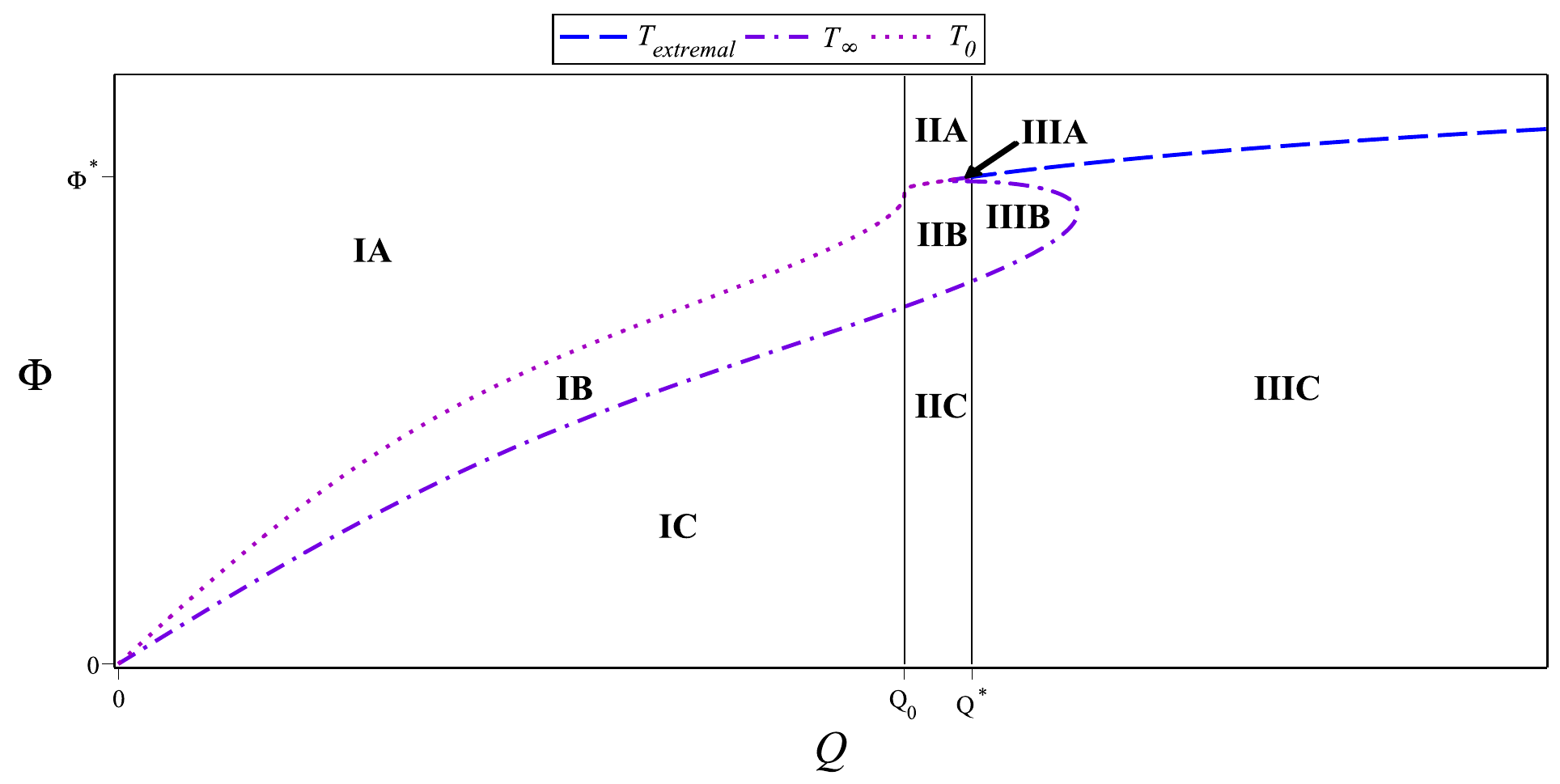}}
	\subfigure[The phase space $S$ vs $T$ separated into the relevant regions]{\includegraphics[width=13cm]{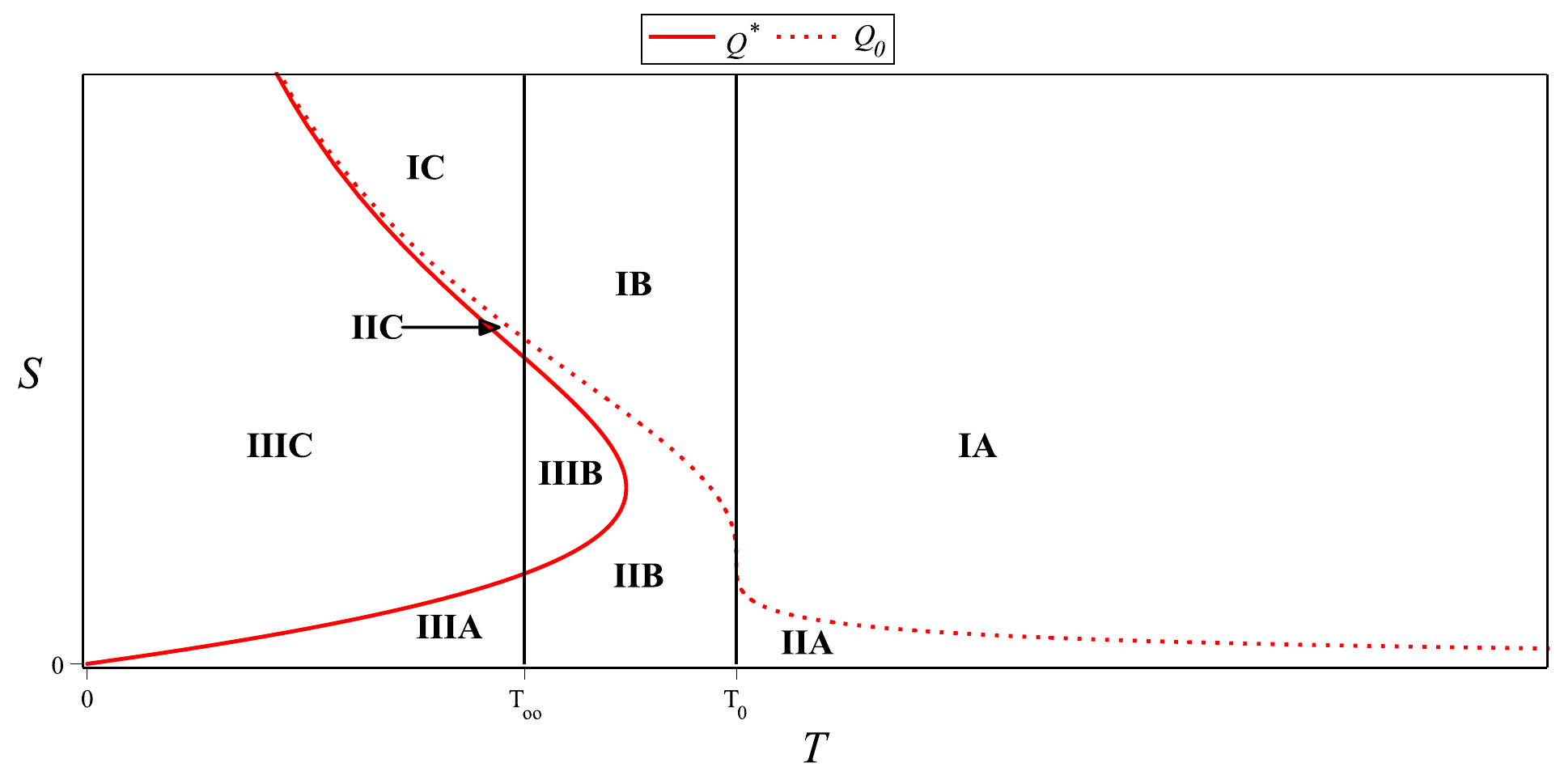}}
	\caption{\small The phase spaces: \textbf{(a)} $\Phi$ vs $Q$ and \textbf{(b)} $S$ vs $T$.}
	\label{fig:CritR}
\end{figure}

To begin the analysis of thermodynamic stability, first note that $T_0$ and $Q_0$ define a critical point (their definitions are provided in Table \ref{table:def_var}). This can be checked by inserting $T_0$ into the equation (\ref{Q_T}) and obtain $Q_0$. 
This observation makes possible to identify in a practical and direct manner the regions where the response functions are positively defined, as it is shown in Fig. \ref{fig:CritR}. Before presenting the details of the analysis, we would like to emphasize that by comparing  the number of times each response function changes its sign in a specific region, we can extract important information with the help of Fig. \ref{fig:CritF}, e.g. the relevant intervals for the thermodynamic stability are $T<T_{\infty}$ and $Q>Q_0$. However, we have to make sure that all regions with unstable configurations are excluded and now we investigate this issue case by case (see Fig.  \ref{fig:CritR}).

\subparagraph*{Regions IA, IB, IC:}
These regions are characterized by $Q<Q_{0}$. The electric permittivity is positive, but the heat capacity is negative and so all the configurations are thermodynamically unstable. 
\subparagraph*{Regions IIA, IIC:}
Within these two regions, characterized by $Q_0<Q<Q^{*}$ and $T>T_0$ (for \textbf{IIA}) and $T<T_\infty$ (for \textbf{IIC}), we observe that $\epsilon_T>0$ but $C_Q<0$, therefore, there are no thermodynamically stable black holes.
\subparagraph*{Region IIB:}
In this region, characterized by $Q_0<Q<Q^{*}$ and $T_\infty<T<T_0$, the response functions change two times their signs. Despite that, it is not straightforward to determine whether both are positive for a given configuration, as required for thermodynamic stability. We should discuss further this case below.
\subparagraph*{Region IIIA:}
Inside this region, $C_{Q}$ changes one time its sign, while $\epsilon_{T}$ changes its sign at least one time. Again, this is not sufficient to prove that both of them are positive for a given configuration and we investigate this case below.
\subparagraph*{Region IIIB:}
Since both response functions change their signs only one time, one can not conclude about the stability. We shall comment on this case below.
\subparagraph*{Region IIIC:}
Inside this region, $\epsilon_{T}$ can change its sign at least one time, while the sign of $C_{Q}$ changes only one time. That is not sufficient to prove that both of them are positive for a given configuration. We shall comment on this case right below.

\begin{figure}[t!]
	\centering
	\includegraphics[width=9.0cm]{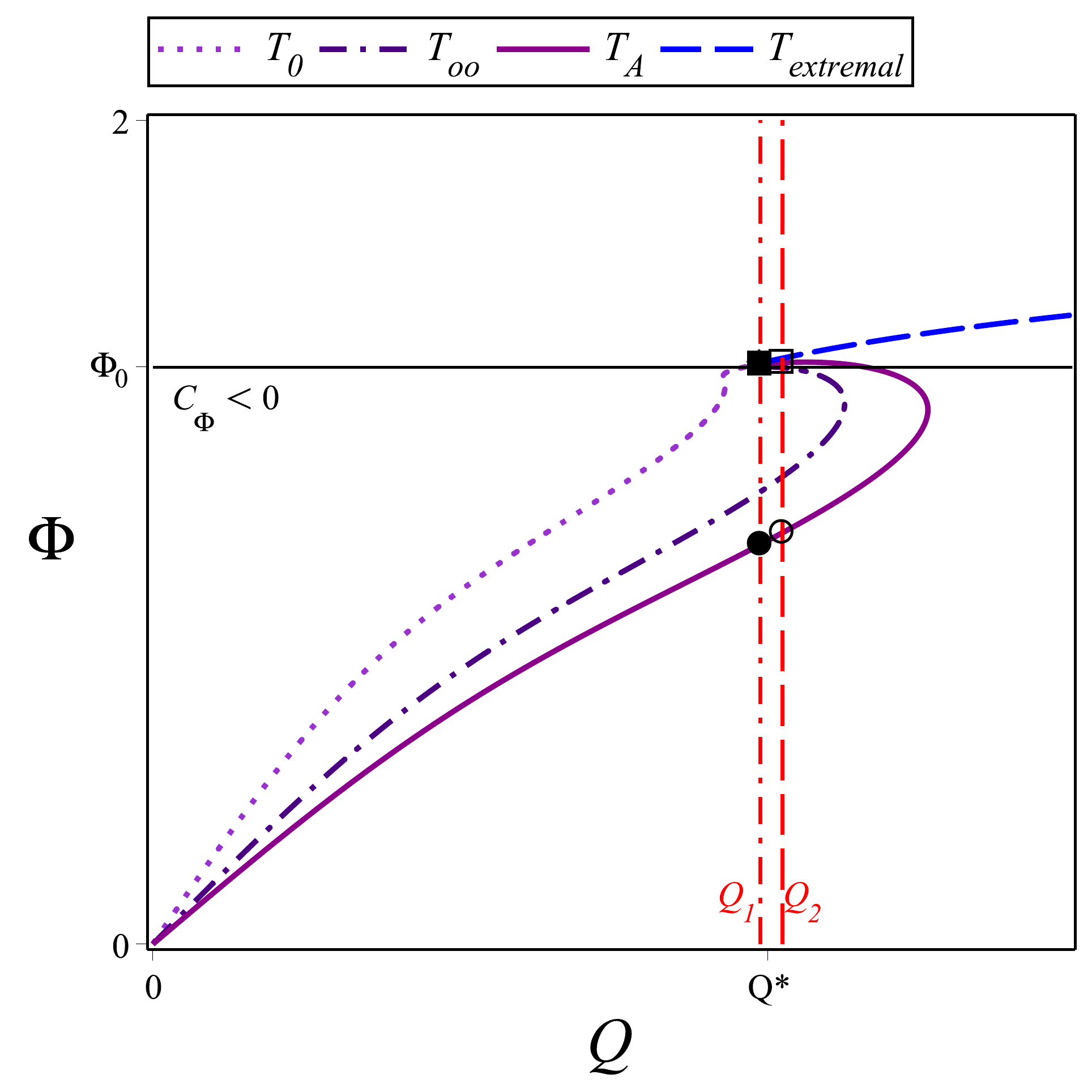} 
	\caption{\small $\Phi$ vs $Q$ at fixed $T$. The relevant isotherms are $T_0$ and $T_\infty$. An arbitrary isotherm $0<T_A<T_\infty$ is traced. Two arbitrary isocharges, $Q_1<Q^{*}$ and $Q^{*}<Q_2$, and their intersections with $T_{A}$ was marked. The line $Q=Q_1$ intersects three times $T_A$ (the two intersections at the top cannot be distinguished at this scale. See Fig. \ref{fig:FA_PQ_ST}{\bf a} for a zoomed image). The line $Q=Q_2$ intersects two times $T_A$.}
	\label{fig:PQ_T_F}
\end{figure}

So far, we can definitely conclude that the regions \textbf{IA}, \textbf{IB}, \textbf{IC}, \textbf{IIA} and \textbf{IIC} do not contain locally stable configurations. Now, to complete the analysis, it will be useful to consider the following thermodynamic relation:
\begin{equation}
C_{\Phi}=C_{Q}+\epsilon_{T}\alpha^{2}_{Q}T
\label{RTCPhi}
\end{equation}
where $\alpha_{Q}\equiv(\partial \Phi/\partial T)_{Q}$. From the results presented in Section \ref{grand} for the grand canonical ensemble, we know that $C_{\Phi}<0$ as long as $\Phi<\Phi_{0}$, which implies that either $C_Q<0$ or $\epsilon_T<0$. {For $\Phi_0<\Phi$, there is a sub-region within \textbf{IIB} where one of the three configurations at a given $T_\infty<T<T_0$ has $C_\Phi>0$ (see Fig. \ref{fig:CritR}{\bf b}), but it has also $\epsilon_T<0$ as inferred after drawing the corresponding isotherm in Fig. \ref{fig:CritR}{\bf a}.} Therefore, the regions \textbf{IIB} and also \textbf{IIIB} contain no thermodynamically stable configurations. This is better appreciated in Fig. \ref{fig:PQ_T_F}, where the line $\Phi=\Phi_0$ was explicitly marked.

\begin{figure}[t!]
	\centering
	\subfigure[Zoom to the stability region (see Fig. \ref{fig:PQ_T_F})]
	{\includegraphics[width=7.0cm]{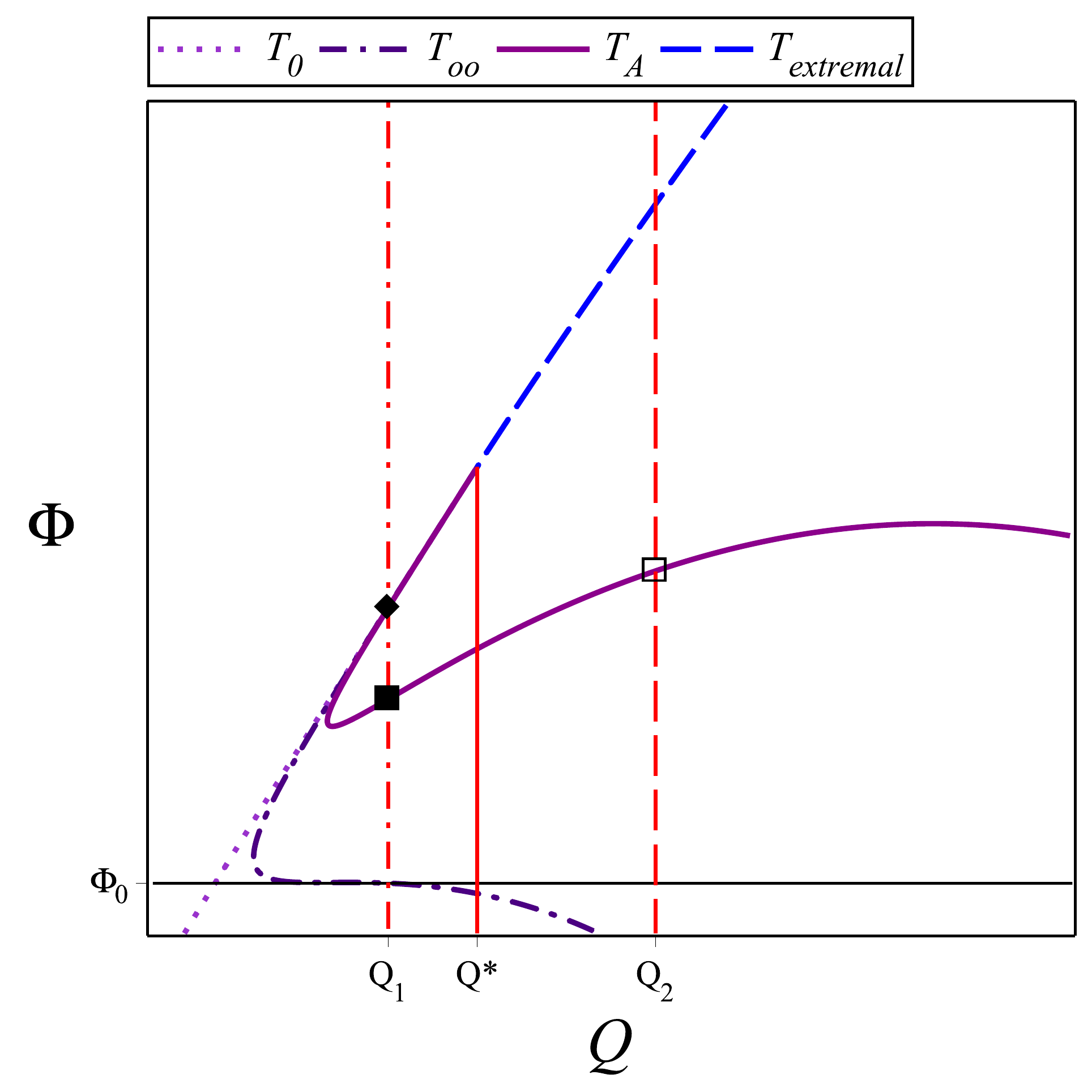}} 
	\quad
	\subfigure[$Q_{0}<Q_{1}<Q^{*}<Q_{2}$] {\includegraphics[width=7.0cm]{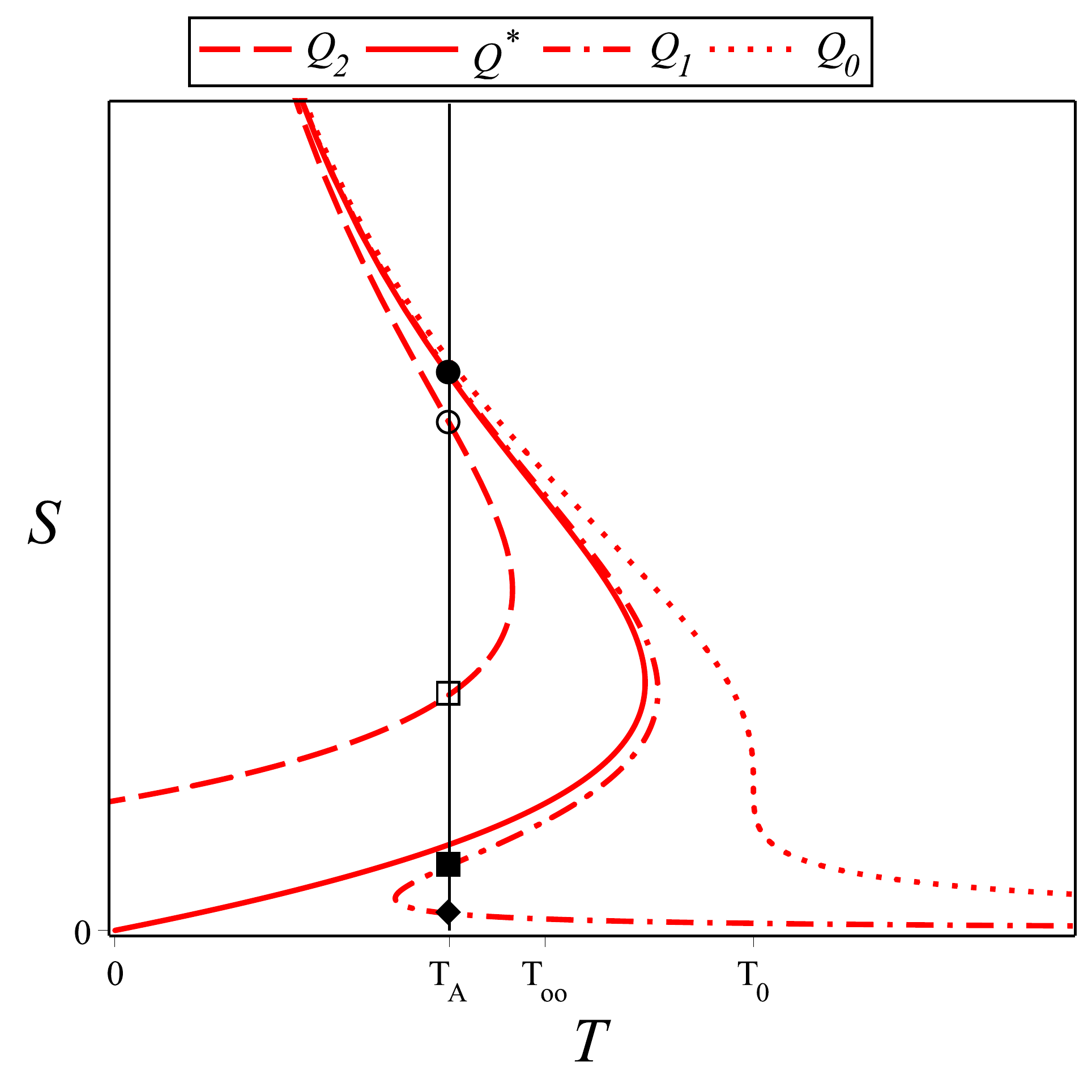}}
	\caption{\small \textbf{(a)} A closer look into the region of stability. The intersections between $Q_{1}$, $Q_{2}$ and $T_{A}$ are represented with different symbols and, in \textbf{(b)}, it is shown the some intersection points by the same symbols. \textbf{(b)} $S$ vs $T$ at fixed $Q$; it is shown the intersection between $T_{A}$ with $Q_{1}$ and $Q_{2}$. }
	\label{fig:FA_PQ_ST}
\end{figure}

In what follows, we are going to consider only the cases $\Phi>\Phi_0$, corresponding to the remaining regions to be analized: \textbf{IIIA} and \textbf{IIIC}. The equation (\ref{RTCPhi}) does not provide enough information to conclude about the stability in the case $\Phi>\Phi_{0}$. The reason is that $C_{\Phi}$ is not always positive inside this interval. In fact, inside both regions \textbf{IIIA} and \textbf{IIIC}, the heat capacity can have negative or positive values (see Figs. \ref{fig:PQ_T_F} and \ref{fig:FA_PQ_ST}{\bf b}, where two curves at constant charges, $Q_1$ and $Q_2$, are plotted).
Then, in order to determine whether these regions contain thermodynamic stable configurations, let us consider the isotherm $T_A<T_\infty$ depicted in Fig. \ref{fig:PQ_T_F} and magnified in Fig. \ref{fig:FA_PQ_ST}{\bf a}. If $Q<Q^{*}$, the isotherm $T_A$ represents the region \textbf{IIIA} and, if $Q>Q^{*}$, represents region \textbf{IIIC}, as follows from Fig. \ref{fig:CritR}{\bf a}.\footnote{The region IIIA is a tiny region which is better appreciated in Fig. \ref{fig:FA_PQ_ST}{\bf a}.}. That is why we have considered two isocharge lines, $Q_1$ and $Q_2$, so that $Q_0<Q_1<Q^*$ and $Q^*<Q_2$. The isotherm $T_A$ intersects three times the isocharge line $Q_1$ and two times the isocharge line corresponding to $Q_2$. 

Let us focus first on the three intersection points between the curves characterized by $Q_1$ and $T_A$. We have used three different symbol for each intersection: a filled diamond (with the highest value of $\Phi$ in $\Phi$ vs $Q$ and the lowest value of $S$ in $S$ vs $T$), a filled squared (with an intermediate value of $\Phi$ and an intermediate value of $S$, respectively), and a filled circle (with lowest $\Phi$ and highest $S$, respectively). From Fig. \ref{fig:FA_PQ_ST}, it follows that, indeed, the branch represented by the filled squared (located within \textbf{IIIA}) contains configurations with both $\epsilon_T>0$ and $C_Q>0$. These thermodynamically stable black holes, however, do not have an extremal limit, because the region $Q<Q^{*}$ does not contain extremal black holes.

Let us now consider $T_A$ and an isocharge line $Q_2>Q^{*}$. From the plot $S$ vs $T$ it is clear that there are two branches of configurations and that one of them has $C_Q>0$. Interestingly, in the plot $\Phi$ vs $Q$, $Q_2$ intersects $T_A$ for two configurations. If $Q_2$ is close enough to $Q^{*}$, then the two intersections correspond to branches where $\epsilon_T>0$. Therefore, inside this region, contained within \textbf{IIIC}, there are also thermodynamically stable black holes which, in addition, have a well defined extremal limit.

Since the charge $Q_{1}$ is arbitrary and the temperature $T_{A}$ is just limited by the condition $T_{A}<T_{\infty}$, we can certainly conclude that the most general region of stability in the plot of $\Phi-Q$ at constant $T$ is found between the local minimum and maximum of every isotherm such that $T<T_{\infty}$ (see Fig. \ref{fig:FA_PQ_ST}{\bf a}), namely, inside a subset of both the region \textbf{IIIA} and \textbf{IIIC}.

\section{Linear spherically symmetric perturbations}
\label{sec5}

In this section we are going to study the stability of the hairy black holes studied in the previous sections under linear spherically symmetric perturbations. The anaylsis follows the steps developed previously for scalarized Reissner-Nordstr\"om black holes \cite{Blazquez-Salcedo:2019nwd,Blazquez-Salcedo:2020nhs,Blazquez-Salcedo:2020jee} (see also \cite{Astefanesei:2019qsg, Leaver:1990zz,Ferrari:2000ep,Myung:2018vug,Myung:2018jvi,Brito:2018hjh,Jansen:2019wag,Blazquez-Salcedo:2018jnn, Anabalon:2013baa}), and that it has also been used for black holes in alternative theories of gravity with scalar fields (see for example \cite{Blazquez-Salcedo:2020rhf,Blazquez-Salcedo:2020caw,Blazquez-Salcedo:2017txk,Blazquez-Salcedo:2016enn,Torii:1998gm,Ayzenberg:2013wua,Blazquez-Salcedo:2018pxo}).

Up to first order in the perturbation parameter $\epsilon$, the metric (\ref{metric}) becomes
\begin{equation}
ds^2=\Omega(x)\[1+\epsilon e^{-i\omega t}F_z(x)\]
\left\{
-f(x)\[1+\epsilon e^{-i\omega t}F_t(x)\]dt^2
+\frac{\eta^2\[1+\epsilon e^{-i\omega t}F_r(x)\]dx^2} {f(x)\[1+\epsilon e^{-i\omega t}F_t(x)\]}+d\Sigma^2
\right\}
\end{equation}
where $d\Sigma^2\equiv d\theta^2+\sin^2\theta d\varphi^2$. Also, at first order, the perturbed electromagnetic field and scalar field are
\begin{equation}
A_t(x,t)=a(x)\[1+\epsilon e^{-i\omega t}F_{a}(x)\],
\qquad
\Phi(x,t)=\phi(x)\[1+\epsilon e^{-i\omega t}F_\phi(x)\]
\end{equation}
where $a(x)=-\frac{q}{\nu x^{\nu}}$ and $\phi(x)=\sqrt{\nu^{2}-1}\ln(x)$ are given by (\ref{fields}). The functions $F_z$, $F_t$, $F_r$, $F_a$, and $F_\phi$ are the perturbation functions all associated to a Fourier mode with frequency $\omega$.
By choosing the gauge $F_z=0$, it can be shown that the equations of motion reduce to a single master equation of Schr\"odinger type
\begin{equation}
\frac{d^2Z}{dR^2}=\(-\omega^2+U\)Z
\end{equation}
where $Z\equiv \phi\sqrt{\Omega}F_\phi$. The new coordinate $R$ is related to the $x$-coordinate by the transformation
$dR=\frac{\eta\,dx}{f(x)}$
and the effective potential $U=U(x)$ is
\begin{align}
U(x)&=\left[{\frac {2\phi'( V\gamma+V_1) {\Omega}^{2}}{\Omega'}} -\left({\frac{\phi'^{2}f'}{\Omega'{\eta}^{2}}} +{\frac{4\gamma\phi'}{\Omega'}}-V_2 +V{\gamma}^{2}\right)\Omega +{\frac{2\gamma\phi'f'}{{\eta}^{2}}} -{\frac{\Omega'f'(2{\gamma}^{2}-1)} {2\Omega{\eta}^{2}}}+2{\gamma}^{2}\right]f  \notag\\
&\quad +\left[{\frac{{\Omega}^{2}\phi'^{4}} {2{\eta}^{2}\Omega'^{2}}} -{\frac{\gamma\phi'^{3}\Omega}{\Omega'{\eta}^{2}}} +{\frac{\left(2{\gamma}^{2}-7\right)\phi'^{2}} {4{\eta}^{2}}} +{\frac {3\gamma\,\Omega'\phi'}{\Omega{\eta}^{2}}} -{\frac{\Omega'^{2}\left(3{\gamma}^{2}-1\right)} {2{\Omega}^{2}{\eta}^{2}}}\right] {f}^{2}
\label{effective}
\end{align}
where $V_{1}=\frac{dV}{d\phi}$ and $V_2=\frac{d^2V}{d\phi^2}$. 

The stability of the solution is given by the positivity of the effective potential $U(x)$ between the event horizon and the boundary of spacetime  (see for instance \cite{Kimura:2017uor,Kimura:2018eiv,Kimura:2018whv}). Since the domain for positive branch is $1<x<x_+<\infty$, it is more useful to plot $U$ vs $x^{-1}$, where $x_+^{-1}<x^{-1}<1$. 

Near the boundary, the effective potential can be expanded as
\begin{equation}
\label{Uexpanded}
U(x)=\[\frac{32Q^2\eta^4}{\nu-1}+\alpha-\eta^2\](x-1)^3 +\mathcal{O}\[(x-1)^4\]
\end{equation}
and so the effective potential asymptotically vanishes. From the equation (\ref{effective}), we also observe that the effective potential vanishes at the event horizon. 

%Since the thermodynamic stability was found for $Q>Q^*\equiv \frac{\nu\sqrt{2(\nu+2)}}{4\sqrt{\alpha(\nu-1)}}$, we would like to establish the mechanical stability at least in that range.
%

\begin{figure}[t!]
	\centering \subfigure[$x_+\approx 4.7358$]
	{\includegraphics[width=4.9cm]{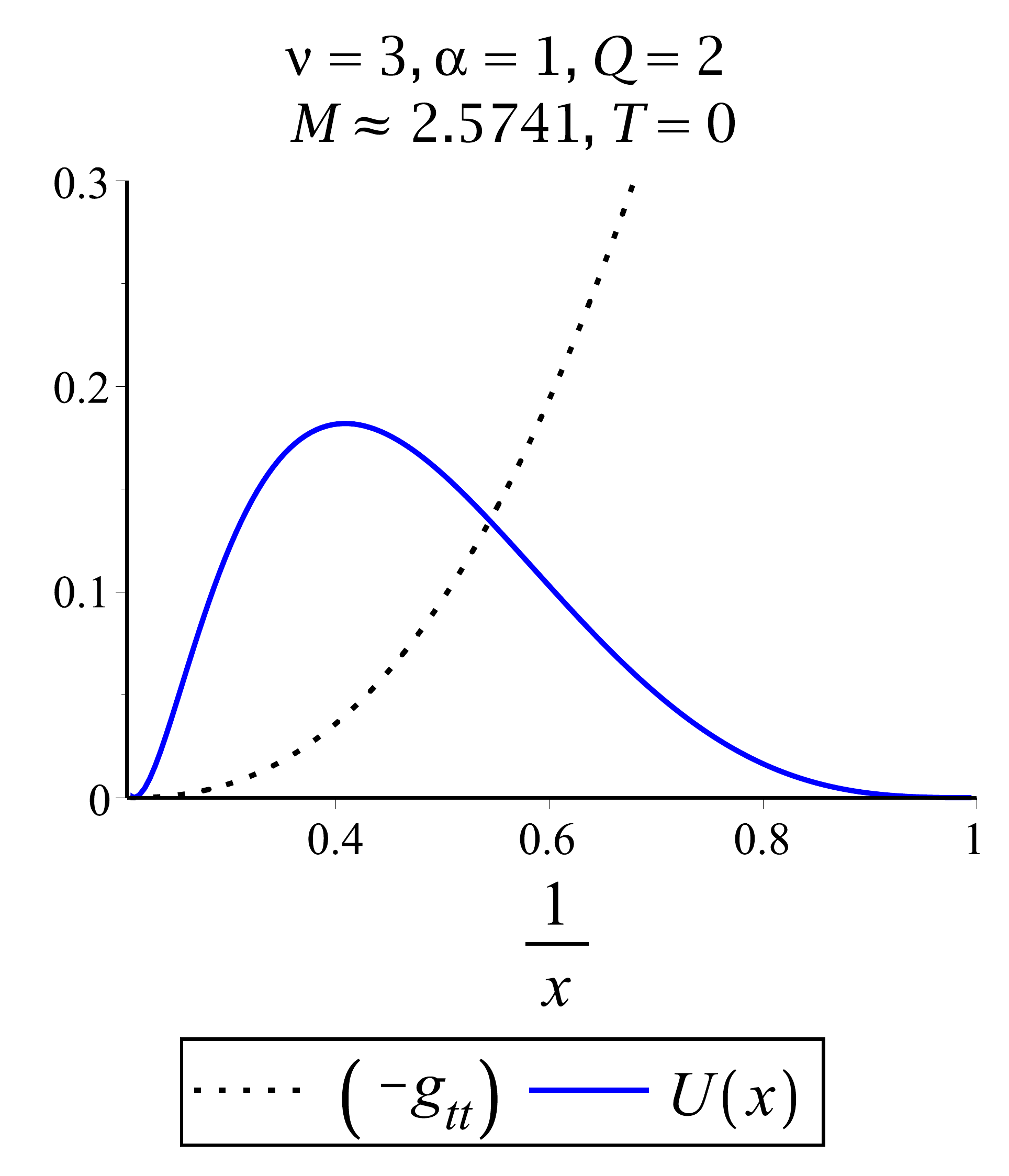}}
	\subfigure[$x_+\approx 2.9960$]
	{\includegraphics[width=4.9cm]{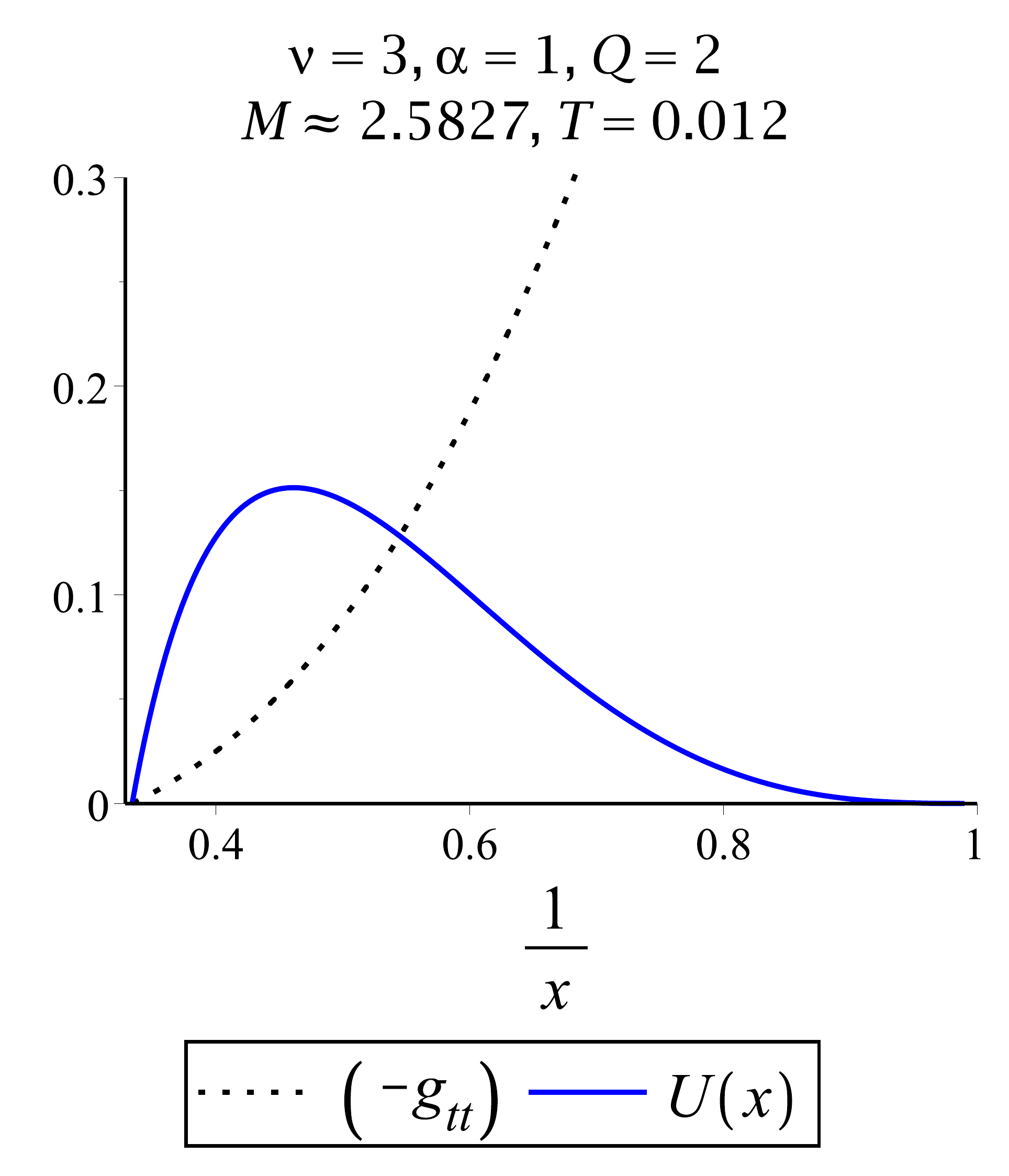}}
	\subfigure[$x_+\approx 1.7531$]
	{\includegraphics[width=4.9cm]{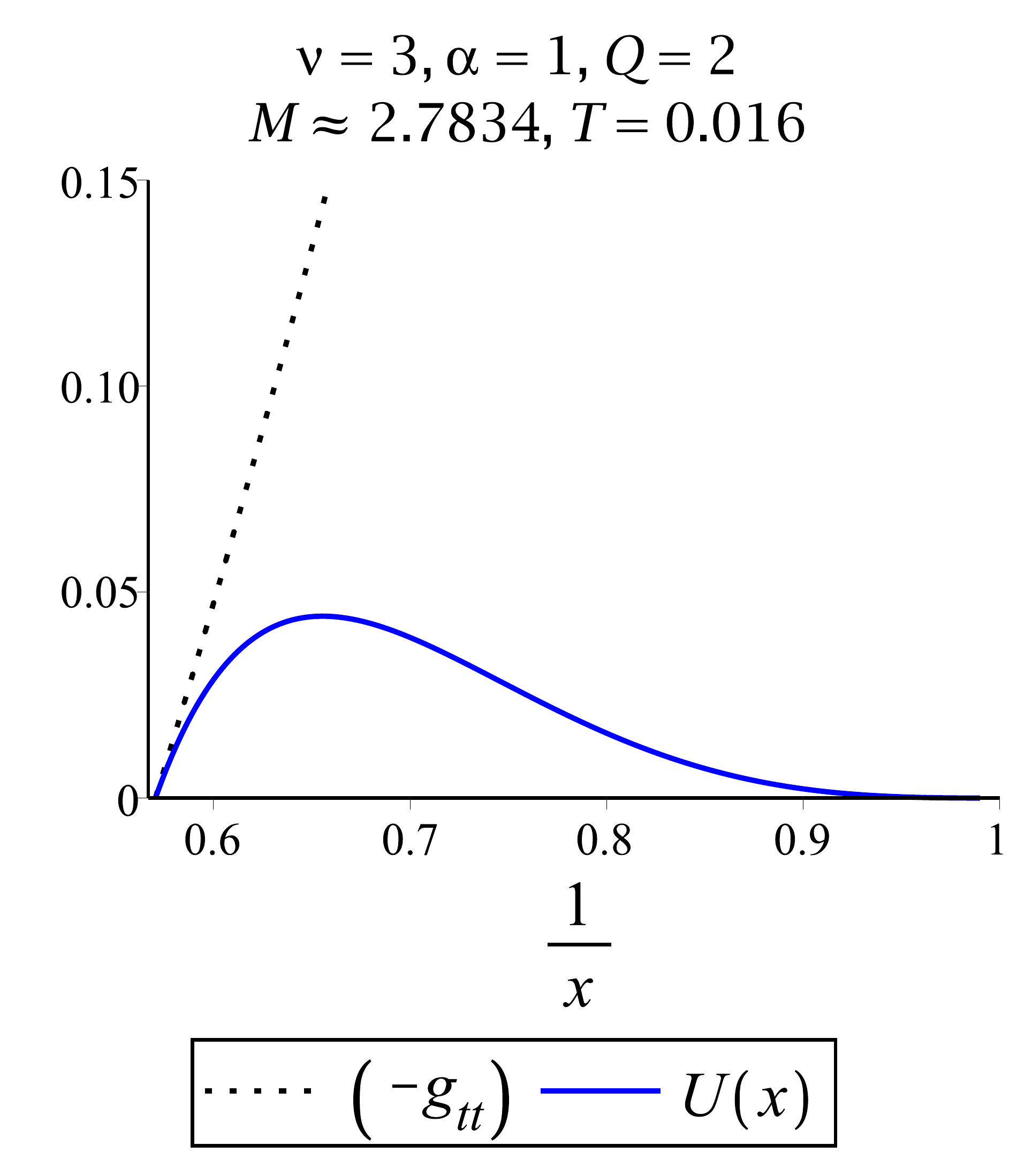}}
	\caption{\small Effective potential $U(x)$ vs $x^{-1}$ for the theory $\alpha=1$ and $\nu=3$. The electric charge of the black hole is $Q=2$. The graphics show the region between the event horizon $x=x_+$ and the boundary $x=1$, where $-g_{tt}>0$. \textbf{(a)} Extremal black hole $T=0$. \textbf{(b)} $T=0.012$. \textbf{(c)} $T=0.016$.} \label{stability1}
\end{figure}

A few typical profiles of the potential, as a function of $1/x$, are presented in Fig. \ref{stability1}. The metric function $-g_{tt}$ is represented by the dashed black line and the effective potential $U(x)$ by the solid blue line. In each panel we have plotted these functions for three distinct solutions in the theory with $\nu=3$ and $\alpha=1$. Notice that in this model, $Q^{*}=1.67705$, and we have considered $Q=2$. According to the previous analysis, when $T<0.018$, these configurations contain thermodynamically stable black holes (when $\Phi>\Phi_0$.). 
In Fig. \ref{stability1}{\bf a}, we present the extremal case with $T=0$. The solution extends from the black hole horizon at $x=x_+$, where both the $g_{tt}$ function and the effective potential vanish, up to the asymptotic boundary at $x=1$, where $-g_{tt}=1$ and the potential vanishes again. In particular, we can see that the potential is always positive outside the horizon, meaning the solution is stable under radial perturbations. The two other solutions in Fig. \ref{stability1}{\bf b} and {\bf c} are non-extremal black holes with $T=0.012$ and $T=0.016$, respectively. Qualitatively, the results are similar to the extremal case and so these solutions are also stable under spherical perturbations.
\begin{figure}[t!]
	\centering \subfigure[$x_+\approx 4.7663$]
	{\includegraphics[width=4.9cm]{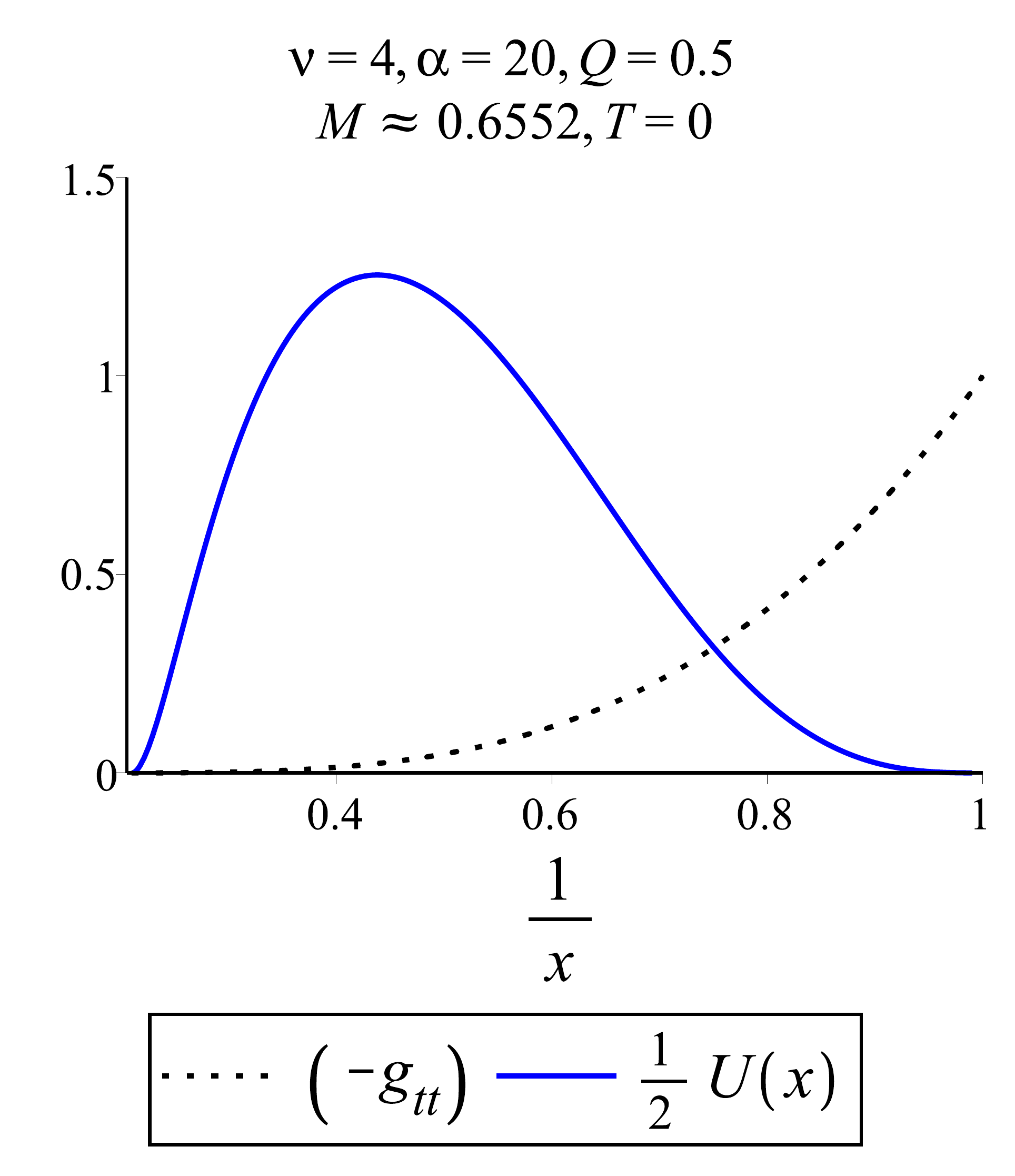}}
	\subfigure[$x_+\approx 3.4227$]
	{\includegraphics[width=4.9cm]{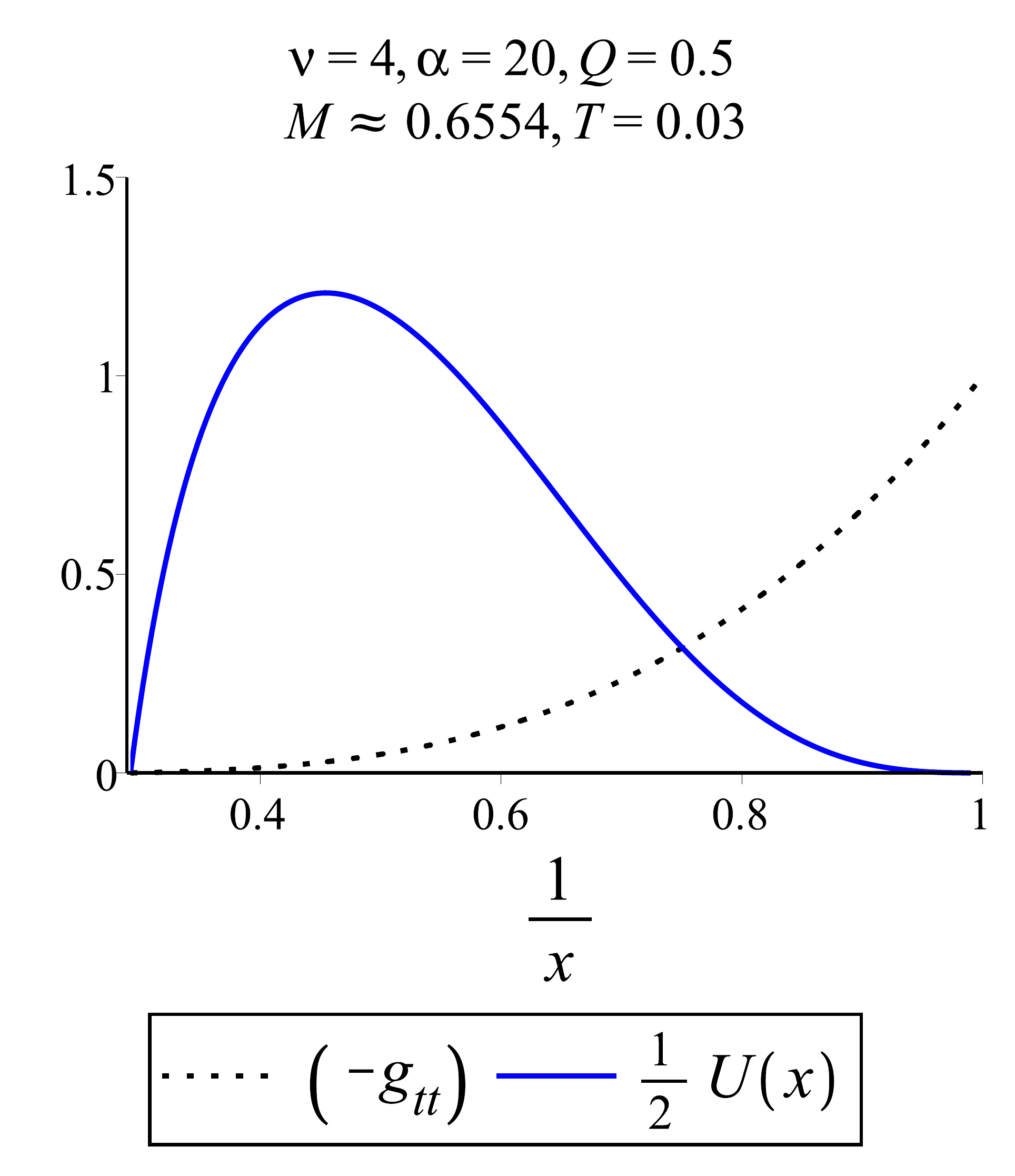}}
	\subfigure[$x_+\approx 2.1723$]
	{\includegraphics[width=4.9cm]{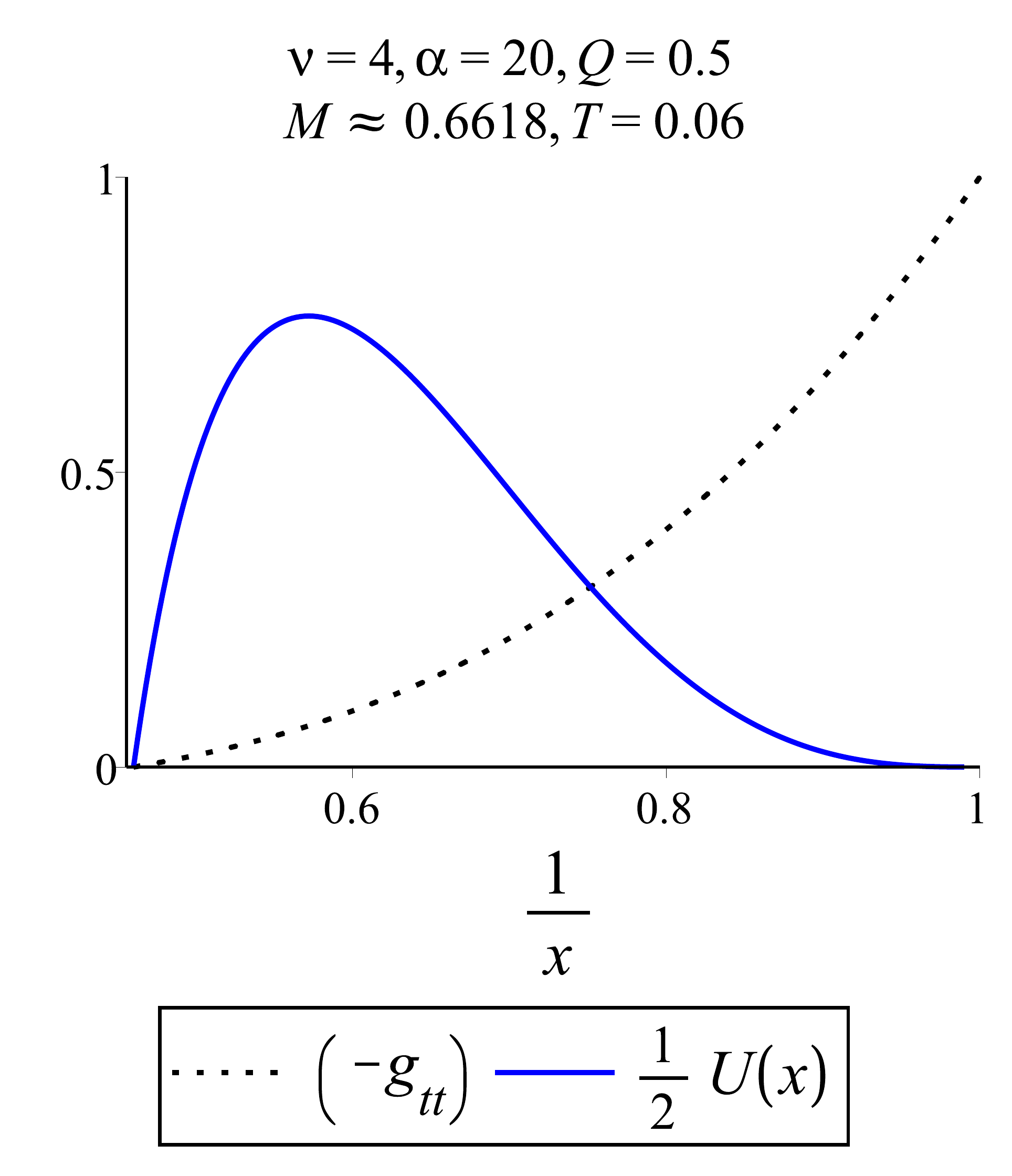}}
	\caption{\small Effective potential $U(x)$ vs $x^{-1}$ for the theory $\alpha=20$ and $\nu=4$. The electric charge of the black hole is $Q=1/2$. The graphics show the region between the event horizon $x=x_+$ and the boundary $x=1$, where $-g_{tt}>0$. \textbf{(a)} Extremal black hole $T=0$. \textbf{(b)} $T=0.03$. \textbf{(c)} $T=0.06$.} \label{stability2}
\end{figure}

Interestingly, the positivity of the effective potential seems to be generic. In Fig. \ref{stability2}, we present similar plots for a few solutions in a different theory with parameters $\nu=4$ and $\alpha=20$, where $Q^*=0.44721$. We consider $Q=1/2$ and so they are also thermodynamically stable. Each panel shows again three solutions with different values of the horizon temperature and, again, the effective potential is always positive definite. Scanning the space of solutions in various theories, it is always found that the positivity of the effective potential $U(x)$ is satisfied. This means that $\omega^2$ is positive and so the hairy black hole solutions, which are thermodynamically stable, are also stable against spherical perturbations.

\section{Conclusions}
\label{sec6}
\begin{figure}[t!]
	\centering
	{\includegraphics[width=6.4cm]{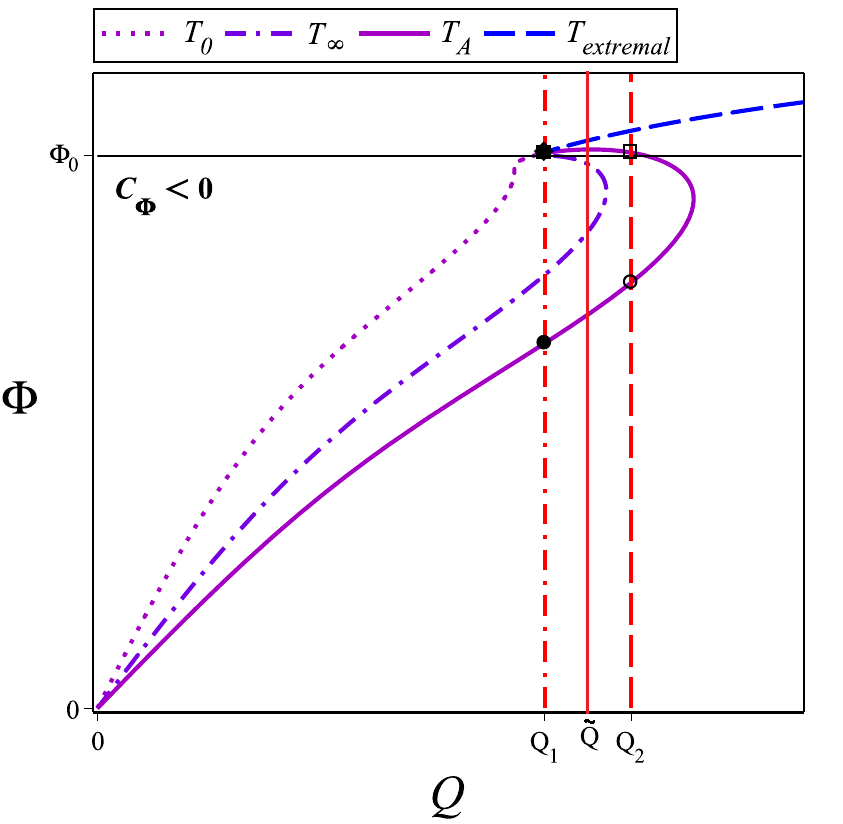}} \qquad\quad
	{\includegraphics[width=6.9cm]{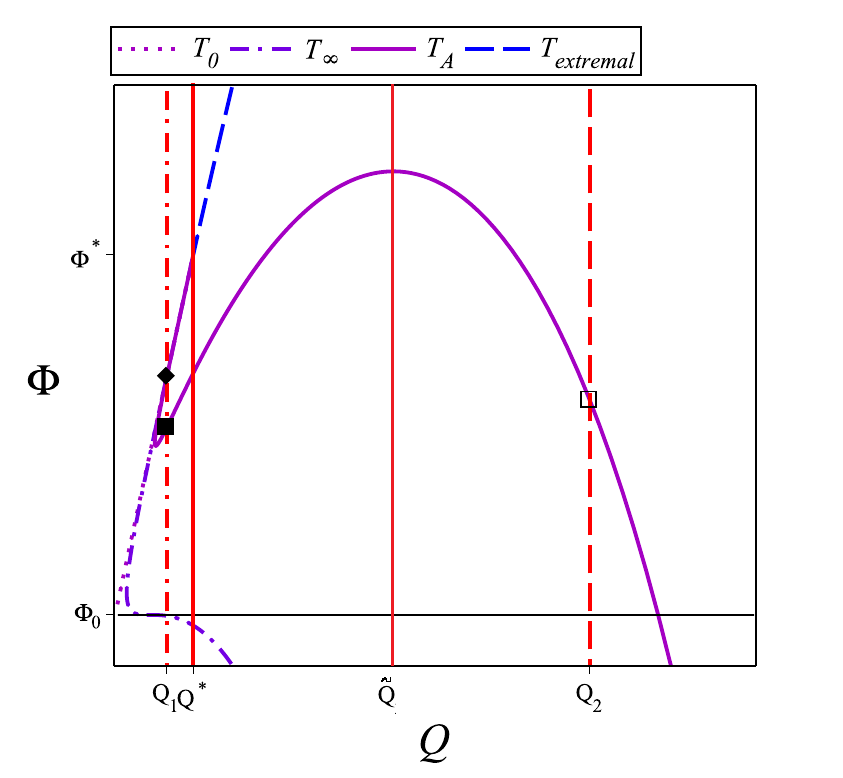}}
	\caption{\small Right hand side: $Q_2>\tilde Q$, where the configuration at a $\Phi>\Phi_0$ becomes unstable ($\epsilon_T<0$). Left hand side: The same image zoomed.} \label{fig:unstable}
\end{figure}

In this paper, we made a detailed examination of the thermodynamic stability of a general class of hairy charged black holes in flat spacetime. To obtain the results in a compact form for any value of the parameter $\nu$ in the dilaton potential, we have used a general criterion to check the relative signs of the relevant response functions (in both the grand-canonical and canonical ensemble) within well delimited regions of the corresponding phase space. Interestingly, we have found that there always exists a sub-class of stable black holes for which the heat capacity and permittivity are positive definite close to the extremality. Similar results were obtained previously in \cite{Astefanesei:2019mds} for a particular case, $\nu \longrightarrow \infty$.\footnote{This case is very special because the limit can not be taken directly in the general ansatz for the solution.} However, the phase structure is much richer when $\nu$ takes finite values. In this case, there exists a new sub-class of thermodynamically stable configurations characterized by only one horizon. Another new feature is the appearance  of swallowtail sections for a specific range of the electric charge, however a detailed analysis will appear elsewhere \cite{Astefanesei:2020toappear}.\footnote{{The scalar fields in AdS also change the black hole behaviour in grand canonical ensemble by generating swallowtail sections \cite{Astefanesei:2019ehu}, a feature that is not present for Reissner-Nordstr\"om-AdS black hole \cite{Chamblin:1999hg}.}} In grand canonical ensemble, the stable configurations exist only when the chemical potential exceeds a critical value $\Phi_{0}$. In the canonical ensemble, though, the region of stability is constrained to temperatures lower than a critical temperature $T_{0}$. Graphically, the stable configurations are those located between the local minimum and the maximum of the isotherm curves in Fig. \ref{fig:FA_PQ_ST}{\bf a}. If $Q$ exceeds $\tilde{Q}$, then $\epsilon_T$ becomes negative even if $C_Q>0$ and so the thermodynamic stability is lost. One concrete example of this is shown in Fig. \ref{fig:unstable}, where the black holes become unstable when the electric charge ($Q_2$) exceeds $\tilde{Q}$.

{In addition to the thermodynamical stability, we have checked stability against spherically symmetric perturbations. For this, we have followed the standard procedure used before for other families of hairy charged black holes \cite{Blazquez-Salcedo:2019nwd,Astefanesei:2019qsg,Blazquez-Salcedo:2020nhs,Blazquez-Salcedo:2020jee}. After perturbing the metric, the electromagnetic field and the scalar field, we parameterize the spherical perturbations in terms of a master equation, a generalized tortoise coordinate, and an effective potential (\ref{effective}). The analysis of this effective potential reveals that the thermodynamically stable solutions are also stable under radial perturbations: all solutions analyzed (extremal and non-extremal) possess a regular and positive definite potential. Hence the spherical perturbations of these black holes are always damped exponentially with time.}

{Let us also comment that, since the black holes that we have considered in this work are spherically symmetric, it is typically expected that perturbations on higher multipolar numbers will also be free of instabilities (see for instance examples of this in \cite{Blazquez-Salcedo:2019nwd,Blazquez-Salcedo:2020jee,Astefanesei:2019qsg,Blazquez-Salcedo:2020rhf,Blazquez-Salcedo:2020caw}), and it is reasonable to expect these configurations to be fully dynamically stable.}

\begin{figure}[t!]
	\centering
	\subfigure[$\Phi<\Phi_{0}$]{\includegraphics[width=4.6cm]{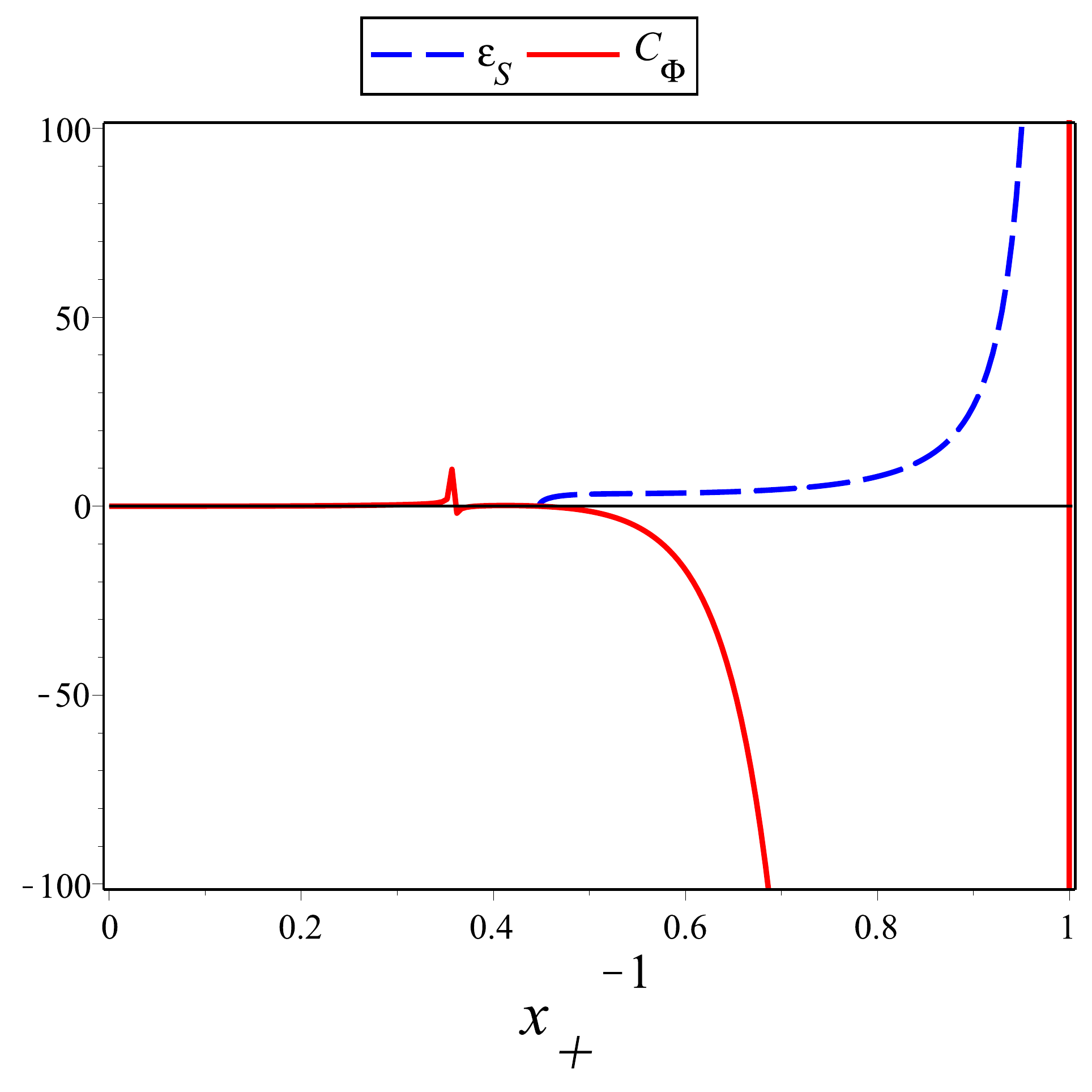}}
	\subfigure[$\Phi_{0}<\Phi<\Phi^{*}$]{\includegraphics[width=4.6cm]{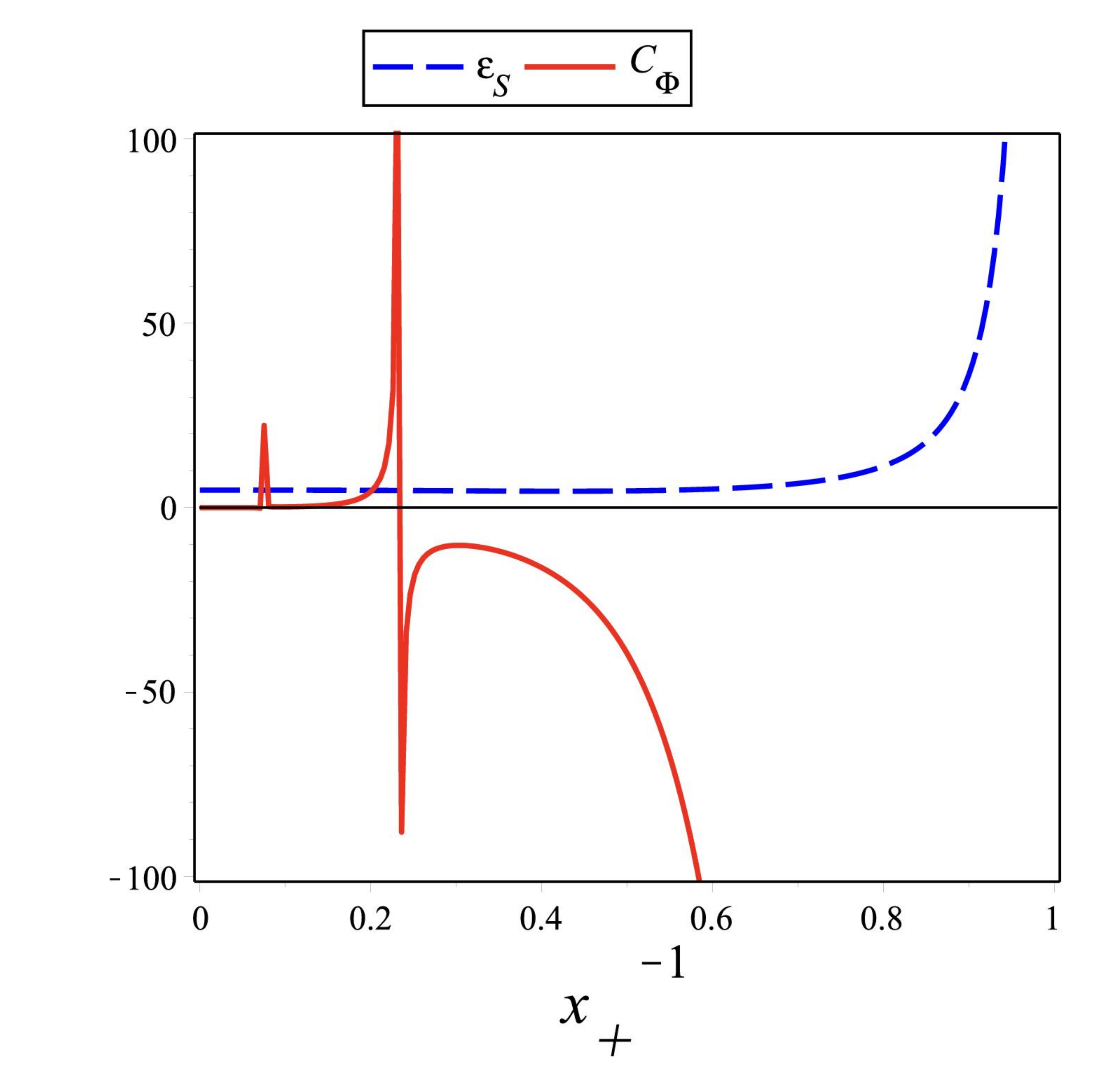}}
	\subfigure[$\Phi^{*}<\Phi$]{\includegraphics[width=4.6cm]{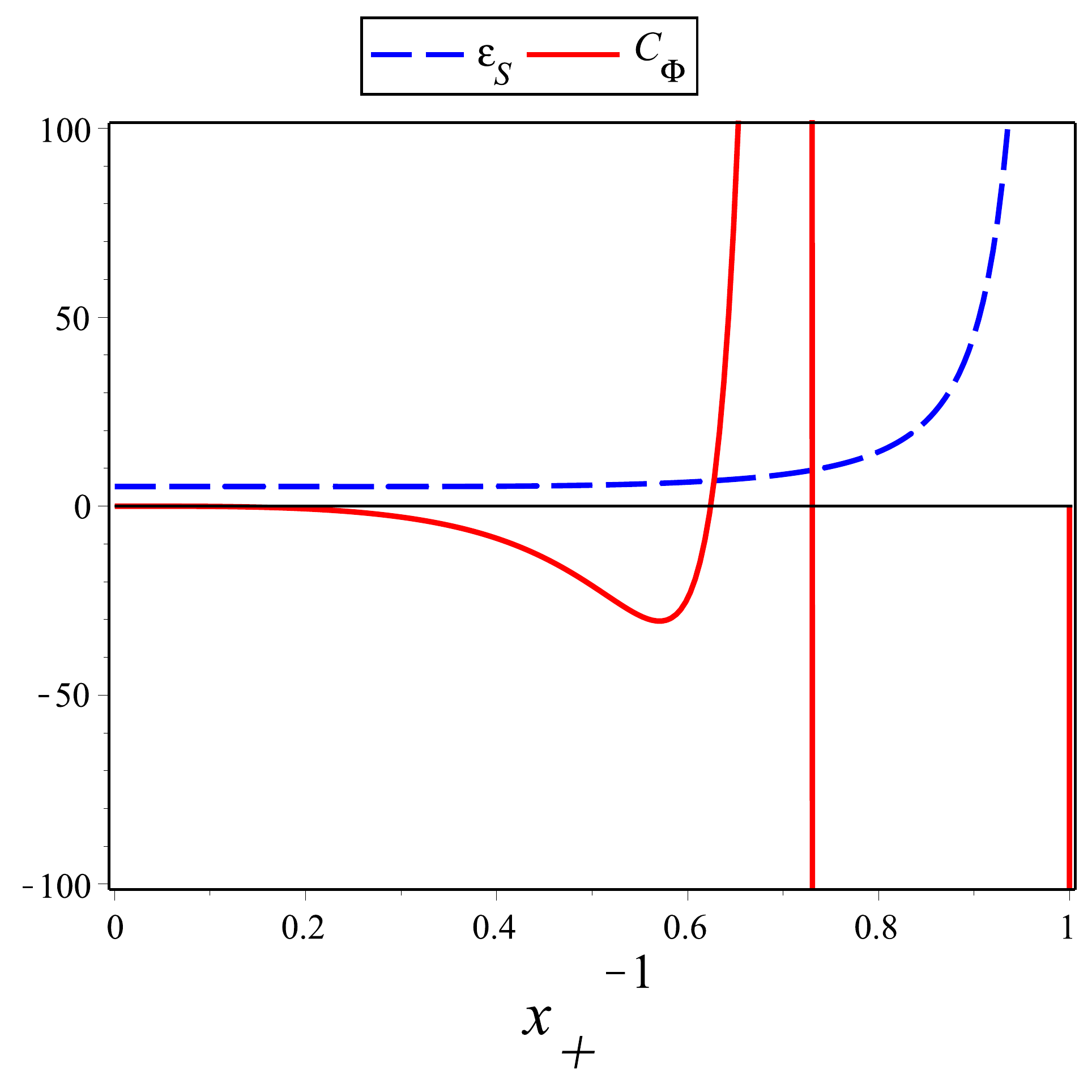}}
	\subfigure[$T<T_{\infty}$]{\includegraphics[width=4.8cm]{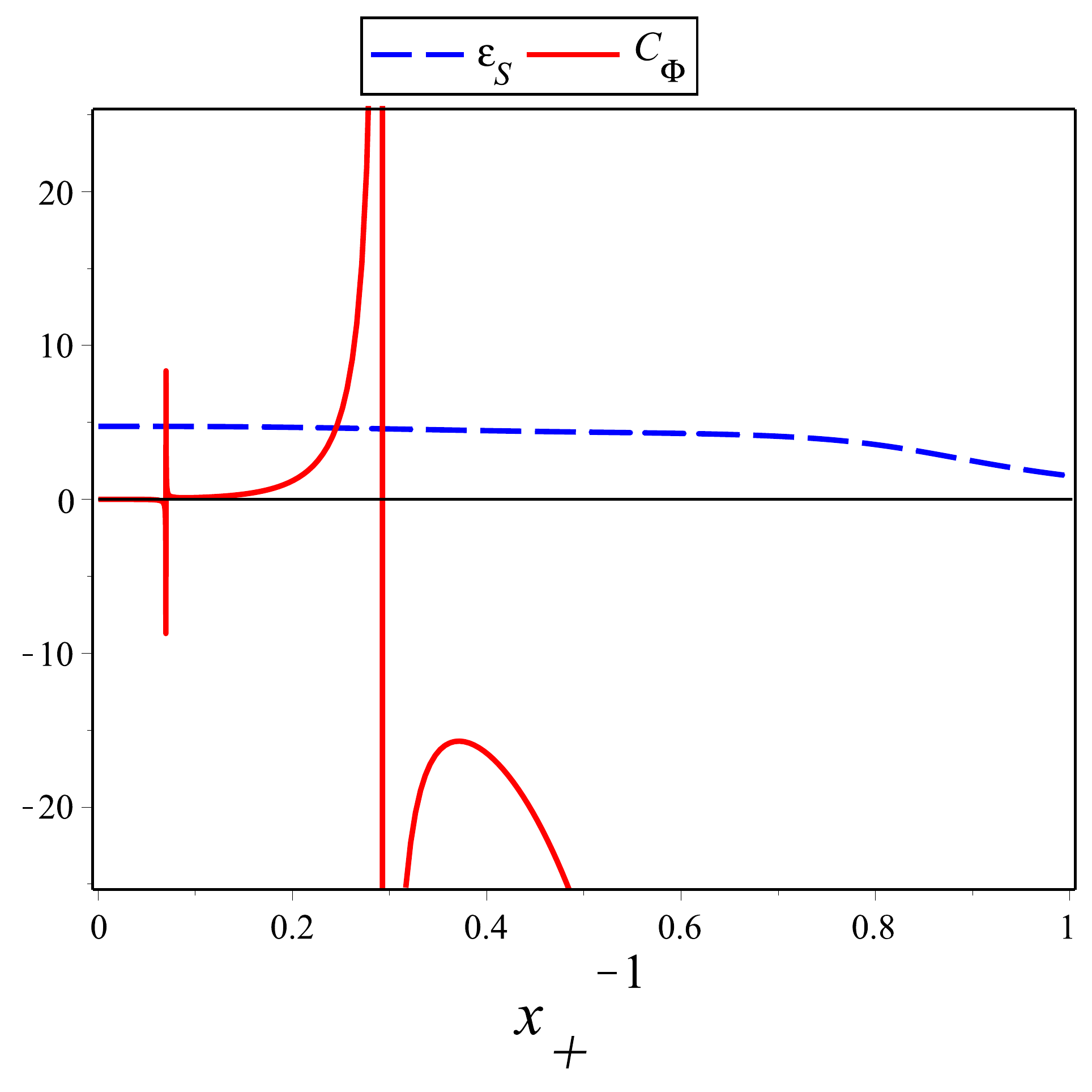}}\qquad\qquad
	\subfigure[$T_{\infty}<T$]{\includegraphics[width=4.8cm]{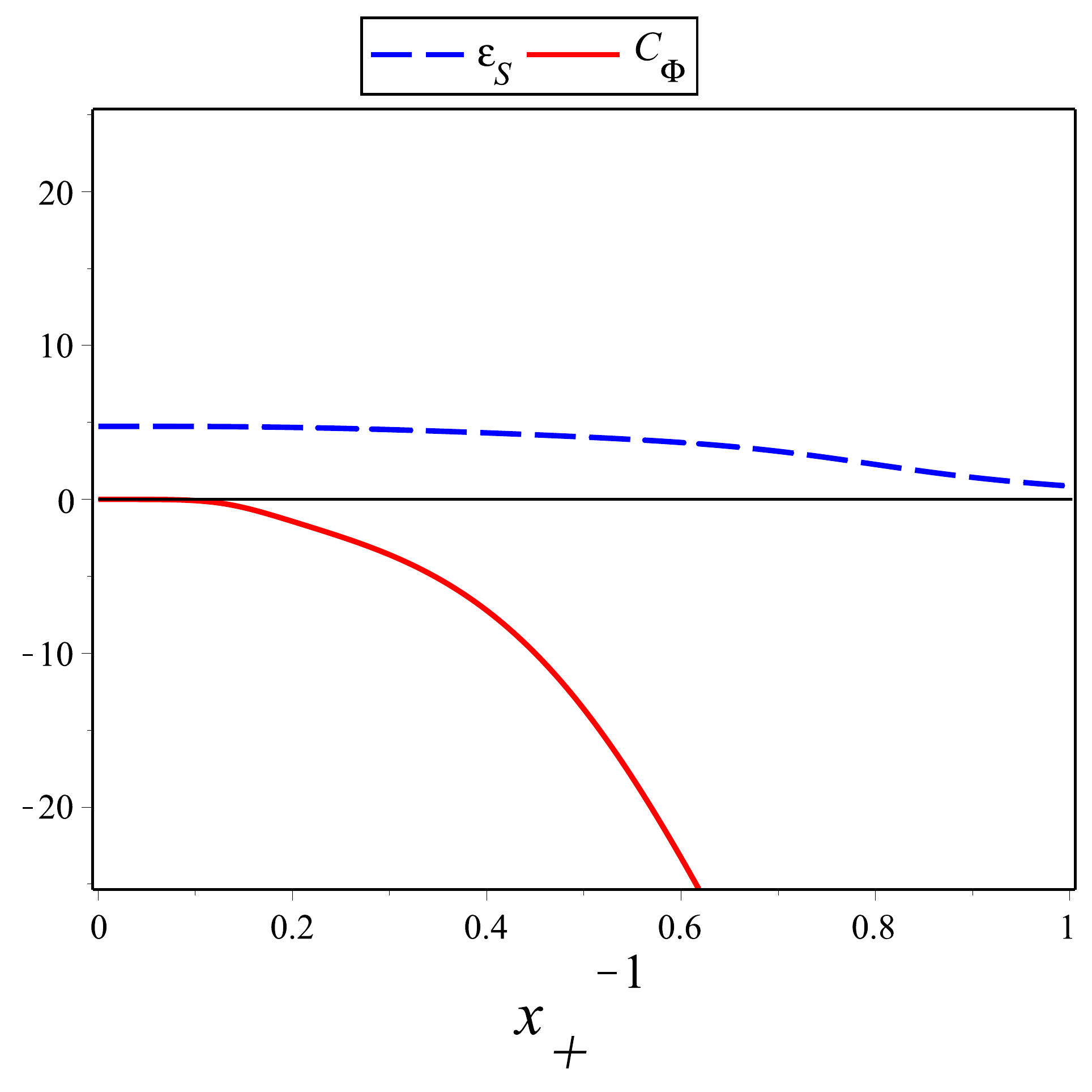}}
	\caption{\small Response functions $\epsilon_{S}$ y $C_{\Phi}$ vs the inverse of the horizon coordinate $x_{+}$ (which accordingly is ranging the positive branch) for different values of the conjugate potential, as in \textbf{(a)}, \textbf{(b)} and \textbf{(c)}, and for temperature below and above the critical temperature $T_\infty$, as in \textbf{(d)} and \textbf{(e)}.}
	\label{fig:fr_gc1}
\end{figure}

\begin{figure}[t!]
	\centering
	\subfigure[$T<T_{\infty}$]{\includegraphics[width=4.6cm]{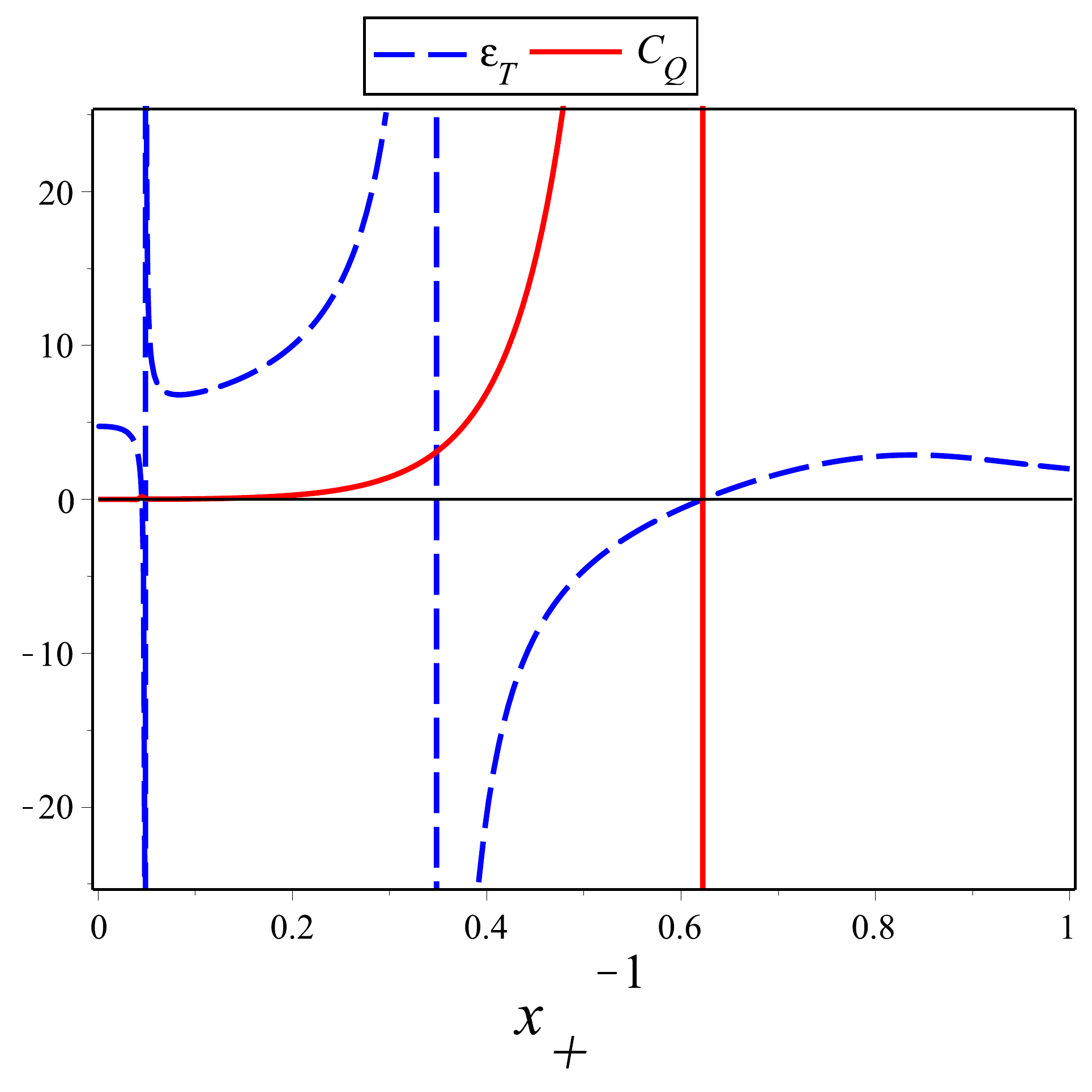}}
	\subfigure[$T_{\infty}<T<T_{0}$]{\includegraphics[width=4.6cm]{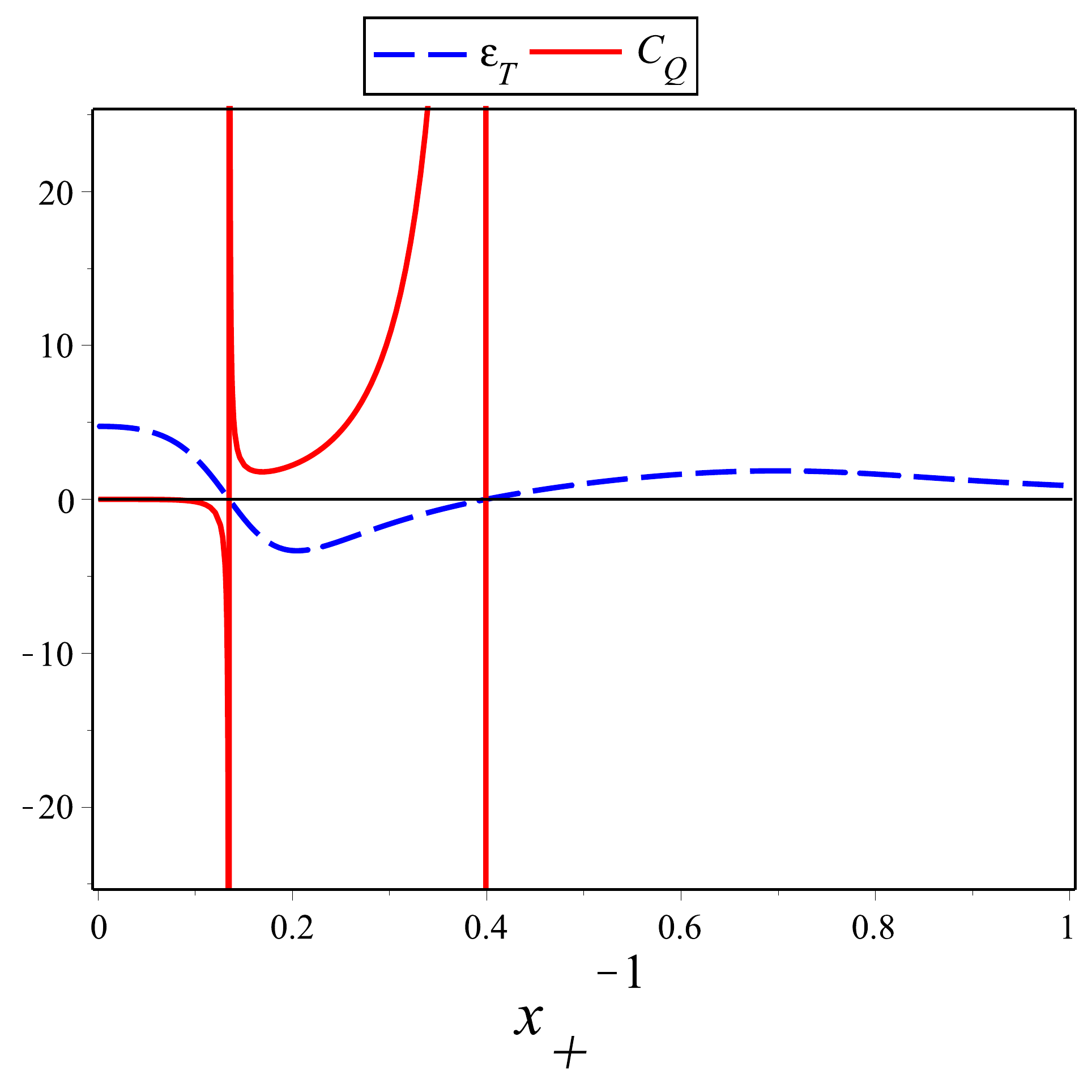}}
	\subfigure[$T_{0}<T$]{\includegraphics[width=4.6cm]{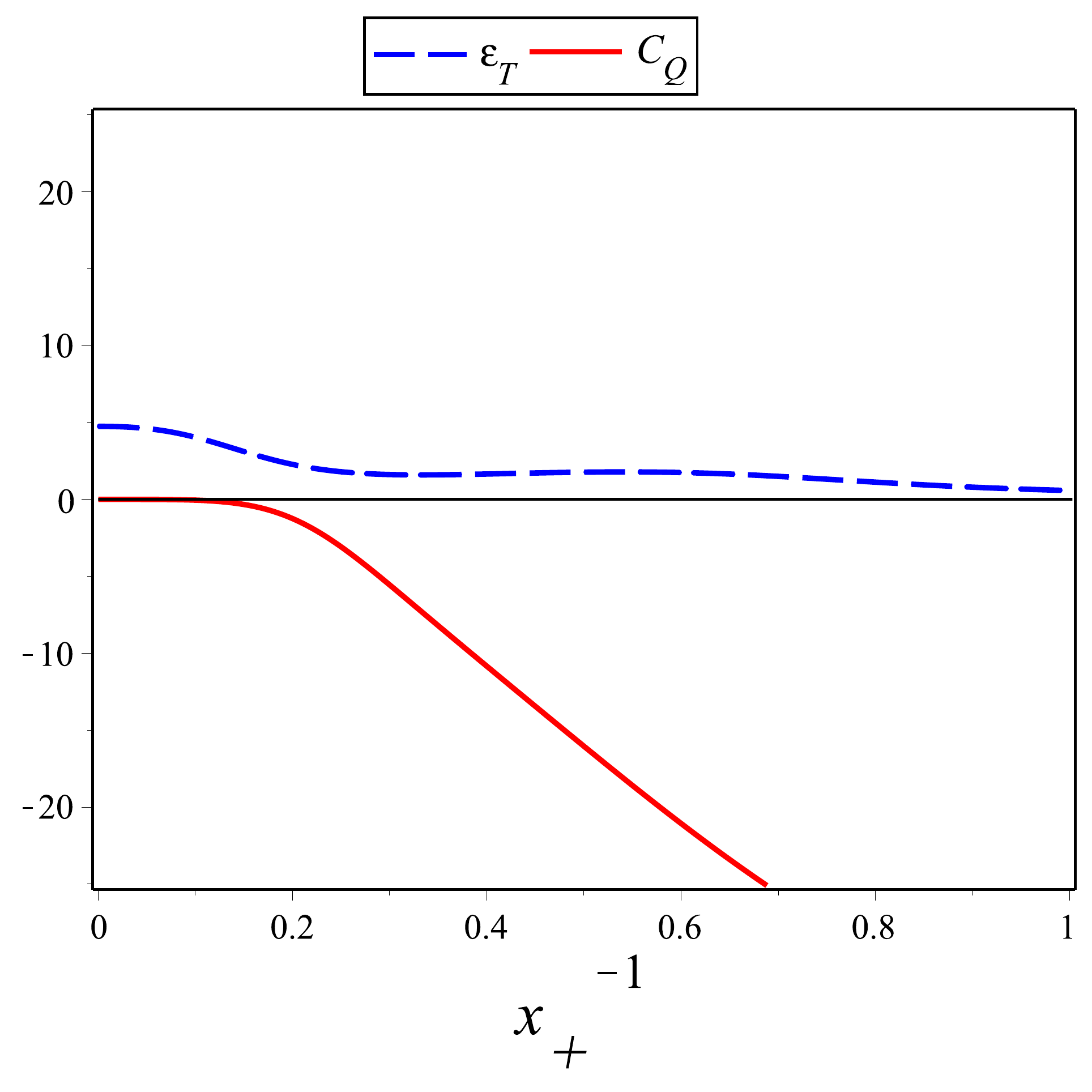}}
	\subfigure[$Q<Q_{0}$]{\includegraphics[width=4.8cm]{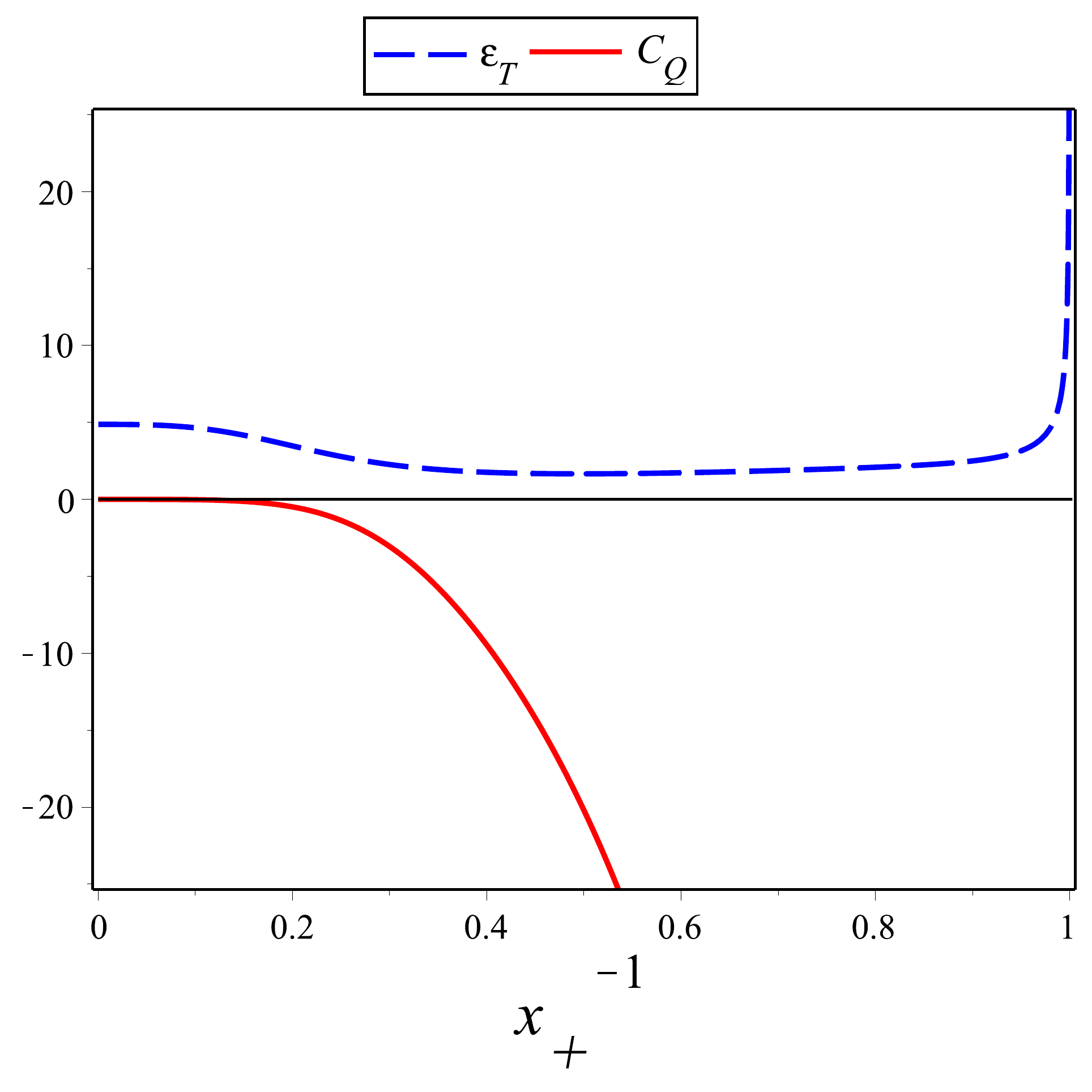}}\qquad\qquad
	\subfigure[$Q_{0}<Q$]{\includegraphics[width=4.8cm]{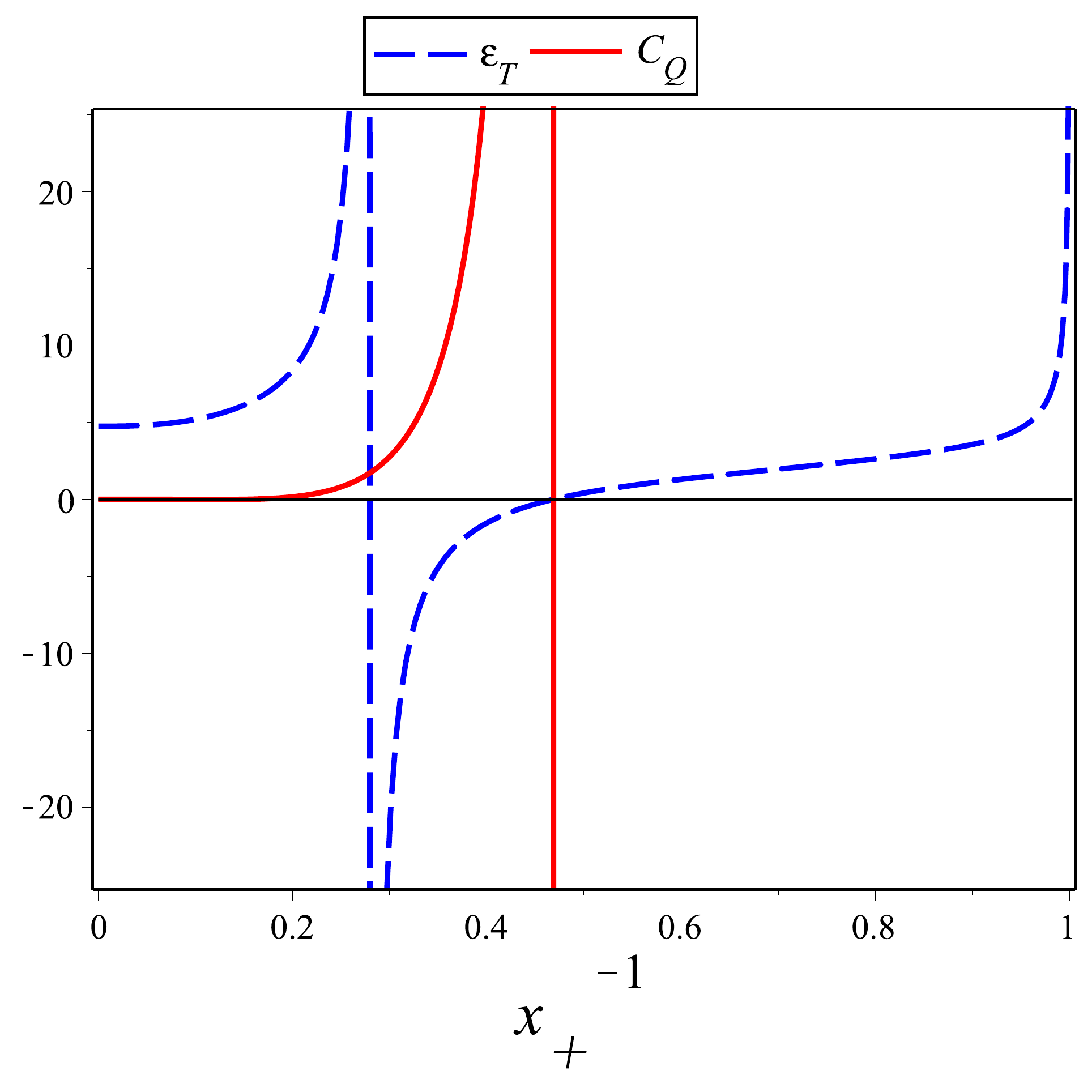}}
	\caption{\small Response functions $\epsilon_{T}$ y $C_{Q}$ vs the inverse of the horizon coordinate $x_{+}$ for different values of the temperature, as in \textbf{(a)}, \textbf{(b)} and \textbf{(c)}, and for electric charge below and above the critical charge $Q_0$, as in \textbf{(d)} and \textbf{(e)}.}
	\label{fig:fr_c1}
\end{figure}

For completeness, let us finally consider the particular case of $\nu=3$ and present, in Fig. \ref{fig:fr_gc1} and Fig. \ref{fig:fr_c1}, the plots of the response functions in a given ensemble with respect to the location of the event horizon. We would like to directly check the existence of stability regions and to compare with the results obtained by the general criterion presented in Section \ref{sec4}.

Let us begin with the grand canonical ensemble, where $\Phi$ is fixed. In this case, we observe that the thermodynamic stable black holes are found within two different intervals for the conjugate potential: $\Phi_0<\Phi<\Phi^{*}$ (see Fig. \ref{fig:fr_gc1}{\bf b}) and $\Phi^{*}<\Phi$ (see Fig. \ref{fig:fr_gc1}{\bf c}). The former corresponds to intermediate-size black holes between the two values of $x_+$ where $C_\Phi$ is discontinuous. This region can also be identified in Fig. \ref{fig:Phicrit_ST}{\bf b} where the slopes in $S$ vs $T$ are positive definite. This interesting new phase, which is not present in the case $\nu\rightarrow\infty$ studied previously in \cite{Astefanesei:2019mds}, contains configurations similar to Schwarzschild black hole in the sense that they have only a horizon. The region of stable black holes with $\Phi^{*}<\Phi$ can also be identified in Fig. \ref{fig:Phicrit_ST}{\bf b} and, clearly, in this case there is a well defined extremal limit. As argued at the end of Section \ref{sec4}, stable black holes are found for $T<T_\infty$, as shown in Fig. \ref{fig:Phicrit_ST}{\bf d} for a particular $0<T<T_\infty$. Therefore, if $T_\infty<T$, there do not exist stable configurations, as shown in Fig. \ref{fig:fr_gc1}{\bf e}.

A similar analysis can be done for the canonical ensemble when $Q$ is fixed. First, we observe again that the stable black holes appear for $T<T_\infty$, as shown in Fig. \ref{fig:fr_c1}{\bf a}. This is consistent with Fig. \ref{fig:FA_PQ_ST}{\bf b} for the isotherm $T_A<T_\infty$. However, for this isotherm, there exist two branches with $C_Q>0$: one for $Q_0<Q<Q^{*}$ with solutions with one horizon and the other one for $Q^{*}<Q$ with stable black holes that have a well defined extremal limit $T=0$. This can also be observed in Fig. \ref{fig:fr_c1}{\bf e}, where $Q_0<Q$. We remark that if $Q<Q_0$, there are no stable black holes.

To  conclude,  let  us  emphasize that not just the AdS arena stabilizes the thermodynamics of black holes, but also the existence of a dilaton potential in flat spacetime can have the same effect.  As a first step, we have shown that the charged hairy black holes in flat space, besides being thermodynamically and perturbatively dynamically stable, have a rich phase structure including critical points and swallow tails as in AdS spacetime \cite{Chamblin:1999tk,Chamblin:1999hg}. A more detailed analysis of their thermodynamics properties is going to be presented in a future work \cite{Astefanesei:2020toappear}.

%%%%%%%%%%%%%%%%%%%%%%%%%%%%%%
\section*{Acknowledgements}
%%%%%%%%%%%%%%%%%%%%%%%%%%%%%%
%
We would like to thank Andr\'es Anabal\'on, David Choque, Carlos Herdeiro, Jutta Kunz, Robert Mann, and  Eugen Radu for interesting discussions at various stages of this work and collaboration on related projects.
The research of DA is supported by the Fondecyt Grants 1200986, 1170279, 1171466, and 2019/13231-7 Programa de Cooperacion Internacional, ANID. JLBS gratefully acknowledges support from DFG Project No. BL 1553, DFG Research Training Group 1620  \textit{Models of Gravity} and the COST Action CA16104 \textit{GWverse}. The research of RR was partially supported by the Ph.D. scholarship CONICYT (now succeeded by ANID) 21140024.
%%%%%%%%%%%%%%%%%%%%%%

\end{document}